\documentclass[%
superscriptaddress,
reprint,
amsmath,amssymb,
aps,
prd,
floatfix,
notitlepage,
showpacs,
]{revtex4-2}
\usepackage{graphicx}
\usepackage{hyperref}
\usepackage{xcolor}
\usepackage{booktabs}
\usepackage{multirow}
\usepackage{siunitx}
\newcommand{\Id}{\mbox{1\hspace{-1.2mm}I}}

\newcommand{\bea}{\begin{eqnarray}}
\newcommand{\eea}{\end{eqnarray}}
\newcommand{\BAN}{\begin{eqnarray*}}
\newcommand{\EAN}{\end{eqnarray*}}

\newcommand{\blue}{\color{blue}}

\usepackage{titlesec}
\titleformat{\paragraph}
{\normalfont\normalsize\bfseries}{\theparagraph}{1em}{}
\titlespacing*{\paragraph}
{0pt}{3.25ex plus 1ex minus .2ex}{1.5ex plus .2ex}

\begin{document}

\newcommand{\ASIOP}
{Institute of Physics, Academia Sinica, Taipei, Taiwan~11529, Republic of China}

\newcommand{\NTU}
{Department of Physics, National Taiwan University, \\ 
Taipei, Taiwan~10617, Republic of China}

\newcommand{\NTNU}
{Department of Physics, National Taiwan Normal University, \\  
Taipei, Taiwan~11677, Republic of China}

\newcommand{\CTP}
{Center for Theoretical Physics, Department of Physics, National Taiwan University, \\ 
Taipei, Taiwan~10617, Republic of China}

\newcommand{\NCTS}
{Physics Division, National Center for Theoretical Sciences,  \\
Taipei, Taiwan~10617, Republic of China}

\title{Symmetries of spatial correlators of light and heavy mesons in high temperature lattice QCD}

\author{Ting-Wai~Chiu}
\email{twchiu@phys.ntu.edu.tw}
\affiliation{\NTNU}
\affiliation{\ASIOP}
\affiliation{\NCTS}
\affiliation{\CTP}

\begin{abstract}

The spatial $z$ correlators of meson operators 
in $N_f=2+1+1$ lattice QCD with optimal domain-wall quarks at the physical point 
are studied for seven temperatures in the range of 190--1540 MeV. 
The meson operators include a complete set of Dirac bilinears   
(scalar, pseudoscalar, vector, axial vector, tensor vector, and axial-tensor vector), 
and each for six flavor combinations 
($\bar u d$, $\bar u s$, $\bar s s$, $\bar u c$, $\bar s c$, and $\bar c c$).
In a recent paper [Phys. Rev. D \textbf{107}, 114501 (2023)], 
I focused on the meson correlators of $u$ and $d$ quarks, 
and discussed their implications for the effective restoration  
of $SU(2)_L \times SU(2)_R$ and $U(1)_A$ chiral symmetries,   
as well as the emergence of approximate $SU(2)_{CS}$ chiral spin symmetry.  
In this work, we extend our study to meson correlators of six flavor contents,      
and first observe the hierarchical restoration of chiral symmetries in QCD, 
from $SU(2)_L \times SU(2)_R \times U(1)_A $  
to $SU(3)_L \times SU(3)_R \times U(1)_A $, and to $SU(4)_L \times SU(4)_R \times U(1)_A $,  
as the temperature is increased from 190 to 1540 MeV. 
Moreover, we compare the temperature windows for the emergence of the approximate 
$SU(2)_{CS}$ symmetry in light and heavy vector mesons, and find that  
the temperature windows are dominated by the $(\bar u c, \bar s c, \bar c c)$ sectors.

\end{abstract}

\maketitle

\section{Introduction}
\label{intro}

Understanding the nature of strongly interacting matter at high temperatures is crucial for 
uncovering the mechanisms governing matter creation in the early universe 
and elucidating the outcomes of relativistic heavy ion collision experiments 
such as those at LHC and RHIC, as well as those of electron ion collision experiments
at the planned electron-ion colliders. A first step in this pursuit is to find out  
the symmetries in quantum chromodynamics (QCD) at high temperatures, 
which are essential in determining the properties and dynamics of matter 
under extreme conditions.

First, consider QCD with $N_f$ massless quarks. Its action possesses the 
$ SU(N_f)_L \times SU(N_f)_R \times U(1)_A $ chiral symmetry. 
At low temperatures $T < T_c^0$ 
(where $T_c^0$ depends on $N_f$, and the superscript "0" denotes zero quark mass),
quarks and gluons are confined in hadrons, and the $ SU(N_f)_L \times SU(N_f)_R $ chiral symmetry 
is spontaneously broken down to $SU(N_f)_V$ by the vacuum of QCD, with nonzero chiral condensate. 
Moreover, the $U(1)_A$ axial symmetry is explicitly broken by the chiral anomaly, 
due to the quantum fluctuations of topologically nontrivial gauge fields. 
As the temperature is increased above $T_c^0$, the chiral condensate becomes zero, 
and the $ SU(N_f)_L \times SU(N_f)_R $ chiral symmetry is restored. 
Furthermore, the $U(1)_A$ axial symmetry is effectively restored at $T \gtrsim T_1^0 \gtrsim T_c^0 $, 
due to the suppression of the quantum fluctuations of topologically nontrivial gauge configurations 
at high temperatures. 
So far, it is still an open question whether $ T_1^0 > T_c^0$ or $T_1^0 \simeq T_c^0 $.
Here we define $T_{c1}^0$ to be the maximum of $T_c^0$ and $T_1^0$,  
\bea
\label{eq:Tc1_0}
T_{c1}^0 \equiv \max(T_c^0, T_1^0), 
\eea
such that the theory possesses the $SU(N_f)_L \times SU(N_f)_R \times U(1)_A$ chiral symmetry
for $T > T_{c1}^0$. 

Next, consider QCD with physical $(u, d, s, c, b)$ quarks. 
Its action does not possess the $SU(N)_L \times SU(N)_R \times U(1)_A $ chiral symmetry 
for any integer $N$ from 2 to 5, due to the explicit breakings of the nonzero quark masses.
However, as $T$ is increased successively, each quark acquires thermal energy of the order of $\pi T$, 
and eventually its rest mass energy becomes negligible when $ \pi T \gg m_q $. 
Also, since the quark masses range from a few MeV to a few GeV, it follows that  
as the temperature is increased successively, the chiral symmetry is restored hierarchically from 
$SU(2)_L \times SU(2)_R \times U(1)_A $ of $(u, d)$ quarks 
to $SU(3)_L \times SU(3)_R \times U(1)_A $ of $(u, d, s)$ quarks,  
then to $SU(4)_L \times SU(4)_R \times U(1)_A $ of $(u, d, s, c)$ quarks, 
and finally to $SU(5)_L \times SU(5)_R \times U(1)_A $ of $(u,d,s,c,b)$ quarks. 
Since the restoration of chiral symmetries is manifested by the degeneracies 
of meson $z$ correlators (as well as other observables),    
we can use the splittings of the meson $z$ correlators of the symmetry multiplets to examine the 
realization of the hierarchical restoration of chiral symmetries in high temperature QCD.
Strictly speaking, these chiral symmetries should be regarded as ``emergent" symmetries rather than 
``restored" symmetries, since the QCD action with physical quark masses does not possess
chiral symmetries at all. In the following, it is understood that ``restoration of chiral symmetries"  
stands for ``emergence of chiral symmetries". Similar to (\ref{eq:Tc1_0}), we define 
\bea
\label{eq:Tc1_qQ}
T_{c1}^{\bar q Q} \equiv \max( T_c^{\bar q Q}, T_1^{\bar q Q}),  
\eea
where $T_c^{\bar q Q}$ ($T_1^{\bar q Q}$) is the temperature for the manifestation of 
$SU(2)_L \times SU(2)_R$ ($U(1)_A$) chiral symmetry via the meson $z$ correlators 
with flavor content $\bar q Q$.  
Then, for $T > T_{c1}^{\bar q Q}$, the theory possesses the $SU(2)_L \times SU(2)_R \times U(1)_A$ 
chiral symmetry of the $\bar q Q$ sector.

Note that since 1987 \cite{Detar:1987kae}, there have been many lattice studies using 
the screening masses of meson $z$ correlators to investigate the effective restoration of 
$U(1)_A$ and $ SU(2)_L \times SU(2)_R$ chiral symmetries 
of $u$ and $d$ quarks in high temperature QCD, see, e.g., Ref. \cite{Bazavov:2019www} 
and references therein. However, so far, there seems no discussions in the literature 
about the hierarchical restoration of chiral symmetries in high temperature QCD, 
except for a brief mention in Ref. \cite{Chiu:2023hnm}. 

In this work, we investigate the hierarchical restoration of chiral symmetries 
in $N_f=2+1+1$ lattice QCD with optimal domain-wall quarks at the physical point.        
We first observe the hierarchical restoration of chiral symmetries 
from $SU(2)_L \times SU(2)_R \times U(1)_A $ of $(u,d)$ quarks,  
to $SU(3)_L \times SU(3)_R \times U(1)_A $ of $(u,d,s)$ quarks, 
and finally to $SU(4)_L \times SU(4)_R \times U(1)_A $ of $(u,d,s,c)$ quarks,  
as the temperature is increased from 190 to 1540 MeV.  
We compute the meson $z$ correlators for a complete set of Dirac bilinears
(scalar, pseudoscalar, vector, axial vector, tensor vector, and axial-tensor vector), 
and each for six combinations of quark flavors 
($\bar u d$, $\bar u s$, $\bar s s$, $\bar u c$, $\bar s c$, and $\bar c c $).
Then we use the degeneracies of meson $z$ correlators 
to investigate the hierarchical restoration of chiral symmetries in high temperature QCD. 

The relationship between the $SU(2)_L \times SU(2)_R$ and $U(1)_A$ chiral symmetries
and the degeneracy of meson $z$ correlators for $(u, d)$ quarks in $N_f=2+1+1 $ QCD 
has been outlined in Ref. \cite{Chiu:2023hnm}, 
and we follow the same conventions/notations therein.  
In this study, following Ref. \cite{Chiu:2023hnm}, we also neglect the disconnected diagrams 
in the meson $z$ correlators.  
%
%
With this approximation, one can straightforwardly deduce the relationship between the 
$SU(N)_L \times SU(N)_R$ and $U(1)_A$ chiral symmetries of $N$ ($ 2 \le N \le N_f$) quarks
and the degeneracy of meson $z$ correlators, in QCD with $N_f$ quarks, as follows.  
The restoration of $SU(N)_L \times SU(N)_R$ chiral symmetry of $N$ quarks    
is manifested by the degeneracies of meson $z$ correlators in the vector and axial-vector channels, 
$C_{V_k}^{\bar q_i q_j}(z) = C_{A_k}^{\bar q_i q_j}(z)$, $(k=1,2,4)$, 
for {\it all} flavor combinations ($\bar q_i q_j$, $i,j=1, \cdots, N$). 
The effective restoration of the $U(1)_A$ symmetry of $N$ quarks     
is manifested by the degeneracies of meson $z$ correlators in the pseudoscalar and scalar channels, 
$C_{P}^{\bar q_i q_j}(z) = C_{S}^{\bar q_i q_j}(z)$, 
as well as in the tensor vector and axial-tensor vector channels, 
$C_{T_k}^{\bar q_i q_j}(z) = C_{X_{k}}^{\bar q_i q_j}(z)$, $(k=1,2,4)$, 
for {\it all} flavor combinations ($\bar q_i q_j$, $i,j=1, \cdots, N$). 
At this point, we recall the studies of the symmetries and meson correlation functions 
in high temperature QCD with $N_f$ massless quarks \cite{Evans:1996wf,Birse:1996dx}, 
in which one of the salient results is that the correlator 
of the flavor nonsinglet pseudoscalar meson $\bar q \gamma_5 \lambda_a q$ 
is equal to that of the flavor singlet pseudoscalar meson $\bar q \gamma_5 q$, 
for QCD with $N_f > 2$ at $ T > T_c$. This implies that the disconnected diagrams 
do not have contributions to meson $z$ correlators in QCD with $N_f > 2$ massless quarks at $ T > T_c$. 
%
%
However, at this moment, it is unknown to what extent the disconnected diagrams are suppressed
in QCD with $N_f=2+1(+1)(+1)$ physical quarks. We will address this question with noise estimation 
of all-to-all quark propagators, and will report our results in the future.

Besides the hierarchical restoration of chiral symmetries,
we are also interested in the question whether there are any (approximate) emergent symmetries 
which are not the symmetries of the entire QCD action but only a part of it,  
e.g., the $SU(2)_{CS}$ chiral spin symmetry 
(with $U(1)_A$ as a subgroup) \cite{Glozman:2014mka,Glozman:2015qva},  
which is only a symmetry of chromoelectric part of the quark-gluon interaction, and also the color charge.   
Since the free fermions as well as the chromomagnetic part of the quark-gluon interaction 
do not possess the $SU(2)_{CS}$ symmetry, its emergence in high temperature QCD 
suggests the possible existence of hadronlike objects which are predominantly 
bound by chromoelectric interactions.
The $SU(2)_{CS}$ symmetry was first observed to manifest approximately 
in the multiplets of $z$ correlators of vector mesons,  
at temperatures $T \sim 220$--$500$~MeV in $N_f=2$ lattice QCD with domain-wall fermions 
\cite{Rohrhofer:2019qwq}. In Ref. \cite{Chiu:2023hnm}, we studied the emergence of 
$SU(2)_{CS}$ chiral-spin symmetry in $N_f=2+1+1$ lattice QCD 
with optimal domain-wall quarks at the physical point, 
and found that the $SU(2)_{CS}$ symmetry breaking in $N_f=2+1+1$ lattice QCD 
is larger than that in $N_f=2$ lattice QCD at the same temperature, 
for both $z$ correlators and $t$ correlators of vector mesons of $u$ and $d$ quarks. 
In this paper, we extend our study to vector meson $z$ correlators 
of all flavor combinations ($\bar u d$, $\bar u s$, $\bar s s $, $\bar u c$, $\bar s c$, $\bar c c $) 
in $N_f=2+1+1$ lattice QCD at the physical point, and compare the emergence of approximate $SU(2)_{CS}$ 
chiral spin symmetry between different flavor sectors.

The outline of this paper is as follows. 
In Sec. \ref{lattice}, the hybrid Monte Carlo simulation of $N_f=2+1+1$ lattice QCD 
with optimal domain-wall quarks at the physical point is briefly outlined, and the 
essential features and parameters of the seven gauge ensembles for this study are summarized. 
In Sec. \ref{kappa}, the symmetry-breaking parameters for measuring the precision of 
various symmetries with the splittings of the $z$ correlators of the symmetry partners are defined. 
The results of meson $z$ correlators for six flavor combinations and seven 
temperatures in the range of 190-1540 MeV are presented in Sec. \ref{meson_Cz}, 
while the corresponding results of symmetry-breaking parameters 
are presented in Sec. \ref{kappa_all}. 
The realization of hierarchical restoration of chiral symmetries 
in $N_f=2+1+1$ QCD, from $SU(2)_L \times SU(2)_R \times U(1)_A $  
to $SU(3)_L \times SU(3)_R \times U(1)_A $, and to $SU(4)_L \times SU(4)_R \times U(1)_A $,  
as the temperature is increased from 190 to 1540 MeV, is demonstrated in Sec. \ref{HRCS}.
The temperature windows for the approximate $SU(2)_{CS}$ symmetry 
of six flavor combinations are presented in Sec. \ref{SU2_CS}, 
which reveal the dominance of heavy vector meson channels of $(\bar u c, \bar s c, \bar c c)$ sectors.
In Sec. \ref{conclusions}, we conclude with some remarks.

\section{Gauge ensembles} 
\label{lattice}

The gauge ensembles in this study are generated by hybrid Monte Carlo (HMC) simulation 
of lattice QCD with $N_f=2+1+1$ optimal domain-wall quarks \cite{Chiu:2002ir} 
at the physical point, on the $32^3 \times (16,12,10,8,6,4,2)$ lattices, 
with the plaquette gauge action at $\beta = 6/g^2 = 6.20 $.
This set of ensembles is generated with the same 
actions \cite{Chiu:2011bm,Chen:2014hyy} and algorithms as their counterparts 
on the $64^3 \times (20,16,12,10,8,6)$ lattices \cite{Chen:2022fid}, 
but with one-eighth of the spatial volume. 
The simulations are performed on a GPU cluster with various Nvidia GPUs. 
For each ensemble, after the initial thermalization, a set of gauge configurations 
is sampled and distributed to 16--32 simulation units, 
and each unit performed an independent stream of HMC simulation. 
For each HMC stream, one configuration is sampled every five trajectories. 
Finally collecting all sampled configurations from all HMC streams gives 
the total number of configurations of each ensemble. 
The lattice parameters and statistics of the gauge ensembles for computing 
the meson $z$ correlators in this study are summarized in Table \ref{tab:7_ensembles}.
The temperatures of these six ensembles are in the range $\sim 190$--$1540$~MeV, all above 
the pseudocritical temperature $ T_c \sim 150 $~MeV.

\begin{table*}[tbp]
\begin{center}
\caption{The lattice parameters and statistics of the seven gauge ensembles 
for computing the meson correlators.
The last three columns are the residual masses of $u/d$, $s$, and $c$ quarks.}
\setlength{\tabcolsep}{3pt}
\vspace{2mm}
\begin{tabular}{|cccccccccccc|}
\hline 
    $\beta$
  & $a$[fm]
  & $ N_x $
  & $ N_t $
  & $ m_{u/d} a $
  & $ m_{s} a $
  & $ m_{c} a $
  & $T$[MeV]
  & $N_{\rm confs}$
  & $ (m_{u/d} a)_{\rm res} $
  & $ (m_{s} a)_{\rm res} $
  & $ (m_{c} a)_{\rm res} $
\\
\hline
\hline
6.20 & 0.0641 & 32 & 16 & 0.00125 & 0.040 & 0.550 & 192 & 591 
& $ 1.9(2) \times 10^{-5}$ & $1.5(2) \times 10^{-5}$ & $ 4.3(7) \times 10^{-6} $\\
6.20 & 0.0641 & 32 & 12 & 0.00125 & 0.040 & 0.550 & 257 & 514 
& $1.9(2) \times 10^{-5}$ & $1.6(1) \times 10^{-5}$ & $ 3.8(5) \times 10^{-6} $\\
6.20 & 0.0641 & 32 & 10 & 0.00125 & 0.040 & 0.550 & 308 & 481 
& $5.7(7) \times 10^{-6}$ & $5.1(6) \times 10^{-6}$ & $ 1.4(2) \times 10^{-6} $\\
6.20 & 0.0641 & 32 & 8  & 0.00125 & 0.040 & 0.550 & 385 & 468 
& $6.3(9) \times 10^{-6}$ & $6.0(7) \times 10^{-6}$ & $ 3.0(9) \times 10^{-6} $\\
6.20 & 0.0641 & 32 & 6  & 0.00125 & 0.040 & 0.550 & 513 & 431 
& $5.8(9) \times 10^{-6}$ & $5.6(8) \times 10^{-6}$ & $ 3.4(7) \times 10^{-6} $\\
6.20 & 0.0641 & 32 & 4  & 0.00125 & 0.040 & 0.550 & 770 & 991 
& $1.2(2) \times 10^{-6}$ & $1.2(2) \times 10^{-6}$ & $ 1.2(2) \times 10^{-6} $\\
6.20 & 0.0641 & 32 & 2  & 0.00125 & 0.040 & 0.550 & 1540 & 770 
& $8.0(5) \times 10^{-7}$ & $7.8(5) \times 10^{-7}$ & $ 8.2(5) \times 10^{-7} $\\
\hline
\end{tabular}
\label{tab:7_ensembles}
\end{center}
\end{table*}

The lattice spacing and the $(u/d, s, c)$ quark masses are determined on the 
the $ 32^3 \times 64 $ lattices with 427 configurations.  
The lattice spacing is determined using the Wilson flow \cite{Narayanan:2006rf,Luscher:2010iy}
with the condition $ \{ t^2 \langle E(t) \rangle \} |_{t=t_0} = 0.3 $
and the input $ \sqrt{t_0} = 0.1416(8) $~fm \cite{Bazavov:2015yea}.
The physical $(u/d, s, c)$ quark masses are obtained by tuning their masses such that 
the masses of the lowest-lying states extracted from 
the time-correlation functions of the meson operators
$ \{ \bar u \gamma_5 d, \bar s \gamma_i s, \bar c \gamma_i c \} $  
are in good agreement with the physical masses of 
$\pi^{\pm}(140) $, $\phi(1020)$, and $J/\psi(3097)$. 

The chiral symmetry breaking due to finite $N_s=16$ (in the fifth dimension)
can be measured by the residual mass of each quark flavor \cite{Chen:2012jya}, 
as given in the last three columns of Table \ref{tab:7_ensembles}.
The residual masses of $(u/d, s, c)$ quarks are
less than $(1.5\%, 0.04\%, 0.001\%)$ of their bare masses, 
amounting to less than (0.06, 0.05, 0.02) MeV/$c^2$, respectively. 
This asserts that the chiral symmetry is well preserved such that the deviation
of the bare quark mass $m_q$ is sufficiently small in the
effective 4D Dirac operator of optimal domain-wall fermion, for both light and heavy quarks.
In other words, the chiral symmetry in the simulations is sufficiently precise
to guarantee that the hadronic observables (e.g., meson correlators) 
can be evaluated to high precision, with the associated uncertainty
much less than those due to statistics and other systematics.

\section{Symmetry breaking parameters}
\label{kappa}

In order to give a quantitative measure for the manifestation of symmetries from the degeneracy
of meson $z$ correlators with flavor content $\bar q Q$, we consider the symmetry breaking parameters 
as follows. To this end, we write the meson $z$ correlators as functions of the dimensionless variable
\bea
zT = (n_z a)/(N_t a) = n_z/N_t,
\label{eq:zT}
\eea
where $ T $ is the temperature.

In general, the degeneracy of any two meson $z$ correlators 
$C_A(zT)$ and $C_B(zT)$ with flavor content $\bar q Q$ (where  
subscripts $A$ and $B$ denote their Dirac matrices with definite transformation properties, 
and the flavor content $\bar q Q$ is suppressed) can be measured by the symmetry breaking parameter 
\bea
\label{eq:kappa_AB}
\kappa_{AB}(zT) = \frac{\left| C_A(zT)-C_B(zT) \right|}{C_A(zT)+C_B(zT)}, \hspace{4mm}  z > 0.
\eea
If $ C_A $ and $ C_B $ are exactly degenerate at $T$, then $\kappa_{AB} = 0 $ for any $z$,
and the symmetry is effectively restored at $T$.
On the other hand, if there is any discrepancy between $C_A$ and $C_B$ at any $z$,
then $\kappa_{AB} $ is nonzero at this $z$, and the symmetry is not exactly restored at $T$. 
Here the denominator of (\ref{eq:kappa_AB}) serves as (re)normalization and 
the value of $\kappa_{AB}$ is bounded between zero and one. 
Obviously, this criterion is more stringent than the equality of the screening masses,  
$m_A^{scr} = m_B^{scr}$, which are extracted from $C_A$ and $C_B$ at large $z$.

Note that $\kappa_{AB}$ in (\ref{eq:kappa_AB}) can be written as 
\BAN
\kappa_{AB}(zT) = \frac{\left| 1 - C_B(zT)/C_A(zT) \right|}{ 1 + C_B(zT)/C_A(zT)}, 
\EAN
which is different from the $\kappa_{AB}(zT) = \left| 1 - C_B(zT)/C_A(zT) \right| $ 
used in Ref. \cite{Chiu:2023hnm}. 
Also, all $z$ correlators in (\ref{eq:kappa_AB}), as well as those shown 
in Figs. \ref{fig:Cz_T192}--\ref{fig:Cz_T1540} are {\it unnormalized}, 
while those in Ref. \cite{Chiu:2023hnm} are {\it normalized} by their values 
at $z/a = 1$ [i.e., $C_\Gamma(zT)=1$ at $z/a = 1$]. 
The former avoids any ``accidental" degeneracies due to the normalization.
In the following, any symmetry breaking parameter to measure the degeneracy of 
two meson $z$ correlators is always defined according to (\ref{eq:kappa_AB}).

\subsection{$SU(2)_L \times SU(2)_R$ and $U(1)$ symmetry breaking parameters}

According to (\ref{eq:kappa_AB}), the $SU(2)_L \times SU(2)_R $ symmetry breaking parameter 
can be written as  
\bea
\label{eq:k_VA_z}
&& \kappa_{VA}(zT) = \frac{\left| C_{V_k}(zT) - C_{A_k}(zT) \right| }{C_{V_k}(zT) + C_{A_k}(zT)}, \nonumber \\
&& \hspace{12mm} z > 0,  \hspace{4mm} (k=1,2,4).  
\eea
Due to the $S_2$ symmetry of the $z$ correlators, it only needs 
to examine $k=1$ and $k=4$ components of (\ref{eq:k_VA_z}). 
In general, the difference between $k=1$ and $k=4$ components of (\ref{eq:k_VA_z}) is negligible,   
thus in the following, we only give the results of (\ref{eq:k_VA_z}) with $k=1$. 
 
In general, to determine to what extent the 
$SU(2)_L \times SU(2)_R$ chiral symmetry is manifested in the $z$ correlators, 
it is necessary to examine whether $\kappa_{VA}$ is sufficiently small. 
To this end, we use the following criterion for the manifestation of 
$SU(2)_L \times SU(2)_R$ chiral symmetry at $T$ for a fixed $zT$
\bea
\label{eq:SU2_crit_z}
\kappa_{VA}(zT) \le \epsilon_{VA},   
\eea
where $ \epsilon_{VA} $ is a small parameter which defines the precision of the chiral symmetry. 
For fixed $zT$ and $\epsilon_{VA}$, the temperature $T_c$ for the manifestation of 
the $SU(2)_L \times SU(2)_R$ symmetry is the lowest temperature satisfying (\ref{eq:SU2_crit_z}), i.e., 
\bea
\label{eq:Tc_epsilon} 
\kappa_{VA}(zT) < \epsilon_{VA} \ \text{for} \ T > T_c.   
\eea
In this study, we set $\epsilon_{VA}$ to two different values, 
0.05 and 0.01, to study how $T_c$ depends on $\epsilon_{VA}$. 

For the $U(1)_A$ symmetry breaking, it can be measured 
by the $z$ correlators in the pseudoscalar and scalar channels, with  
\bea
\label{eq:k_PS_z}
\kappa_{PS}(zT) = \frac{\left| C_P(zT)-C_S(zT) \right|}{C_P(zT)+C_S(zT)}, \hspace{4mm}  z > 0,
\eea
as well as in the tensor vector and axial-tensor vector channels, with 
\bea
\label{eq:k_TX_z}
&& \kappa_{TX}(zT) = \frac{\left| C_{T_k}(zT) - C_{X_k}(zT) \right| }{C_{T_k}(zT) + C_{X_k}(zT)}, \nonumber \\
&& \hspace{12mm} z > 0, \hspace{4mm} (k=1,2,4). 
\eea
Due to the $S_2$ symmetry of the $z$ correlators, it only needs 
to examine $k=1$ and $k=4$ components of (\ref{eq:k_TX_z}). 
In practice, the difference between $k=1$ and $k=4$ components of (\ref{eq:k_TX_z}) is almost zero,  
up to the statistical uncertainties, thus in the following, we only give the results 
of (\ref{eq:k_TX_z}) with $k=4$. 

In general, $\kappa_{PS}(zT) \ne \kappa_{TX}(zT)$. For consistency, we always use $\kappa_{TX}(zT)$  
(\ref{eq:k_TX_z}) with $k=4$ to measure the $U(1)_A$ symmetry breaking. 

Similar to (\ref{eq:SU2_crit_z}), we use the following criterion for the manifestation of 
$U(1)_A$ symmetry at $T$ for a fixed $zT$
\bea
&& \kappa_{TX}(zT) \le \epsilon_{TX},   
\label{eq:U1_TX_crit_z}
\eea
where 
$\epsilon_{TX}$ is a small parameter which defines the precision of $U(1)_A$ symmetry. 
For fixed $zT$ and $\epsilon_{TX}$, the temperature $T_1$ for the manifestation of $U(1)_A$ symmetry 
is the lowest temperature satisfying (\ref{eq:U1_TX_crit_z}), i.e.,
\bea
\label{eq:T1_epsilon} 
\kappa_{TX}(zT) < \epsilon_{TX} \ \text{for} \ T > T_1.   
\eea
In this study, we set $\epsilon_{TX}$ to two different values, 
0.05 and 0.01, to study how the temperature of restoration of $U(1)_A$ symmetry depends on $\epsilon_{TX}$. 

Next, consider QCD with $N_f=2+1(+1)(+1)$ quarks $(q_1, q_2, \cdots, q_{N_f})$.
As discussed in Sec. \ref{intro}, upon neglecting the disconnected diagrams in the meson $z$ correlators, 
the $SU(N)_L \times SU(N)_R$ chiral symmetry of $N$ ($2 \le N \le N_f$) quarks 
is manifested by the degeneracies of 
meson $z$ correlators in the vector and axial-vector channels, 
$C_{V_k}^{\bar q_i q_j}(z) = C_{A_k}^{\bar q_i q_j}(z)$, $(k=1,2,4)$, 
for {\it all} flavor combinations of $N$ quarks ($\bar q_i q_j$, $i,j=1, \cdots, N$). 
Thus, to determine the temperature $T_c$ for the manifestation of the 
$SU(N)_L \times SU(N)_R$ chiral symmetry of $N$ quarks, 
it needs to measure $\kappa_{VA}^{\bar q_i q_j}$ for {\it all} flavor combinations of $N$ quarks,  
and check whether they {\it all} satisfy the criterion (\ref{eq:SU2_crit_z}) 
for fixed $zT$ and $\epsilon_{VA}$.
This amounts to finding the largest $T_c^{\bar q_i q_j}$ satisfying (\ref{eq:SU2_crit_z}) 
among all flavor combinations of $N$ quarks, i.e.,  
\bea
\label{Tc_N}
T_c = \max( T_c^{\bar q_i q_j}, i,j=1, 2, \cdots, N).
\eea  

About the $U(1)_A$ chiral symmetry of $N$ ($2 \le N \le N_f$) quarks,  
upon neglecting the disconnected diagrams in the meson $z$ correlators, 
it is manifested by the degeneracies of meson $z$ correlators in the pseudoscalar and scalar channels, 
$C_{P}^{\bar q_i q_j}(z) = C_{S}^{\bar q_i q_j}(z)$, 
as well as in the tensor vector and axial-tensor vector channels, 
$C_{T_k}^{\bar q_i q_j}(z) = C_{X_{k}}^{\bar q_i q_j}(z)$, $(k=1,2,4)$, 
for {\it all} flavor combinations of $N$ quarks ($\bar q_i q_j$, $i,j=1, \cdots, N$). 
Thus, to determine the temperature $T_1$ for the manifestation of the 
$U(1)_A $ symmetry via the $k=4$ component of the tensor vector and axial-tensor vector channels, 
it needs to measure $\kappa_{TX}^{\bar q_i q_j}$ for {\it all} flavor combinations of $N$ quarks,  
and check whether they {\it all} satisfy the criterion (\ref{eq:U1_TX_crit_z}) 
for fixed $zT$ and $\epsilon_{TX}$.
This amounts to finding the largest $T_1^{\bar q_i q_j}$ satisfying (\ref{eq:U1_TX_crit_z}) 
among all flavor combinations of $N$ quarks, i.e.,  
\bea
\label{T1_N}
T_1 = \max( T_1^{\bar q_i q_j}, i,j=1, 2, \cdots, N).
\eea

\subsection{$SU(2)_{CS}$ symmetry breaking and fading parameters}

Following the discussion and notations in Ref. \cite{Chiu:2023hnm}, 
the $SU(2)_{CS}$ multiplets for the $z$ correlators with flavor content $\bar q Q$ are 
\bea
\label{eq:SU2CS_z_a}
(V_1); \ (A_1, T_4, X_4), \\
(V_4); \ (A_4, T_1, X_1),
\label{eq:SU2CS_z_b}
\eea
where the "2" components due to the $S_2$ symmetry have been suppressed. 
Thus the degeneracies in the above triplets signal the emergence of $SU(2)_{CS}$ chiral spin symmetry.
For $ T \ge T_{c1}^{\bar q Q} $, 
the $SU(2)_L \times SU(2)_R \times U(1)_A$ chiral symmetry is
effectively restored, and $ C_{V_k} = C_{A_k} $ and $ C_{T_k} = C_{X_k} $ for $k=1,2,4$, 
and the multiplets in Eqs. (\ref{eq:SU2CS_z_a}) and (\ref{eq:SU2CS_z_b}) become:
\bea
\label{eq:SU2CS_SU2A_z_a}
(V_1, A_1, T_4, X_4),  \\
(V_4, A_4, T_1, X_1).
\label{eq:SU2CS_SU2A_z_b}
\eea
This suggests the possibility of a larger symmetry group $ SU(4) $ for $T > T_{c1}^{\bar q Q}$
which contains $ SU(2)_L \times SU(2)_R \times SU(2)_{CS} $ as a subgroup. 
For the full $SU(4)$ symmetry, each of the multiplets
in Eqs. (\ref{eq:SU2CS_SU2A_z_a}) and (\ref{eq:SU2CS_SU2A_z_b}) is enlarged
to include the flavor-singlet partners of $A_k$, $T_k$ and $X_k$,
while the flavor-singlet partners of $V_1$ and $V_4$ are $SU(4)$ singlets, i.e.,
\bea
\label{eq:SU4_z_a}
V_1^0; \ (V_1, A_1, T_4, X_4, A_1^0, T_4^0, X_4^0),   \\
V_4^0; \ (V_4, A_4, T_1, X_1, A_4^0, T_1^0, X_1^0), 
\label{eq:SU4_z_b}
\eea
where the superscript "0" denotes the flavor singlet.

In general, to examine the emergence of $SU(2)_{CS}$ symmetry, one needs to measure 
the splittings in both ($A_1, X_4$) and ($T_4, X_4$) of (\ref{eq:SU2CS_z_a}). 
To measure the splitting of $A_1$ and $X_4$, we use   
\bea
\label{eq:k_AT_z}
\kappa_{AT}(zT) = \frac{\left| C_{A_1}(zT) - C_{X_4}(zT) \right|}{C_{A_1}(zT) + C_{X_4}(zT)}, 
\hspace{4mm} z > 0,   
\eea
while the splitting of $T_4$ and $X_4$ is measured by $\kappa_{TX}$ (\ref{eq:k_TX_z}) with $k=4$.
Then we use the maximum of $\kappa_{AT}$ and $\kappa_{TX}$ to measure
the $SU(2)_{CS}$ symmetry breaking, with the parameter 
\bea
\label{eq:k_CS_z}
\kappa_{CS} = \max(\kappa_{AT}, \kappa_{TX}). 
\eea
Note that for $(\bar u d, \bar u s, \bar s s, \bar u c)$ sectors, 
$\kappa_{AT}(zT) > \kappa_{TX}(zT)$ for all $z$ and the seven temperatures in this study,  
thus $\kappa_{CS} = \kappa_{AT}$. 

As the temperature $T$ is increased, the separation between the multiplets of 
$SU(2)_{CS}$ and $U(1)_A$ is decreased.
Therefore, at sufficiently high temperatures, the $U(1)_A$ multiplet $M_0 = (P, S)$ and the 
$SU(2)_{CS} \times SU(2)_L \times SU(2)_R$ multiplet  
$M_2=(V_1, A_1, T_4, X_4)$ merge together,  
then the approximate $SU(2)_{CS}$ symmetry becomes washed out, 
and only the $SU(2)_L \times SU(2)_R \times U(1)_A $ chiral symmetry remains.  
Note that the $SU(2)_{CS} \times SU(2)_L \times SU(2)_R$ multiplet  
$ M_4 = (V_4, A_4, T_1, X_1) $ never merges with 
$M_0$ and $M_2$ even in the limit $T \to \infty$, as discussed in Ref. \cite{Chiu:2023hnm}. 
Thus $M_4$ is irrelevant to the fading of the approximate $SU(2)_{CS} $ symmetry.  

Here we use the $SU(2)_{CS}$ symmetry fading parameter similar to that defined in Ref. \cite{Chiu:2023hnm}, 
except for taking the absolute value and using the unnormalized $z$ correlators, i.e., 
\bea
\kappa(zT) = \left|\frac{C_{A_1}(zT) - C_{X_4}(zT)}{C_{M_0}(zT) - C_{M_2}(zT)} \right|, \hspace{4mm} z > 0, 
\label{eq:kappa_z}
\eea
where
\BAN
&& C_{M_0}(zT) \equiv \frac{1}{2} \left[ C_P(zT) + C_S(zT) \right],  \\
&& C_{M_2}(zT) \equiv \frac{1}{4} \left[ C_{V_1}(zT) + C_{A_1}(zT) \right. \\
&& \hspace{22mm} \left. + \, C_{T_4}(zT) + C_{X_4}(zT) \right].   
\EAN

In general, $\kappa(zT) $ behaves like an increasing function of $T$ for a fixed $zT$. 
If $\kappa(zT) \ll 1 $ for a range of $T$, then  
the approximate $SU(2)_{CS}$ symmetry is well-defined for this window of $T$.
On the other hand, if $ \kappa(zT) > 0.3$ for $T > T_f$, 
then the approximate $SU(2)_{CS}$ symmetry is regarded to be washed out,
and only the $U(1)_A \times SU(2)_L \times SU(2)_R $ chiral symmetry remains. 
Thus, to determine to what extent the $SU(2)_{CS}$ symmetry
is manifested in the $z$ correlators, it is necessary to examine whether 
both $\kappa(zT)$ and $\kappa_{CS}(zT)$ are sufficiently small. 
For a fixed $zT$, the following condition
\bea
\label{eq:SU2_CS_crit_z}
\left(~\kappa_{CS}(zT) < \epsilon_{cs} ~\right)~\land ~\left(~\kappa(zT) < \epsilon_{fcs} ~\right)
\eea
serves as a criterion for the approximate $SU(2)_{CS}$ symmetry 
in the $z$ correlators, where $ \epsilon_{cs} $ is for the $SU(2)_{CS}$ symmetry breaking, 
while $ \epsilon_{fcs} $ for the $SU(2)_{CS}$ symmetry fading. 
For fixed $zT$, (\ref{eq:SU2_CS_crit_z}) gives a window of $T$ for the 
approximate $SU(2)_{CS}$ symmetry. Obviously, the size of this window depends on  
$\epsilon_{cs}$ and $\epsilon_{fcs}$. That is, larger $\epsilon_{cs}$ or $\epsilon_{fcs}$
gives a wider window of $T$, and conversely, smaller $\epsilon_{cs}$ or $\epsilon_{fcs}$ 
gives a narrower window of $T$.


\begin{figure*}[tbp]
  \centering
  \begin{tabular}{@{}c@{}c@{}}
  \includegraphics[width=7.2cm,clip=true]{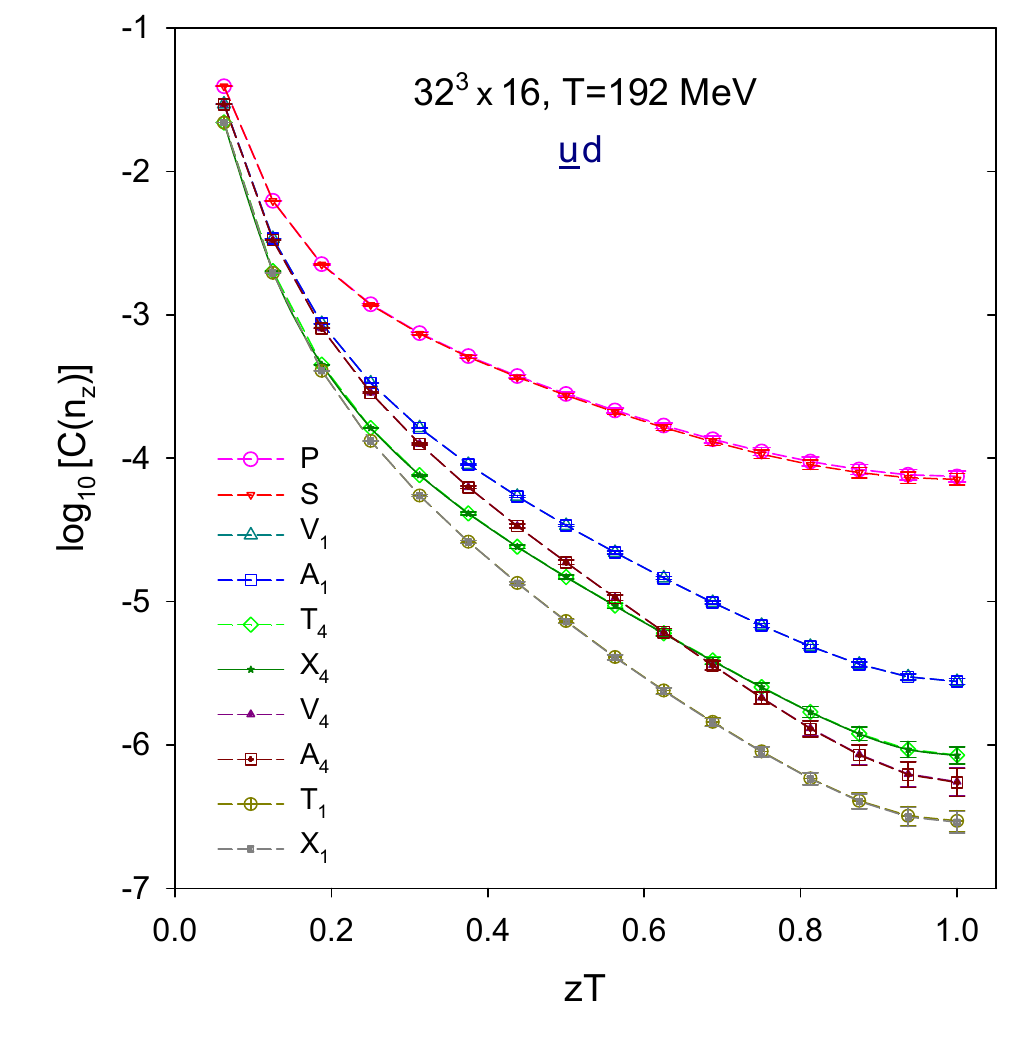}
&
  \includegraphics[width=7.2cm,clip=true]{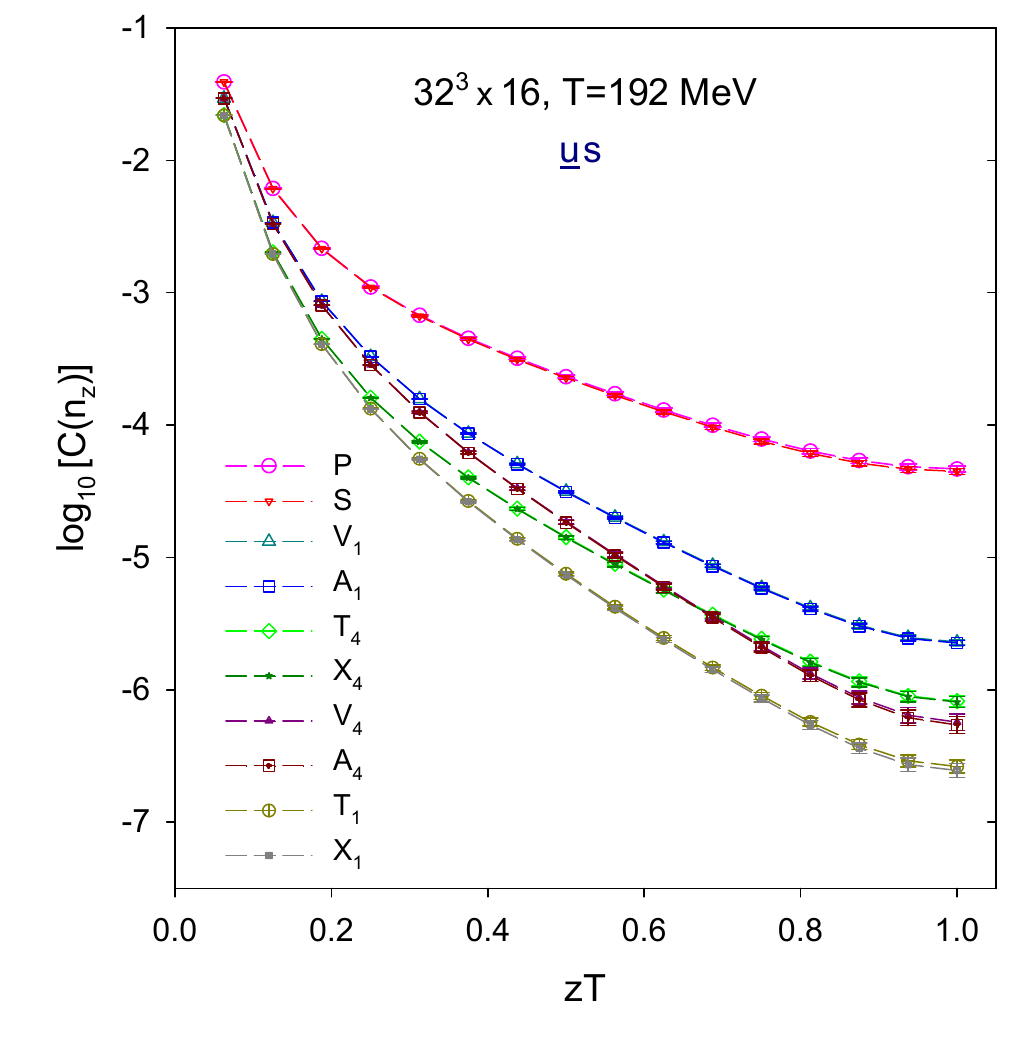}
\\
  \includegraphics[width=7.2cm,clip=true]{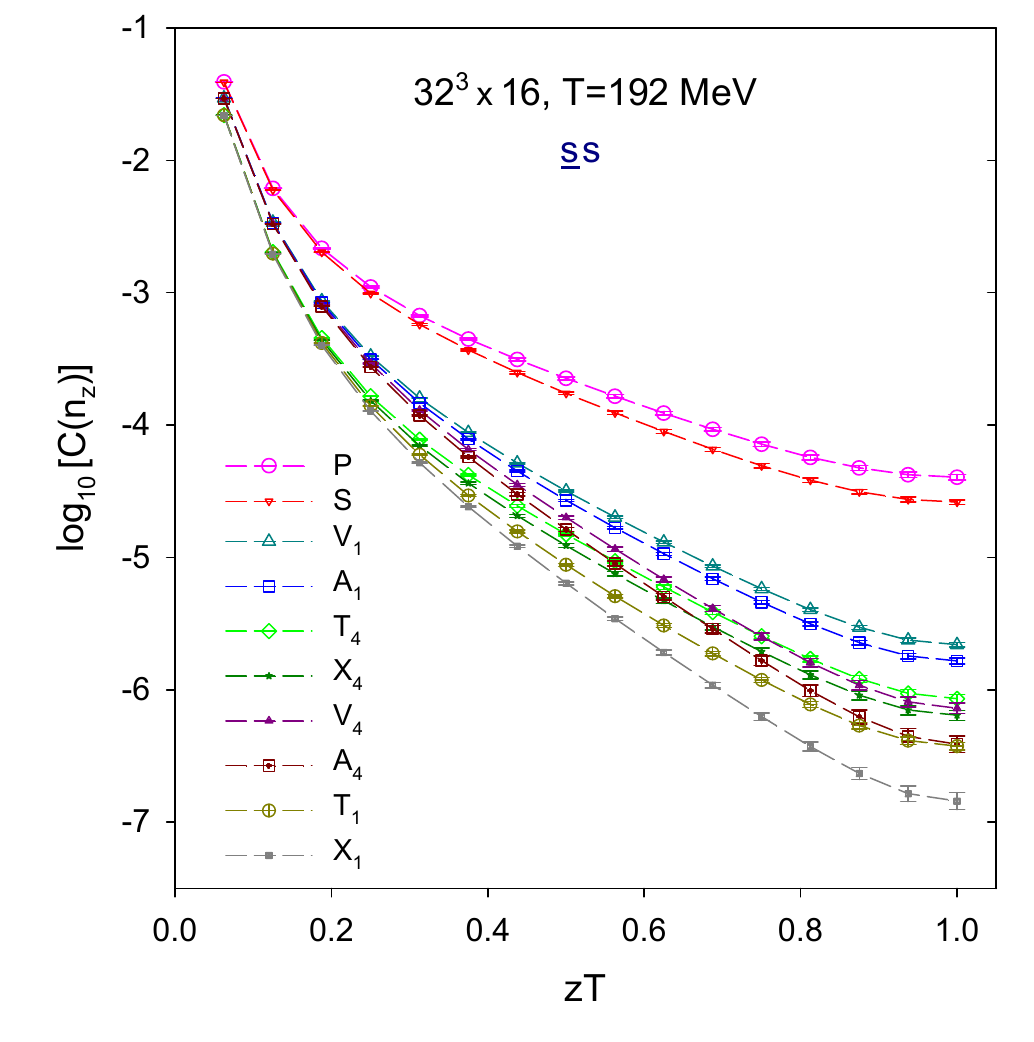}
&
  \includegraphics[width=7.2cm,clip=true]{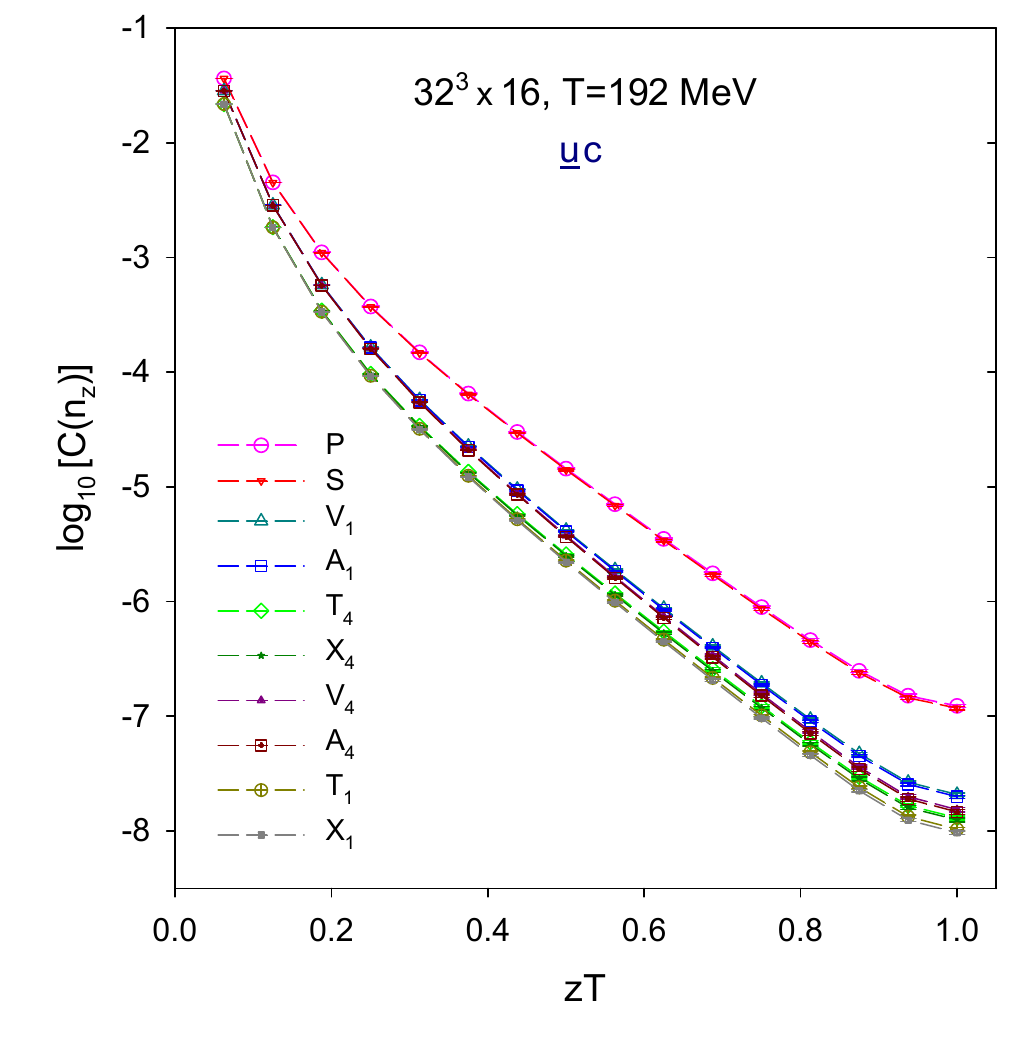}
\\
  \includegraphics[width=7.2cm,clip=true]{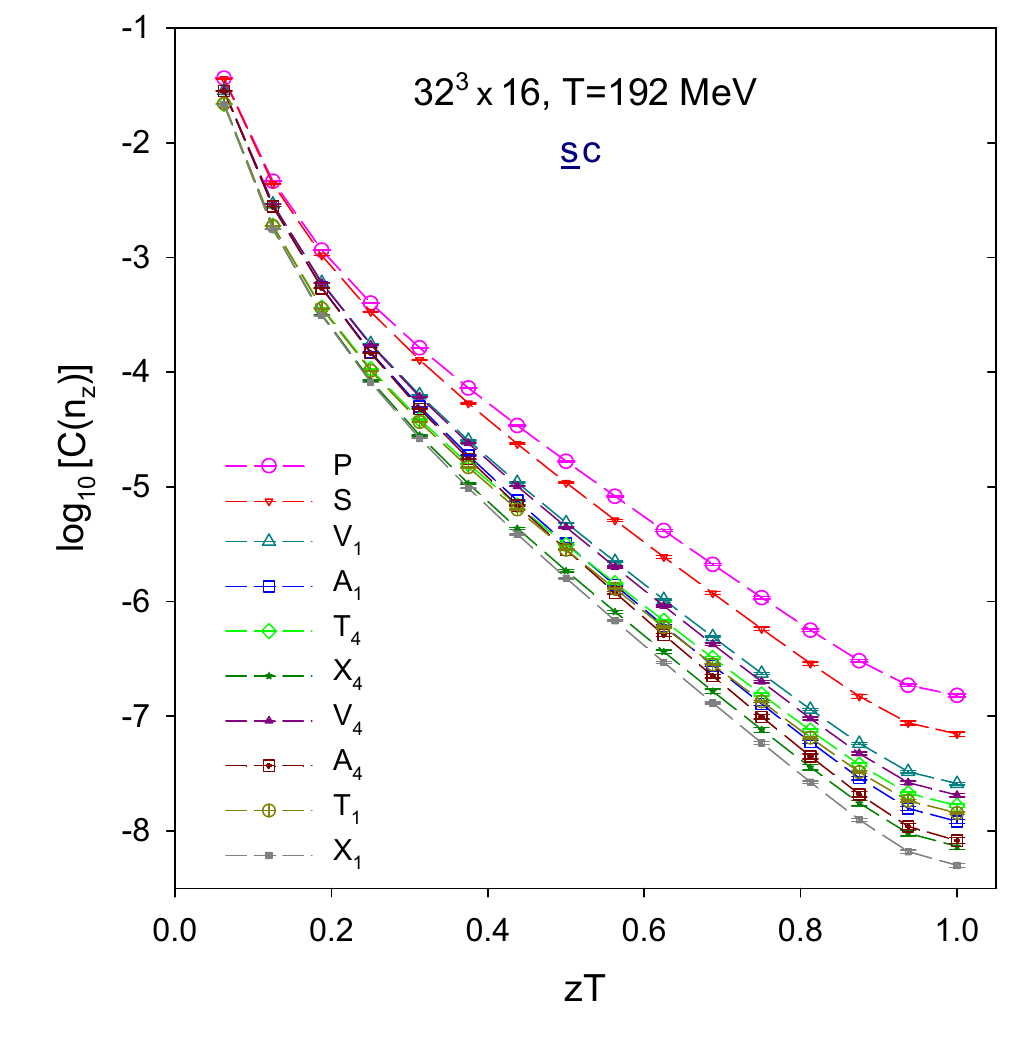}
&
  \includegraphics[width=7.2cm,clip=true]{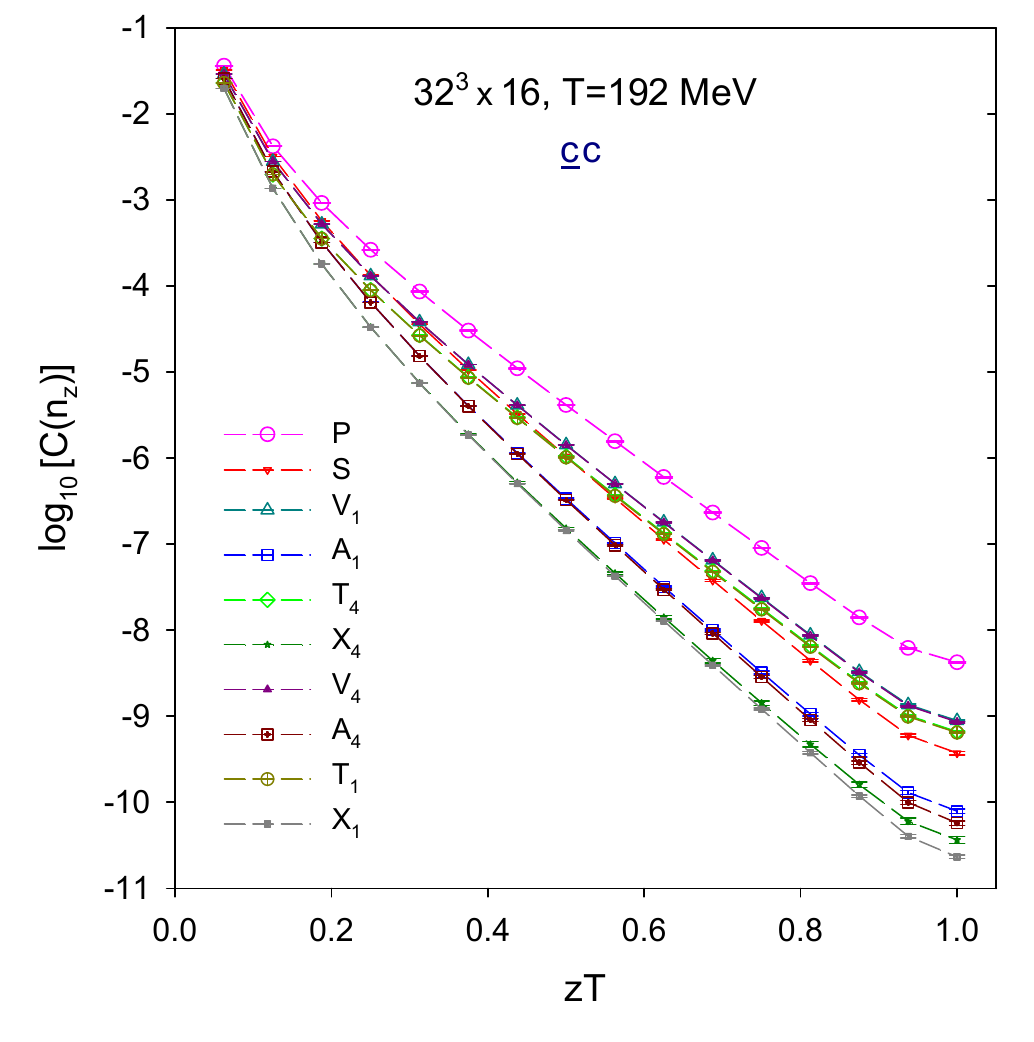}
  \end{tabular}
  \caption{
   The spatial $z$ correlators of meson interpolators for six flavor combinations  
($\bar u d$, $\bar u s$, $\bar s s$, $\bar u c$, $\bar s c$, and $\bar c c $)
in $N_f=2+1+1$ lattice QCD at $T \simeq 192$~MeV.
  }
\label{fig:Cz_T192}
\end{figure*}

\begin{figure*}[tbp]
  \centering
  \begin{tabular}{@{}c@{}c@{}}
  \includegraphics[width=7.2cm,clip=true]{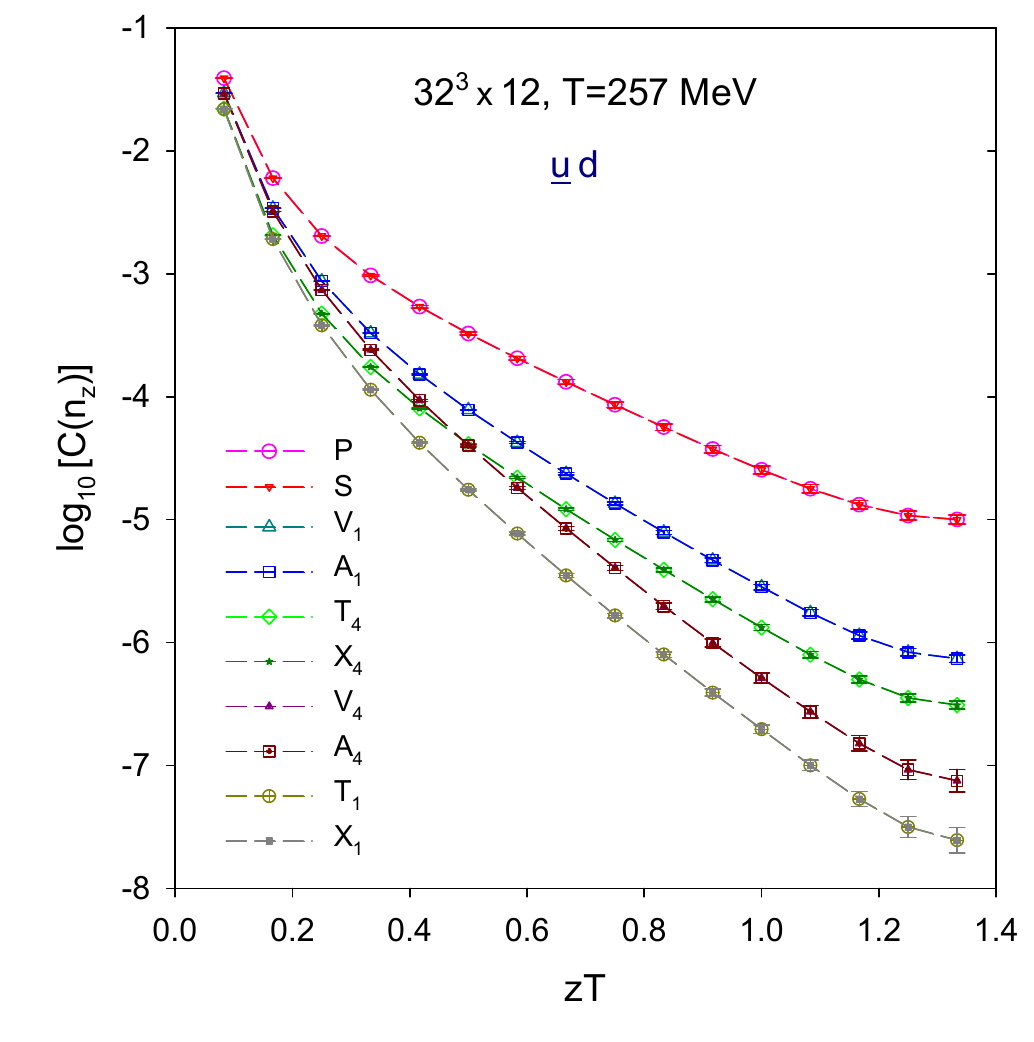}
&
  \includegraphics[width=7.2cm,clip=true]{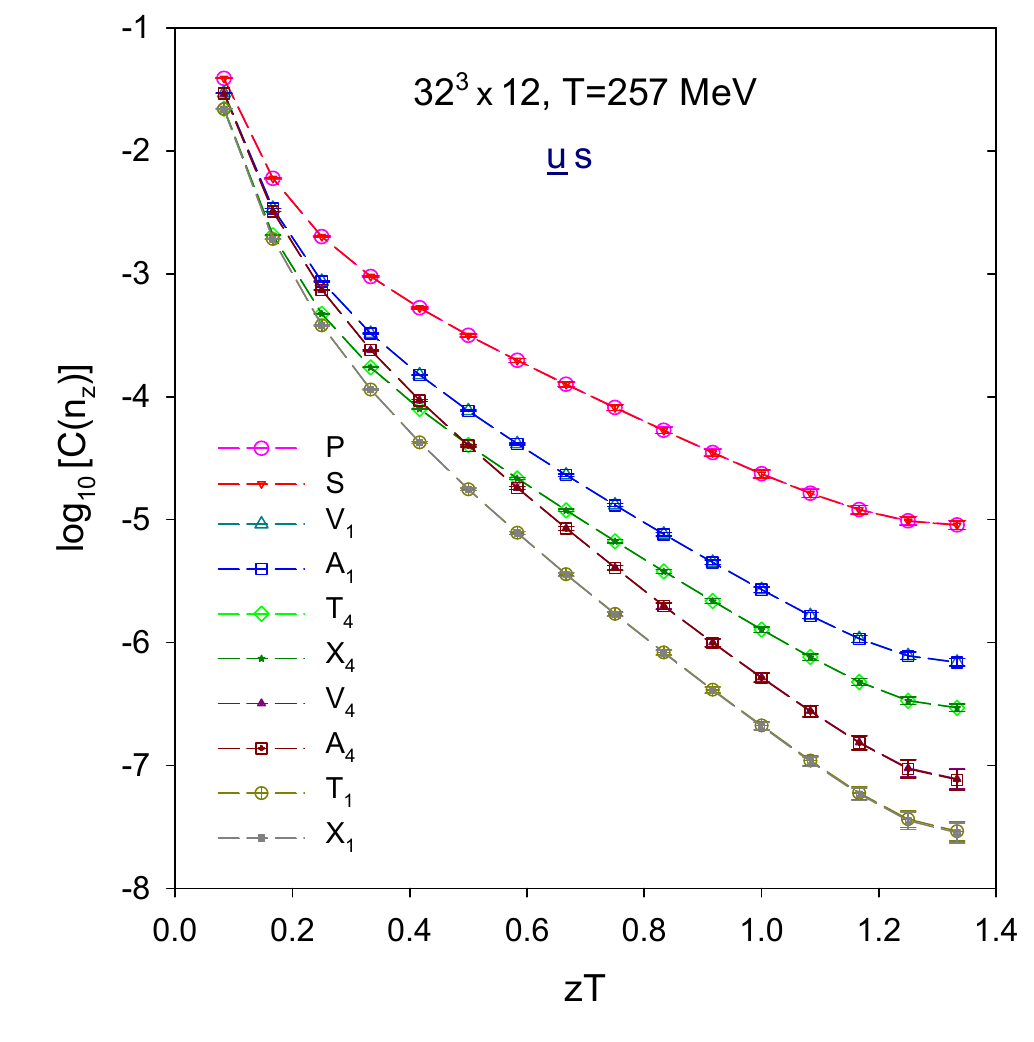}
\\
  \includegraphics[width=7.2cm,clip=true]{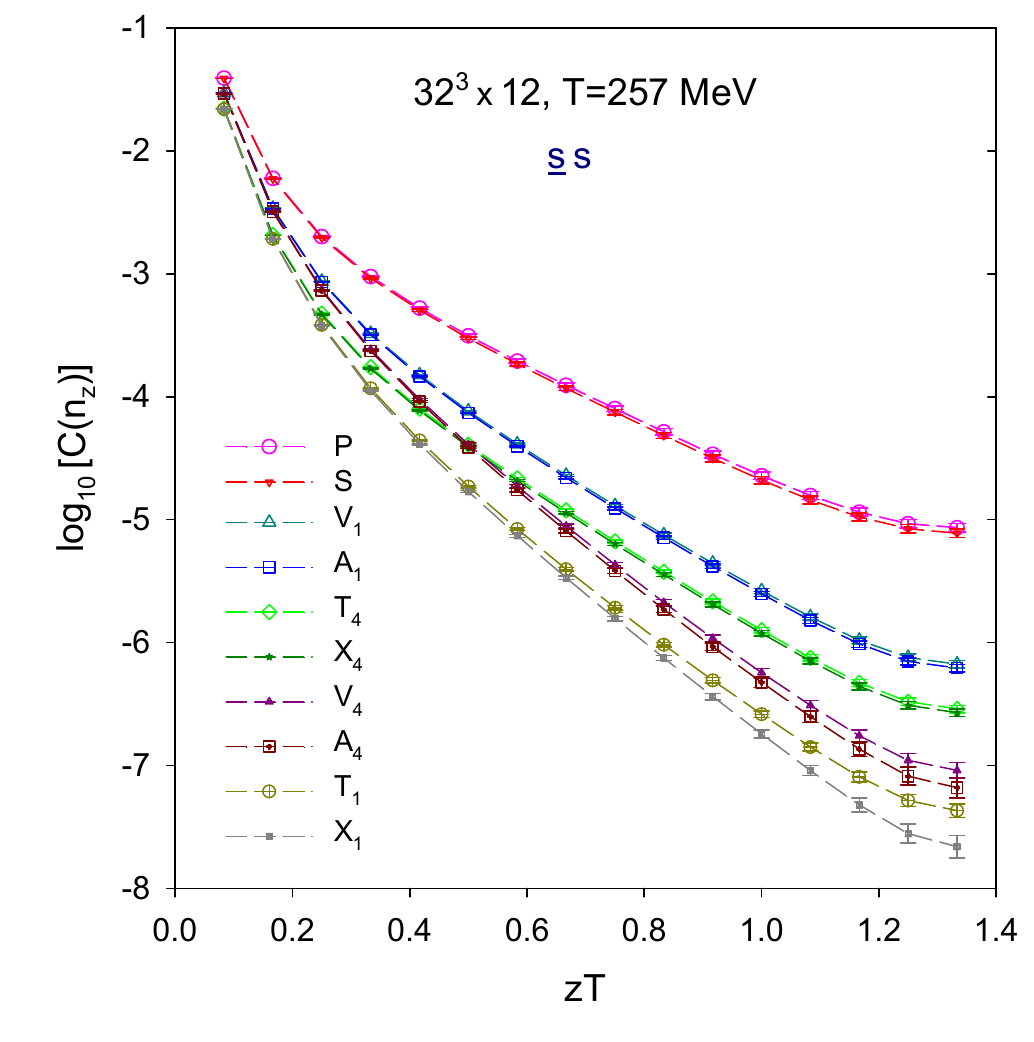}
&
  \includegraphics[width=7.2cm,clip=true]{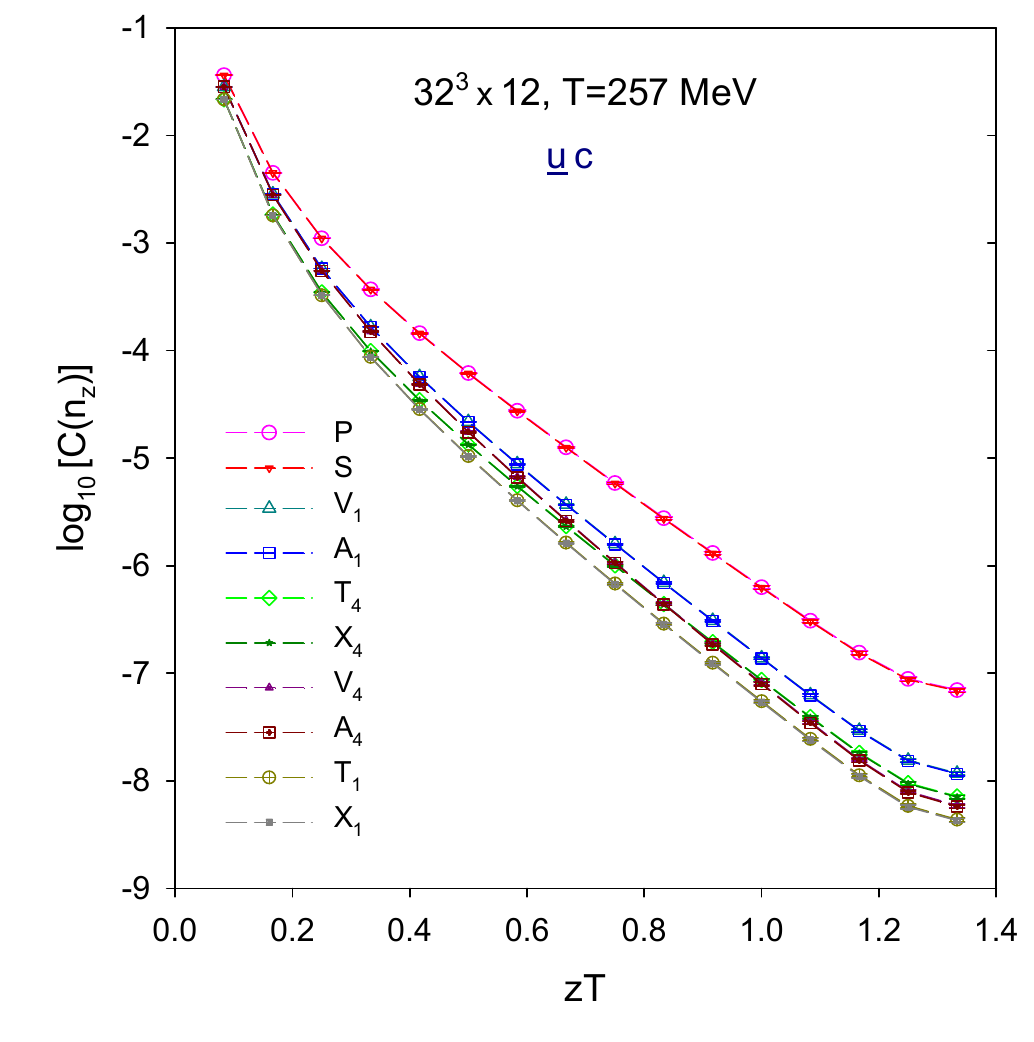}
\\
  \includegraphics[width=7.2cm,clip=true]{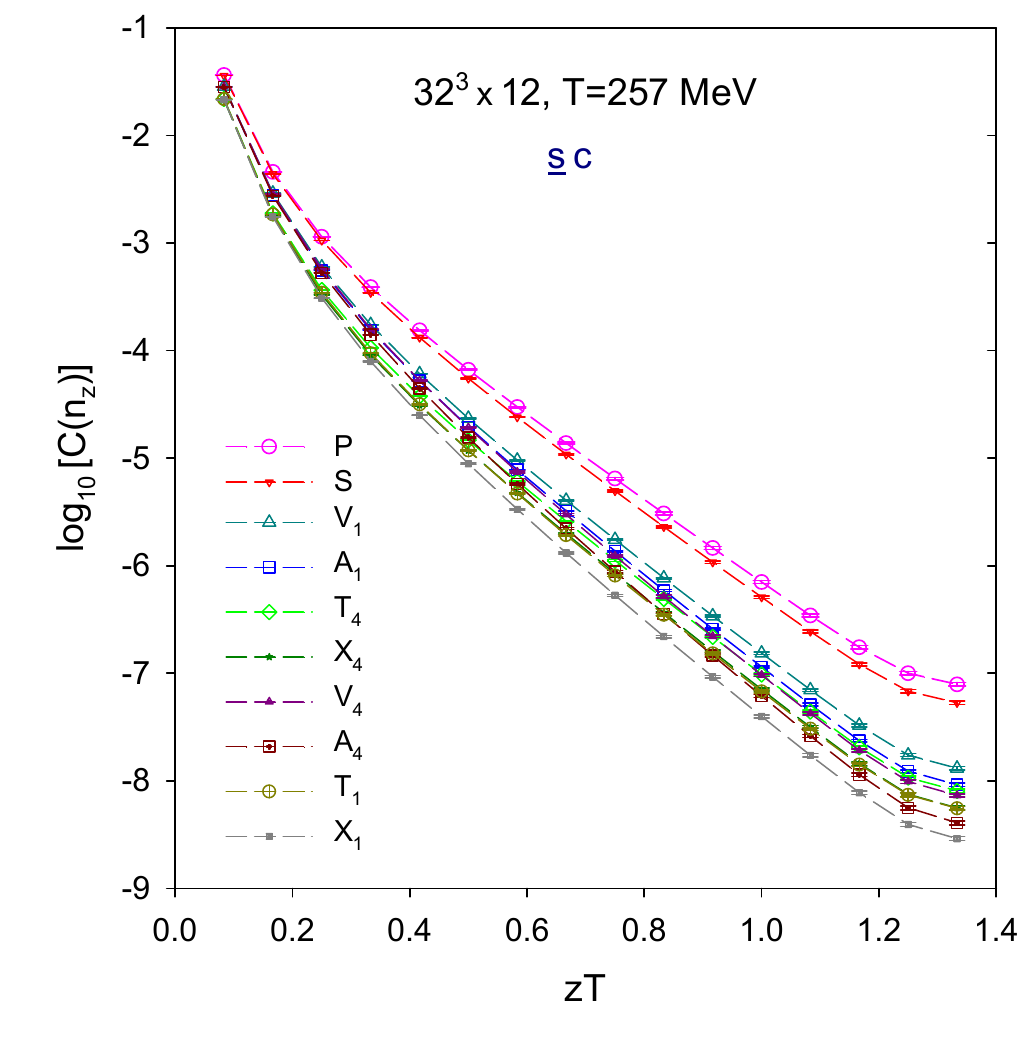}
&
  \includegraphics[width=7.2cm,clip=true]{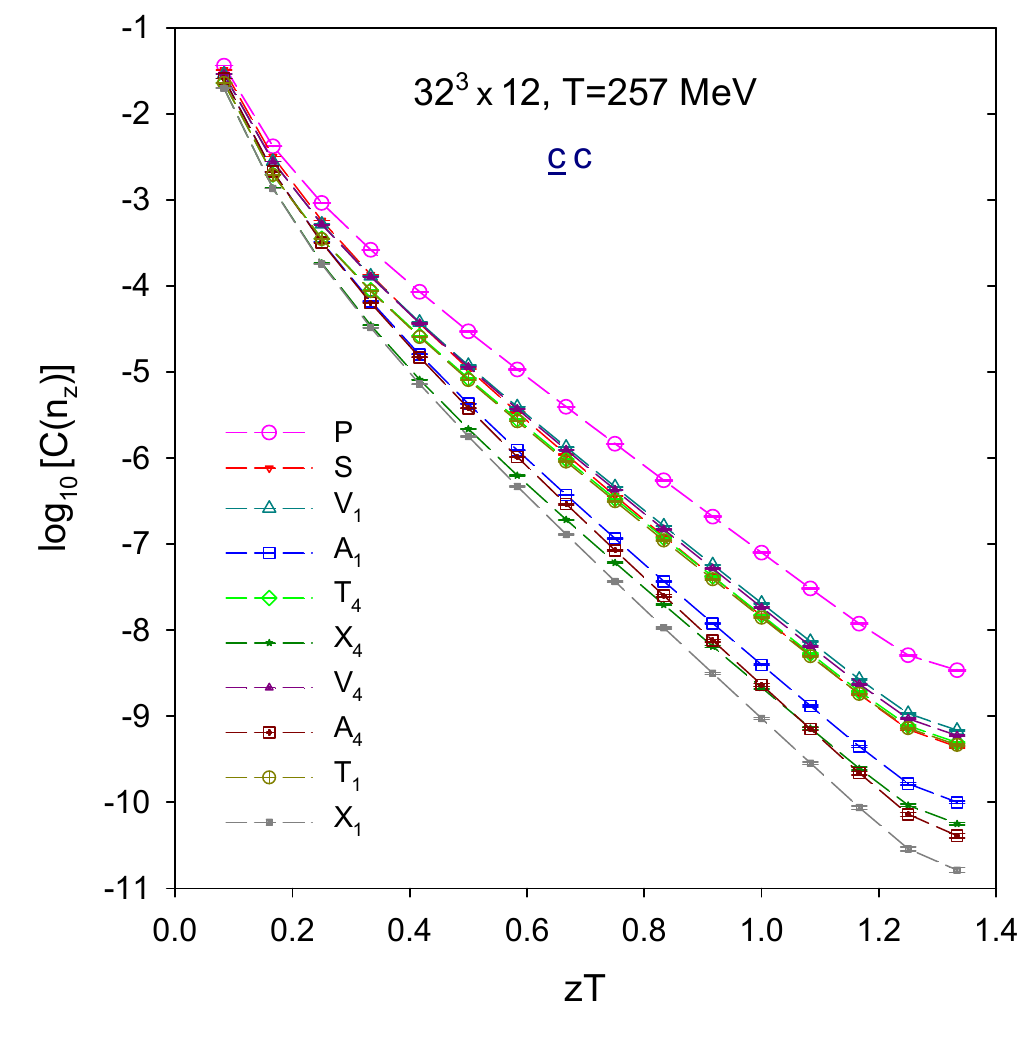}
  \end{tabular}
  \caption{
   The spatial $z$ correlators of meson interpolators for six flavor combinations  
($\bar u d$, $\bar u s$, $\bar s s$, $\bar u c$, $\bar s c$, and $\bar c c $)
in $N_f=2+1+1$ lattice QCD at $T \simeq 257$~MeV.
  }
\label{fig:Cz_T257}
\end{figure*}

\begin{figure*}[tbp]
  \centering
  \begin{tabular}{@{}c@{}c@{}}
  \includegraphics[width=7.2cm,clip=true]{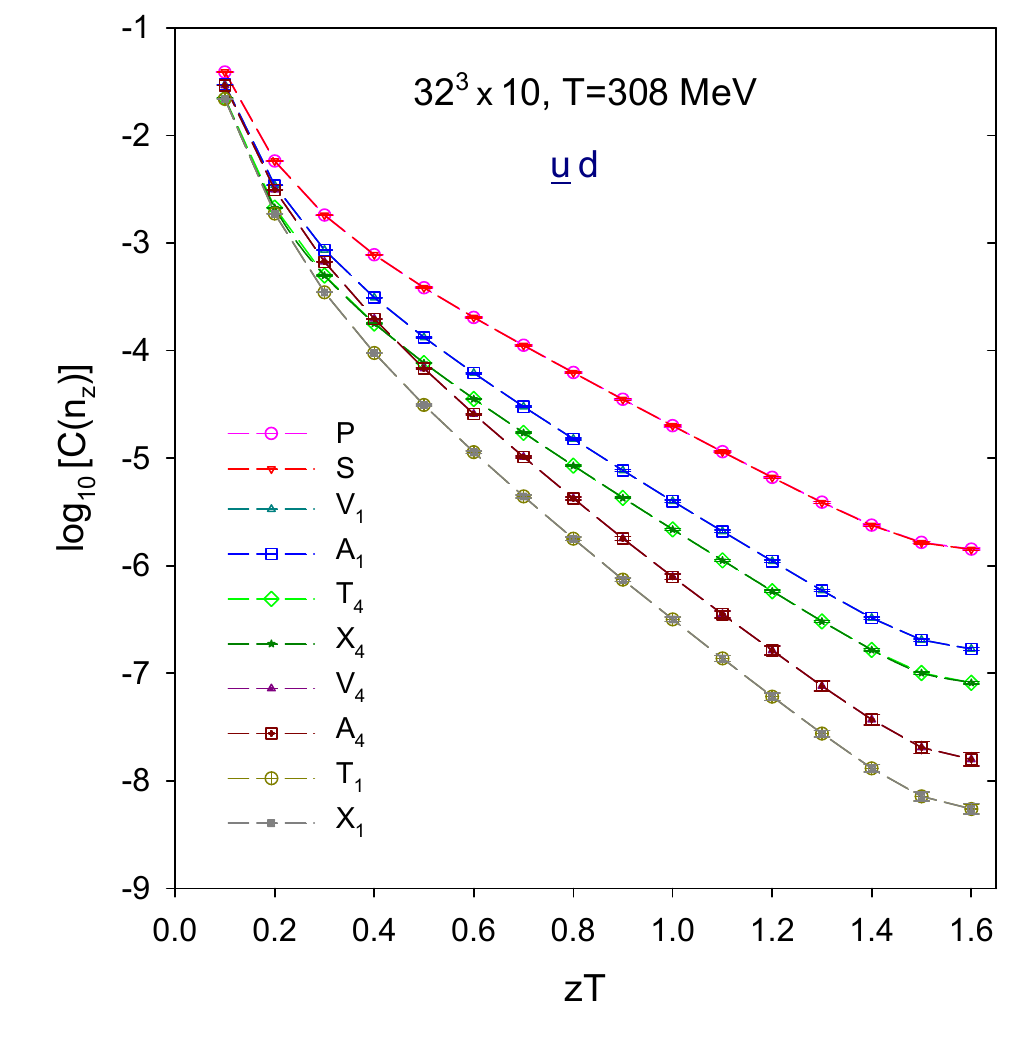}
&
  \includegraphics[width=7.2cm,clip=true]{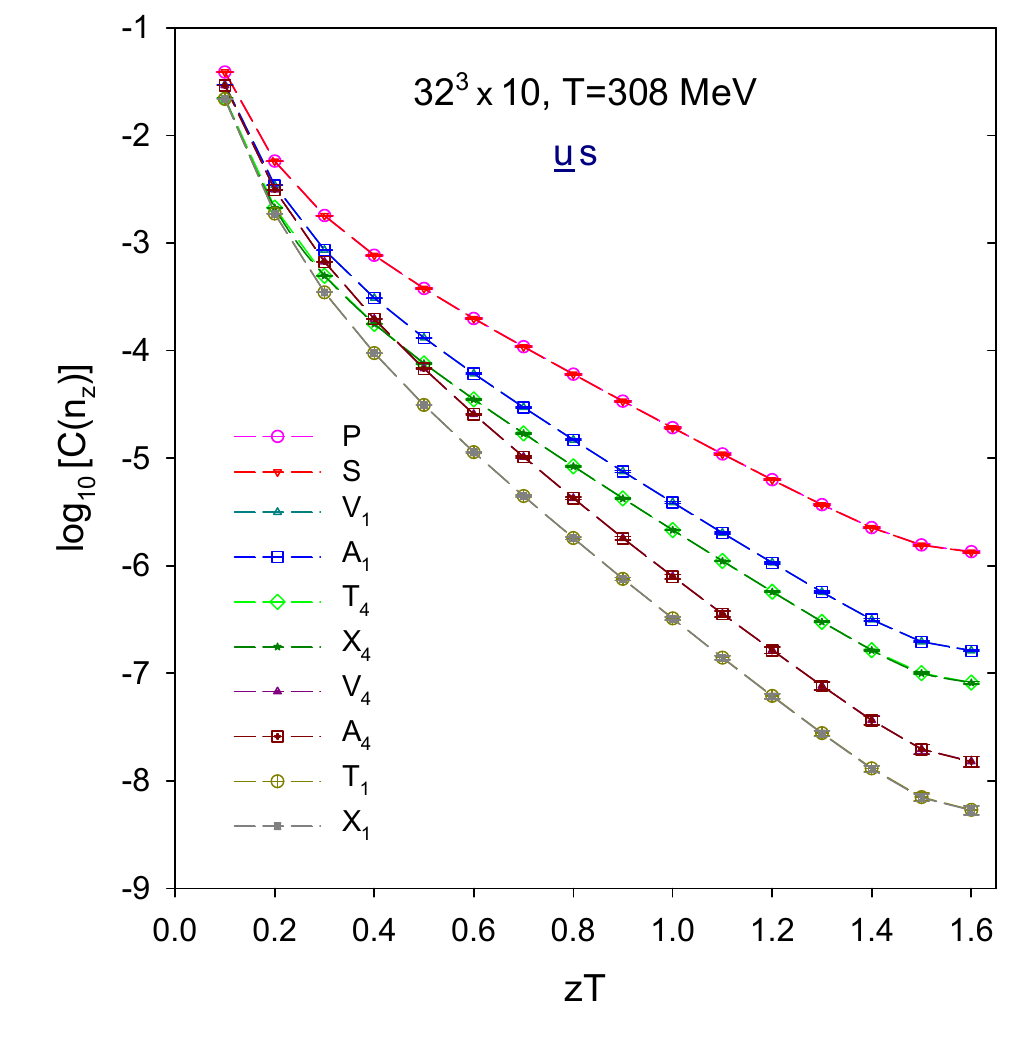}
\\
  \includegraphics[width=7.2cm,clip=true]{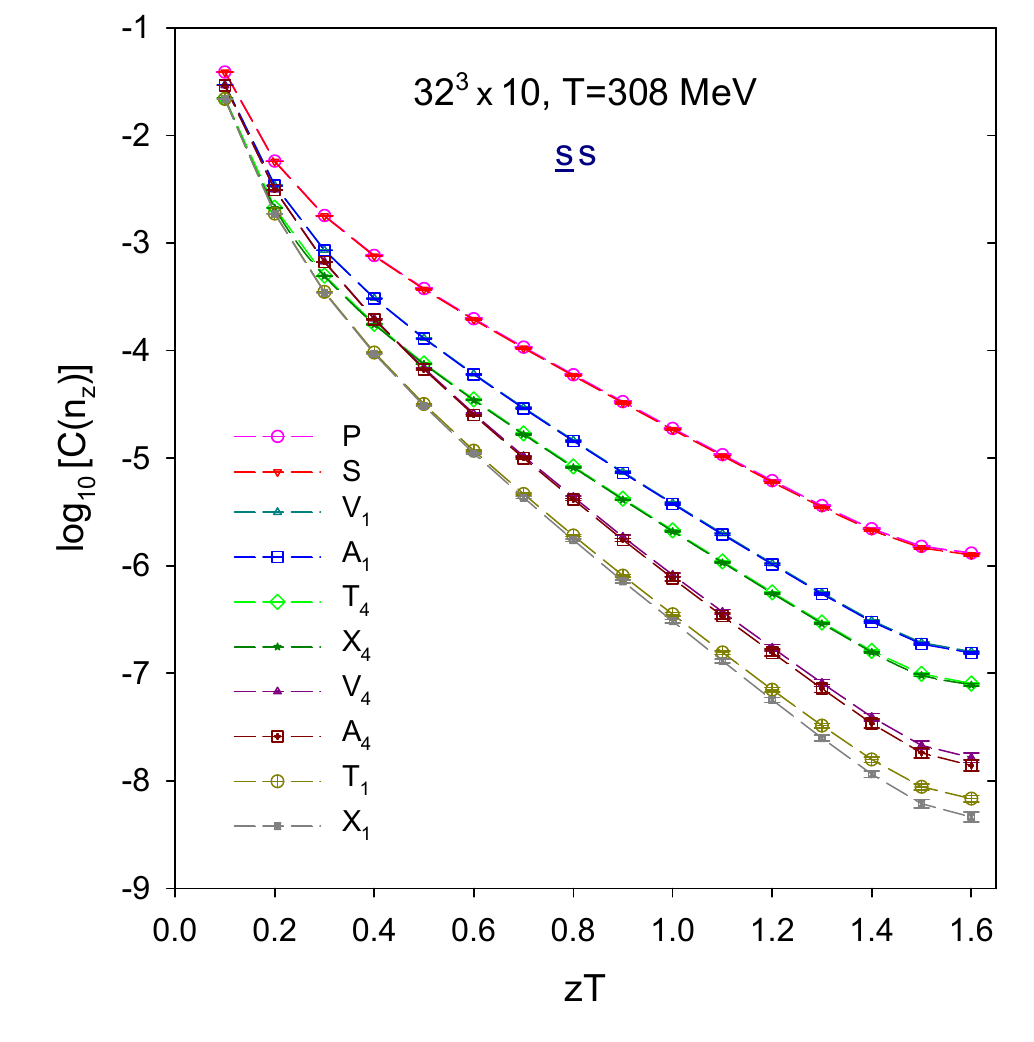}
&
  \includegraphics[width=7.2cm,clip=true]{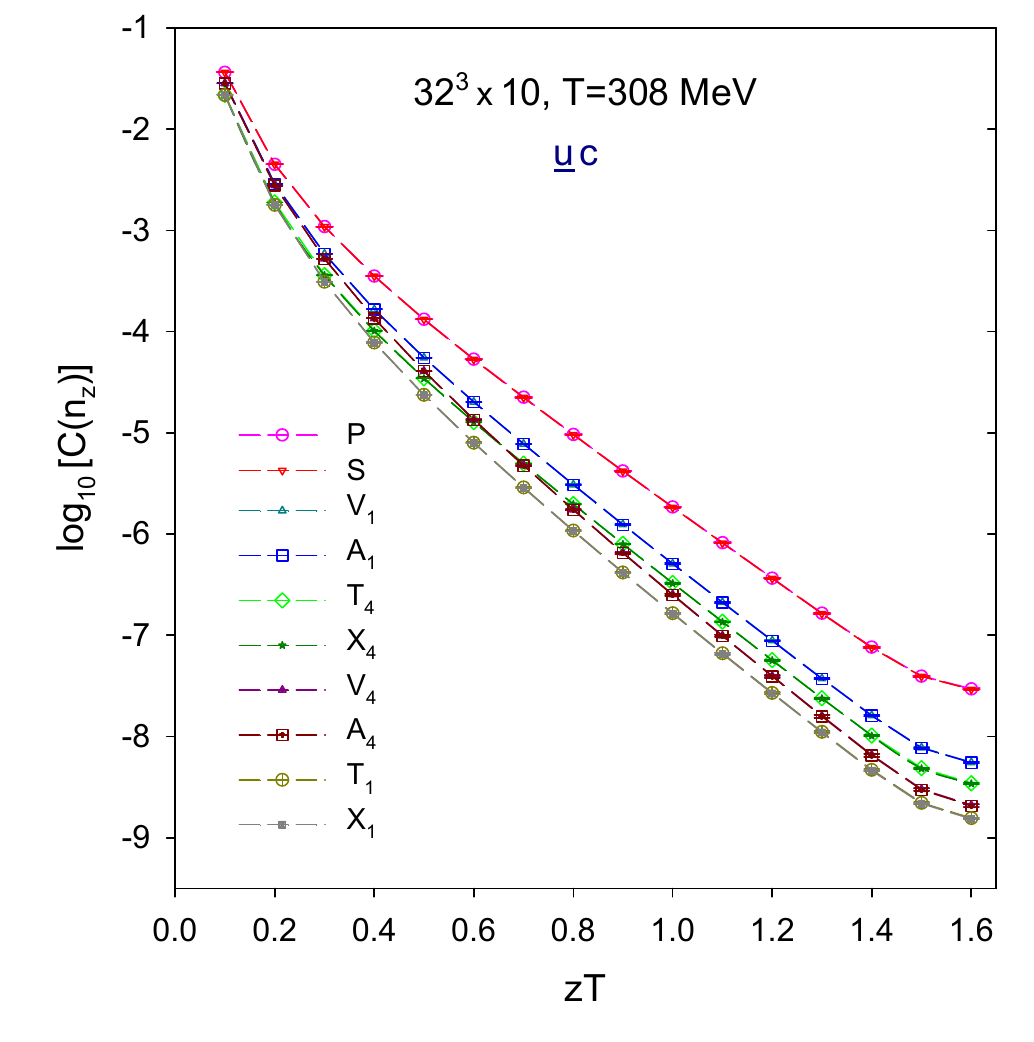}
\\
  \includegraphics[width=7.2cm,clip=true]{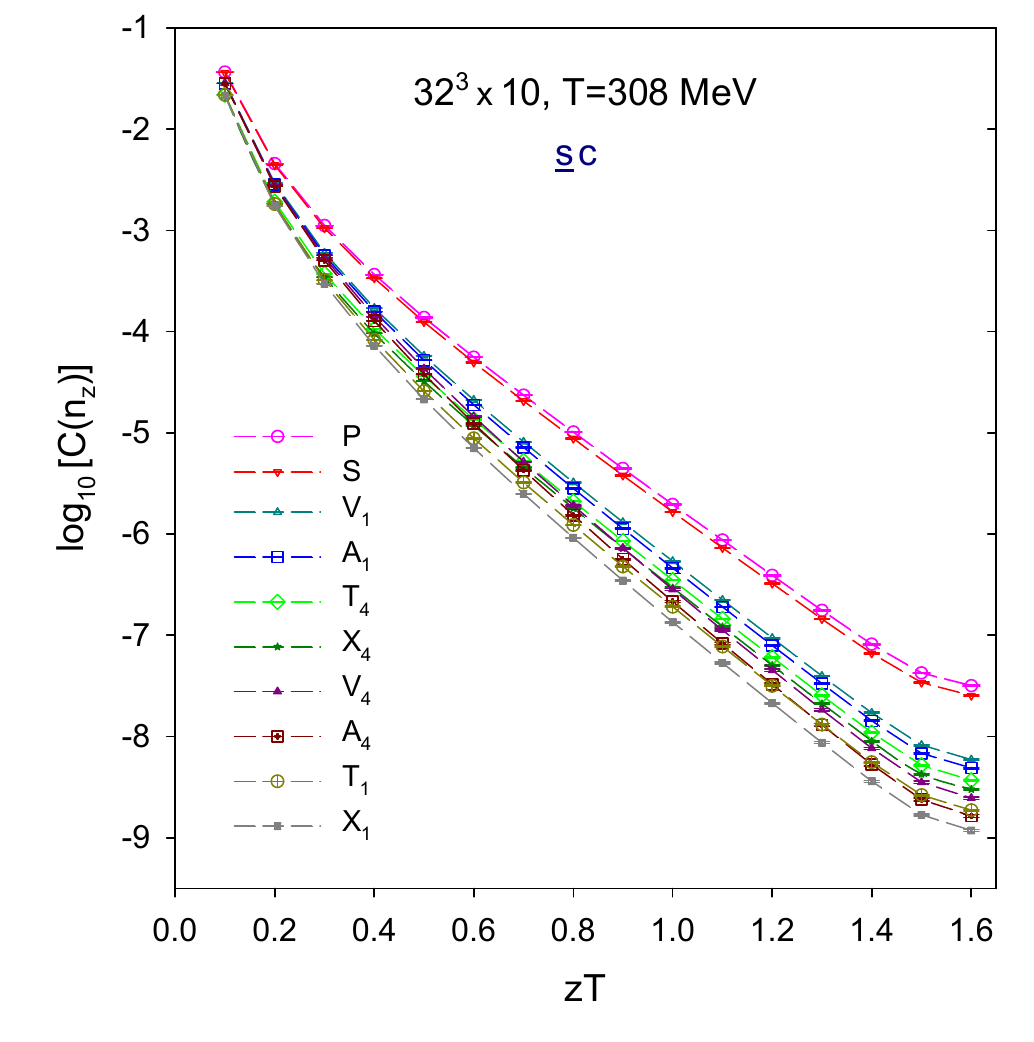}
&
  \includegraphics[width=7.2cm,clip=true]{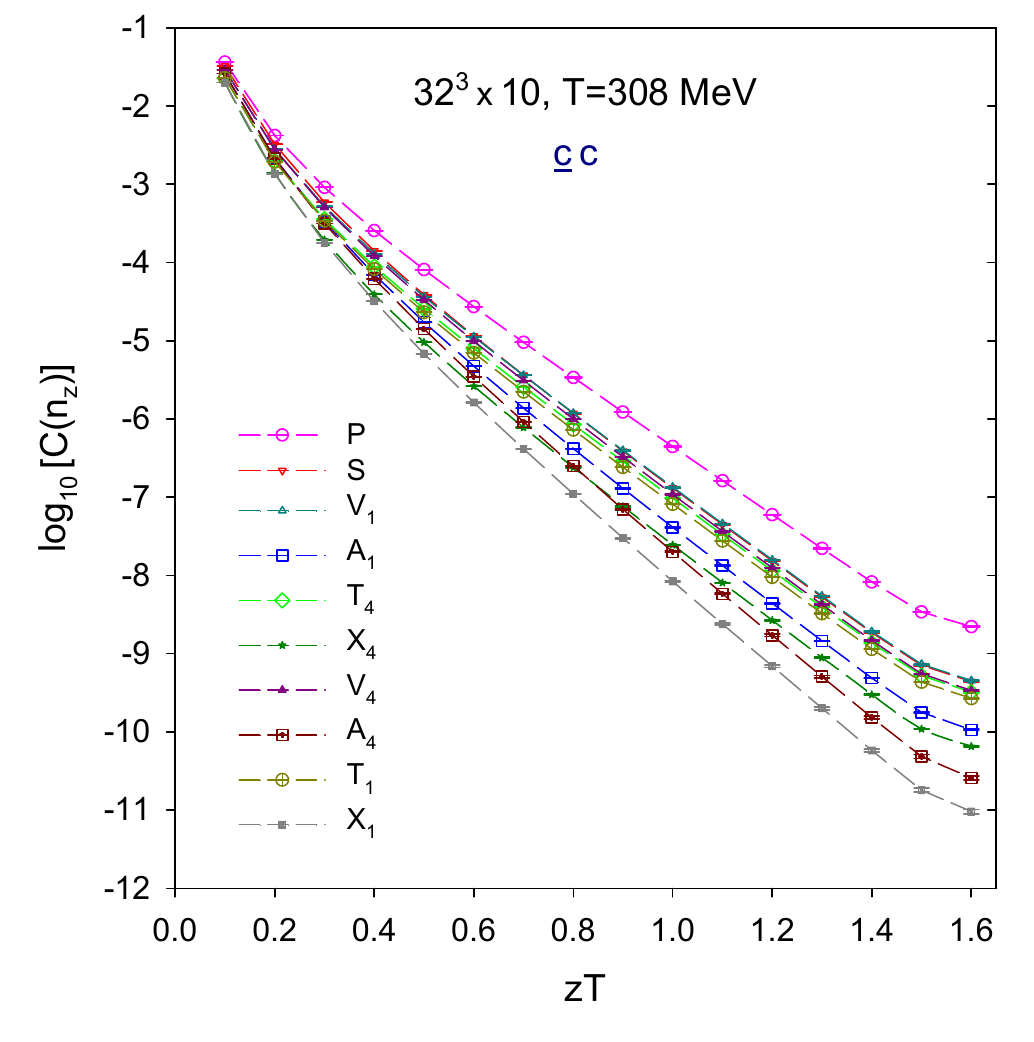}
  \end{tabular}
  \caption{
   The spatial $z$ correlators of meson interpolators for six flavor combinations
($\bar u d$, $\bar u s$, $\bar s s$, $\bar u c$, $\bar s c$, and $\bar c c $)
in $N_f=2+1+1$ lattice QCD at $T \simeq 308$~MeV.
  }
\label{fig:Cz_T308}
\end{figure*}

\begin{figure*}[tbp]
  \centering
  \begin{tabular}{@{}c@{}c@{}}
  \includegraphics[width=7.2cm,clip=true]{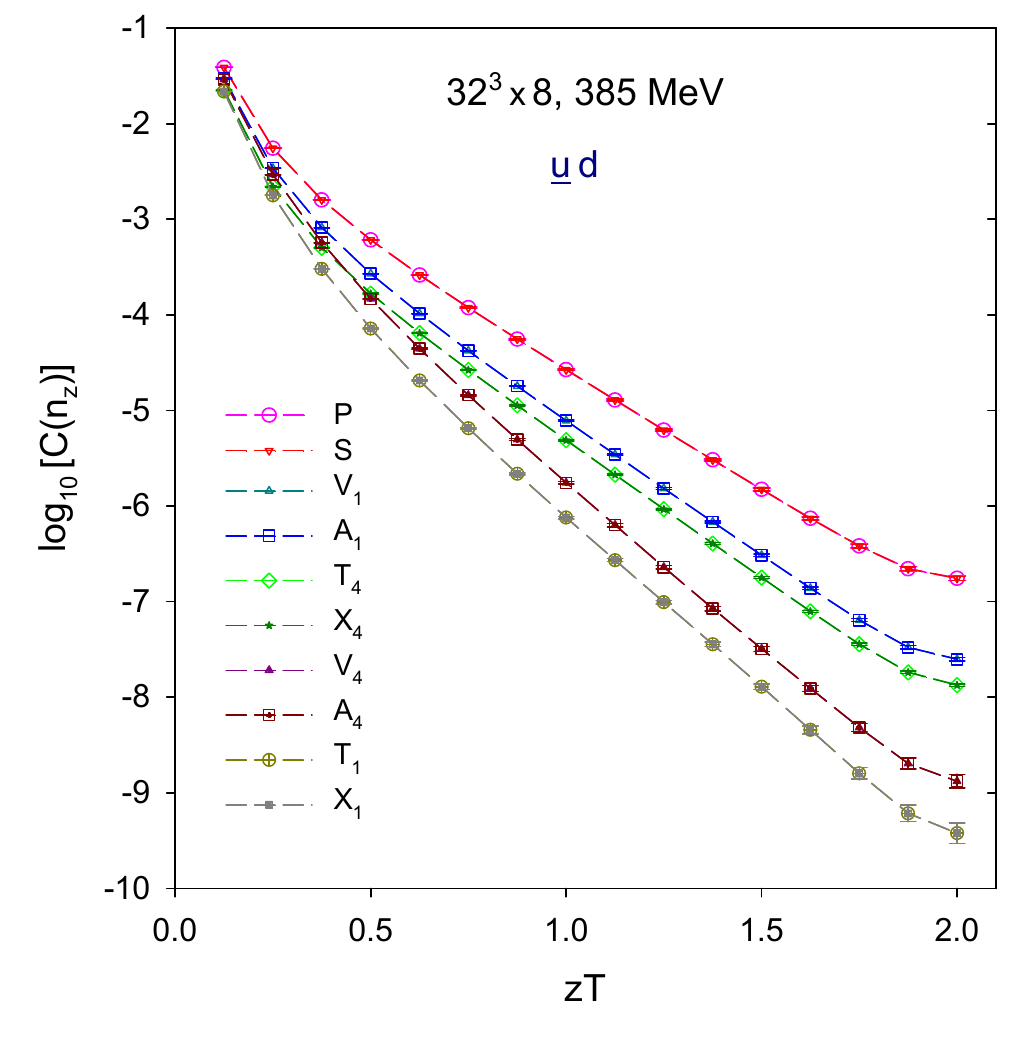}
&
  \includegraphics[width=7.2cm,clip=true]{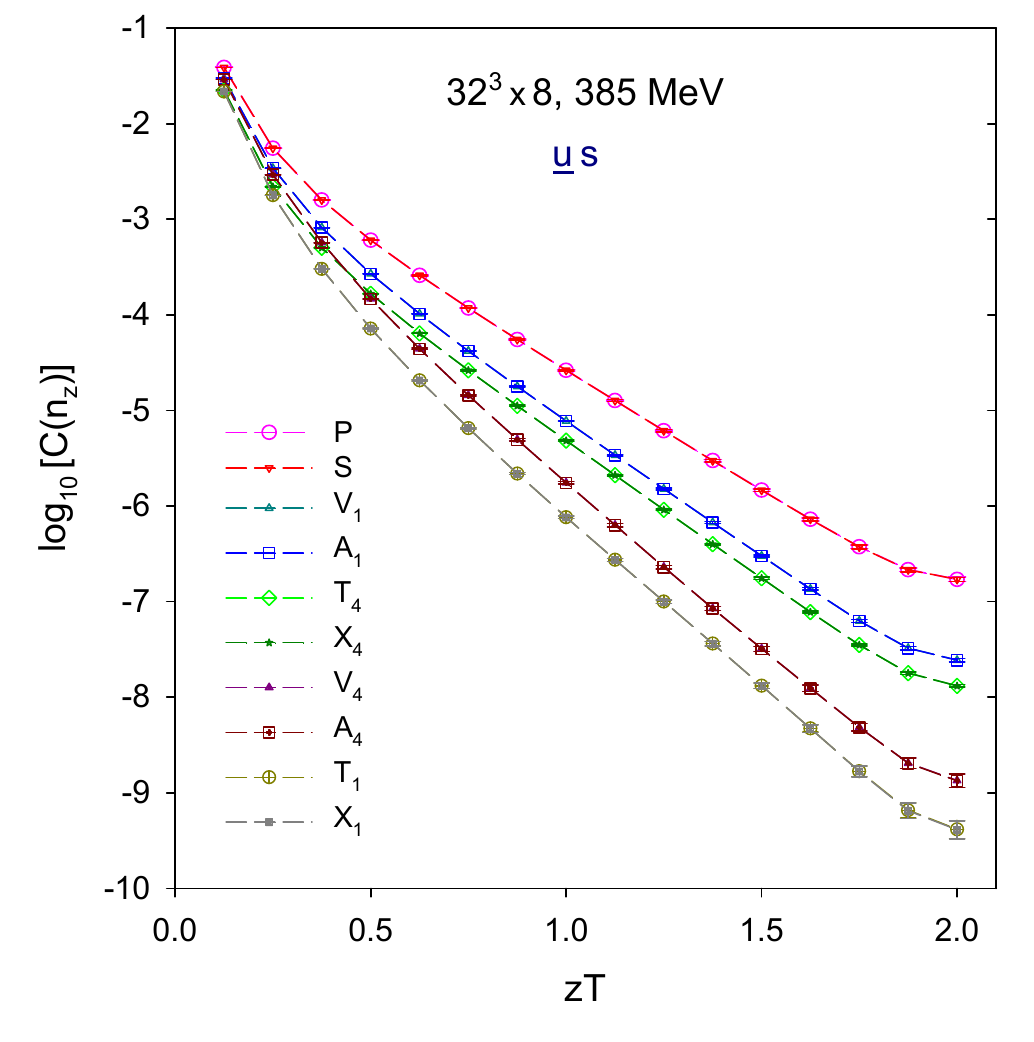}
\\
  \includegraphics[width=7.2cm,clip=true]{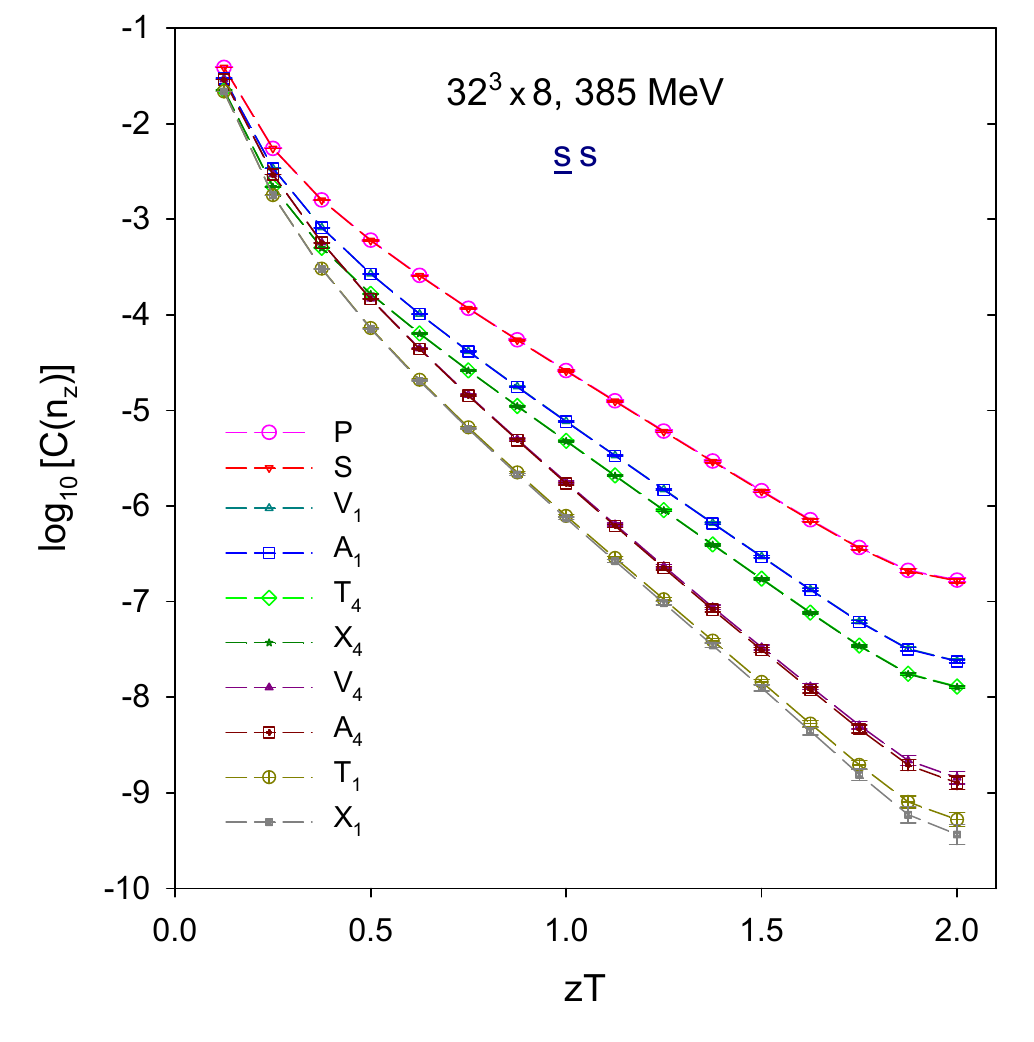}
&
  \includegraphics[width=7.2cm,clip=true]{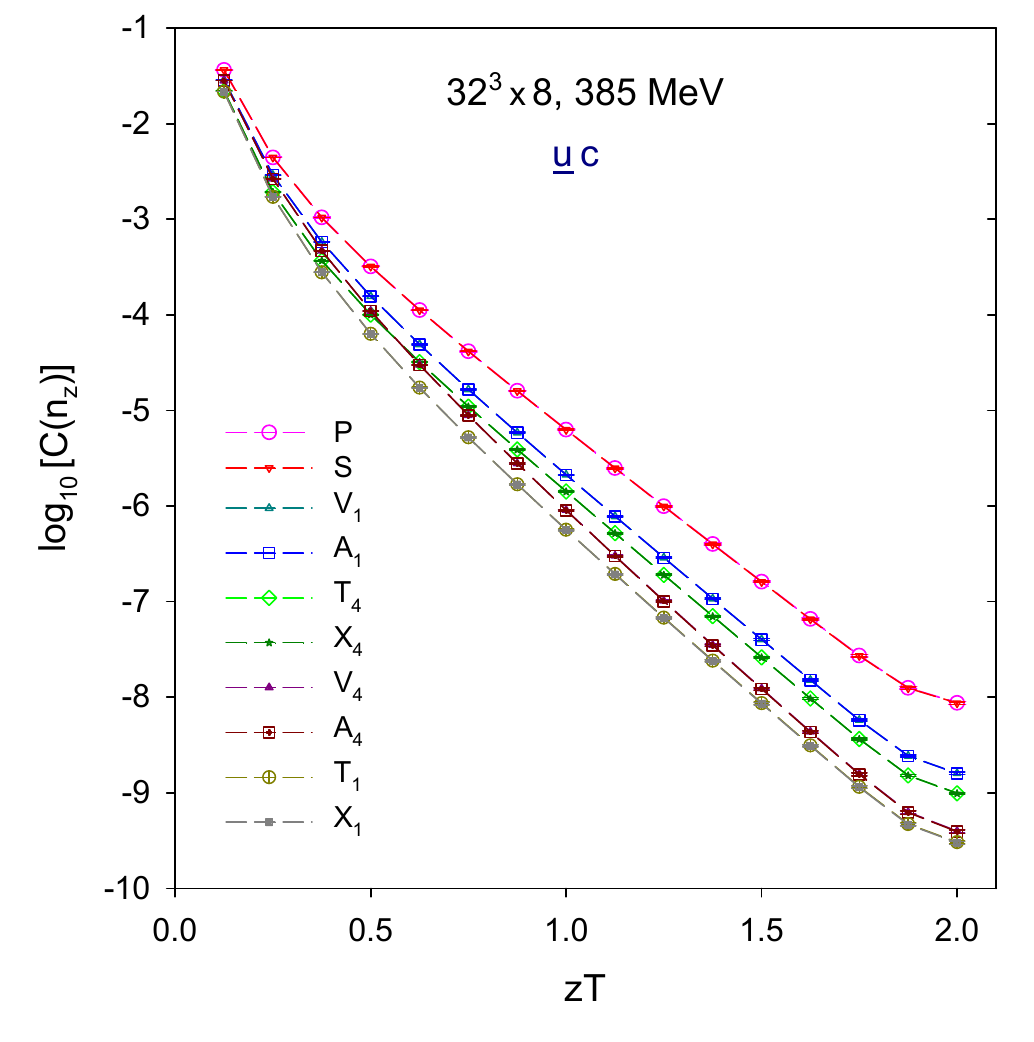}
\\
  \includegraphics[width=7.2cm,clip=true]{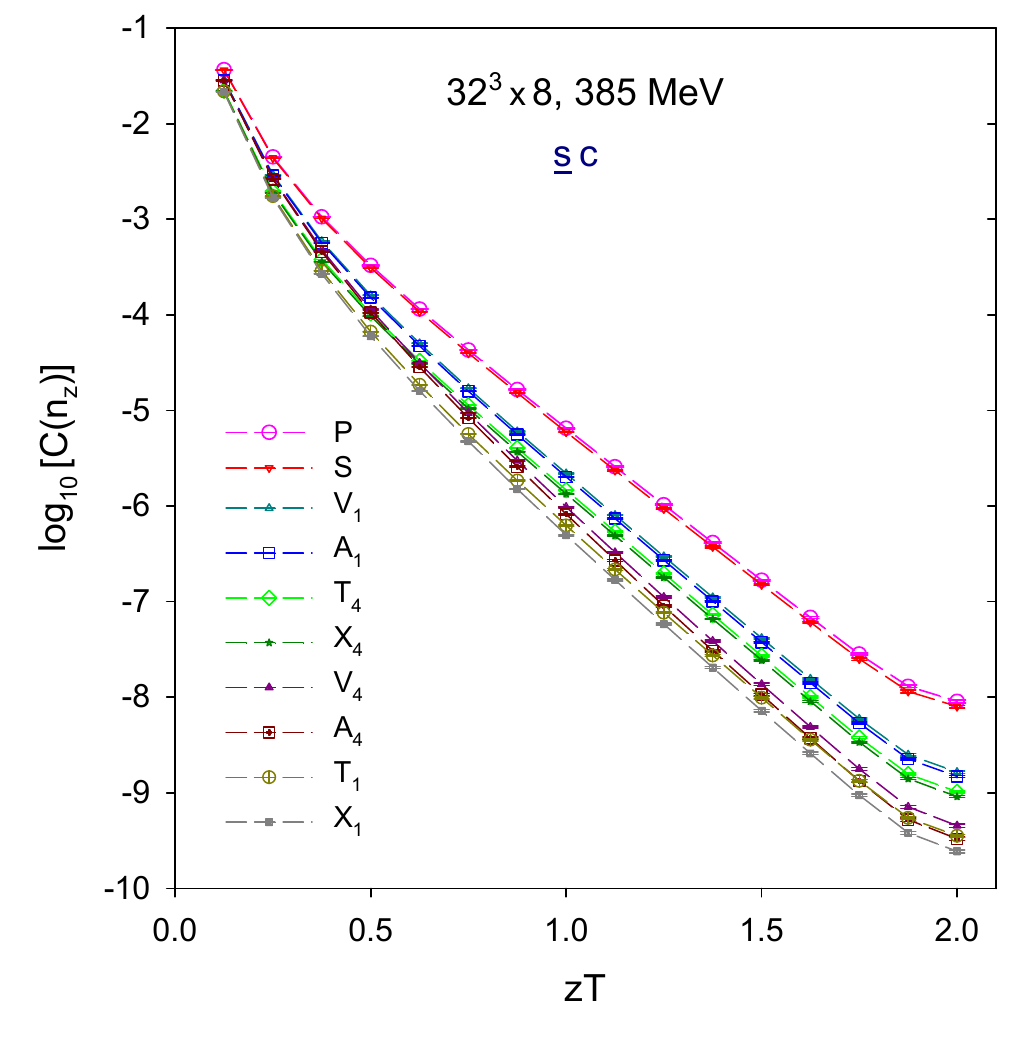}
&
  \includegraphics[width=7.2cm,clip=true]{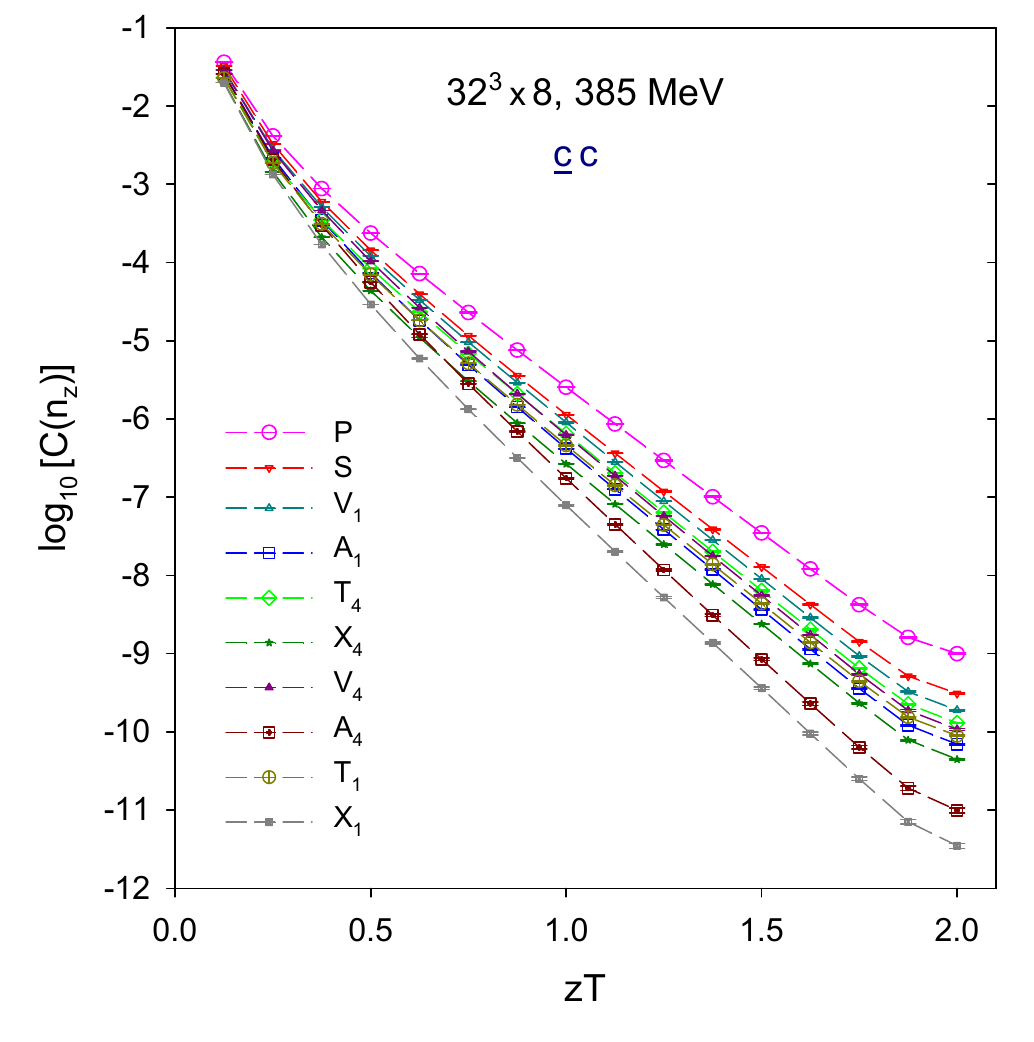}
  \end{tabular}
  \caption{
   The spatial $z$ correlators of meson interpolators for six flavor combinations
($\bar u d$, $\bar u s$, $\bar s s$, $\bar u c$, $\bar s c$, and $\bar c c $)
in $N_f=2+1+1$ lattice QCD at $T \simeq 385$~MeV.
  }
\label{fig:Cz_T385}
\end{figure*}

\begin{figure*}[tbp]
  \centering
  \begin{tabular}{@{}c@{}c@{}}
  \includegraphics[width=7.2cm,clip=true]{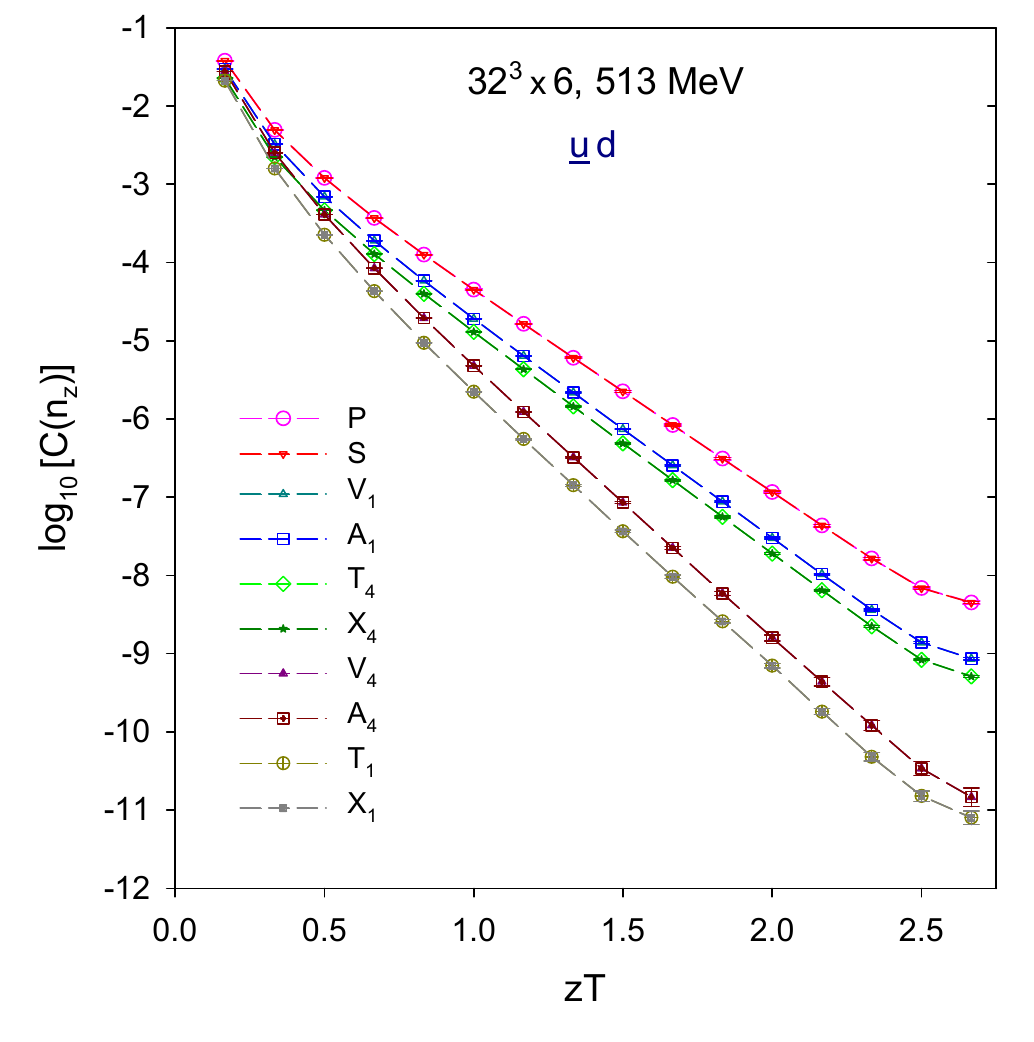}
&
  \includegraphics[width=7.2cm,clip=true]{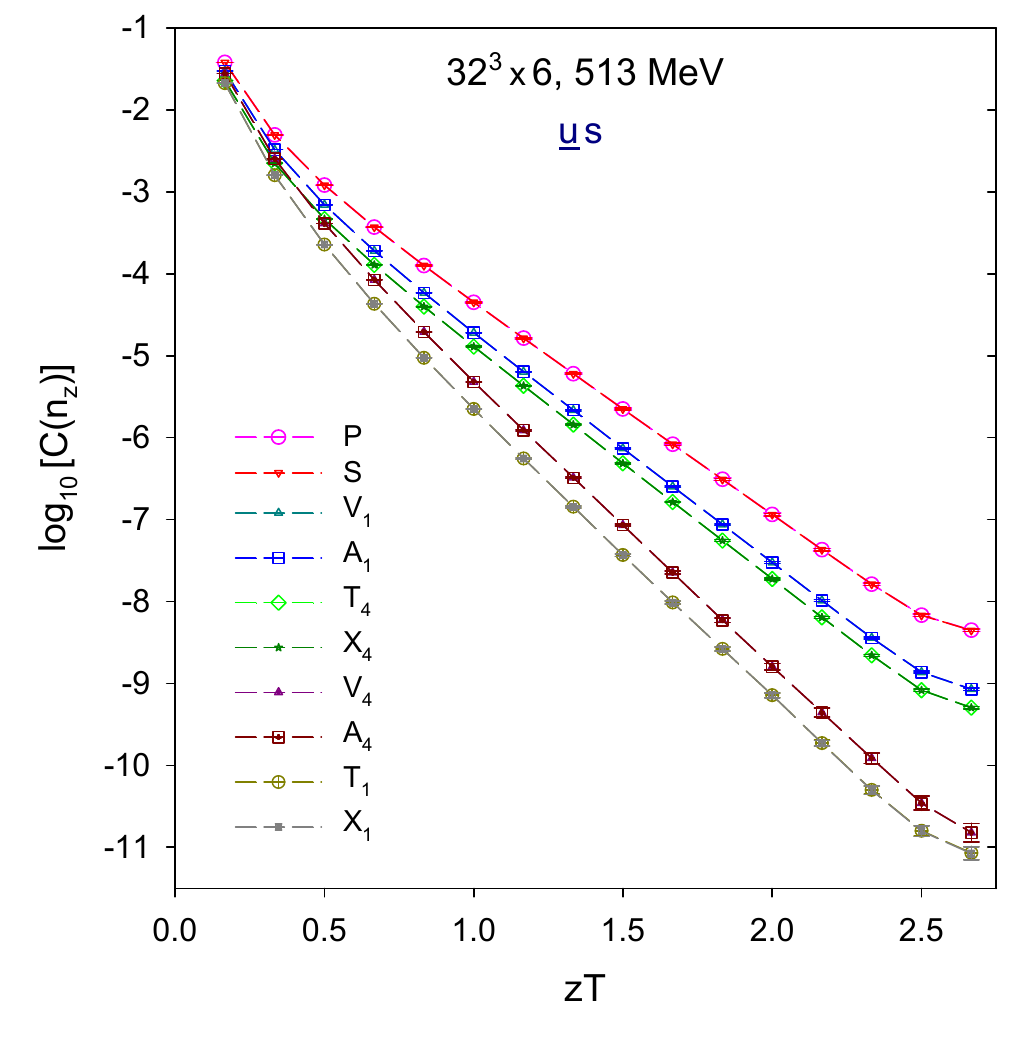}
\\
  \includegraphics[width=7.2cm,clip=true]{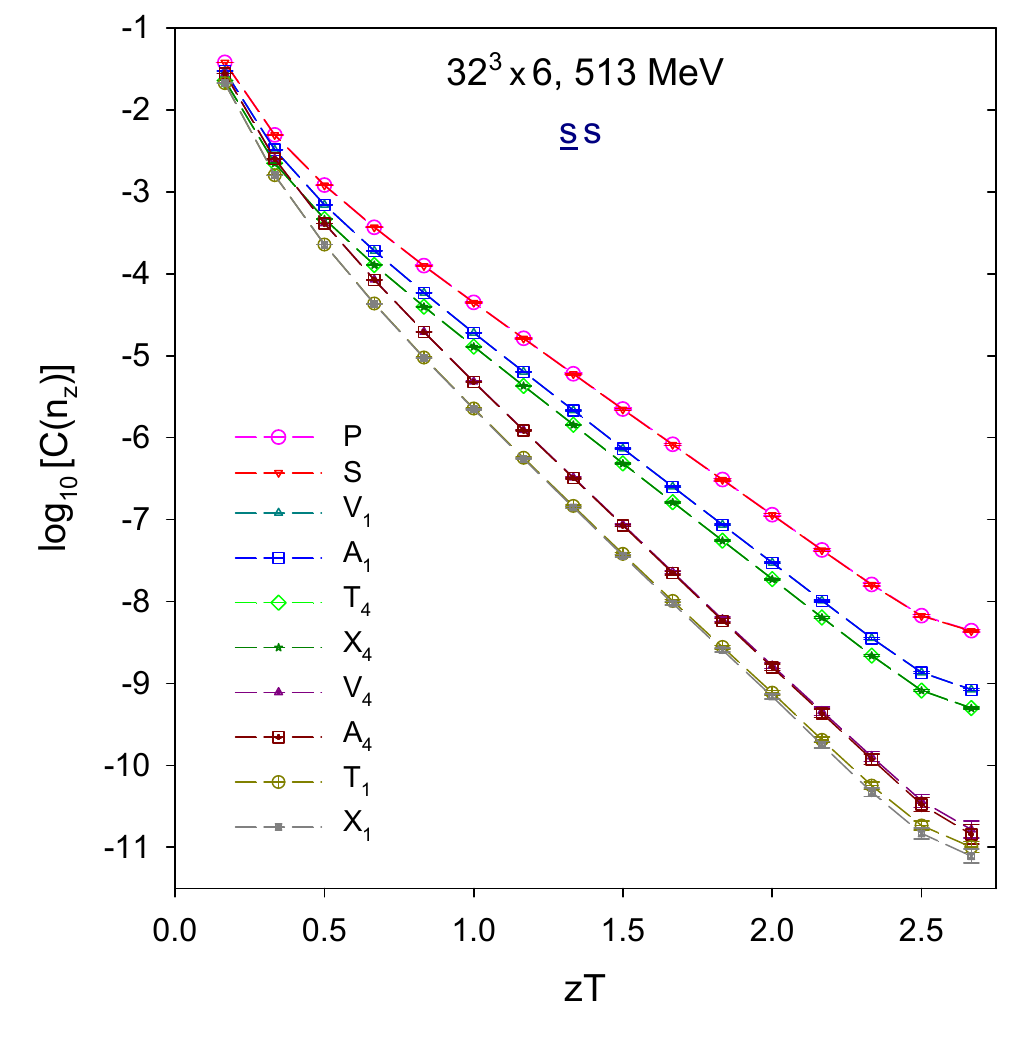}
&
  \includegraphics[width=7.2cm,clip=true]{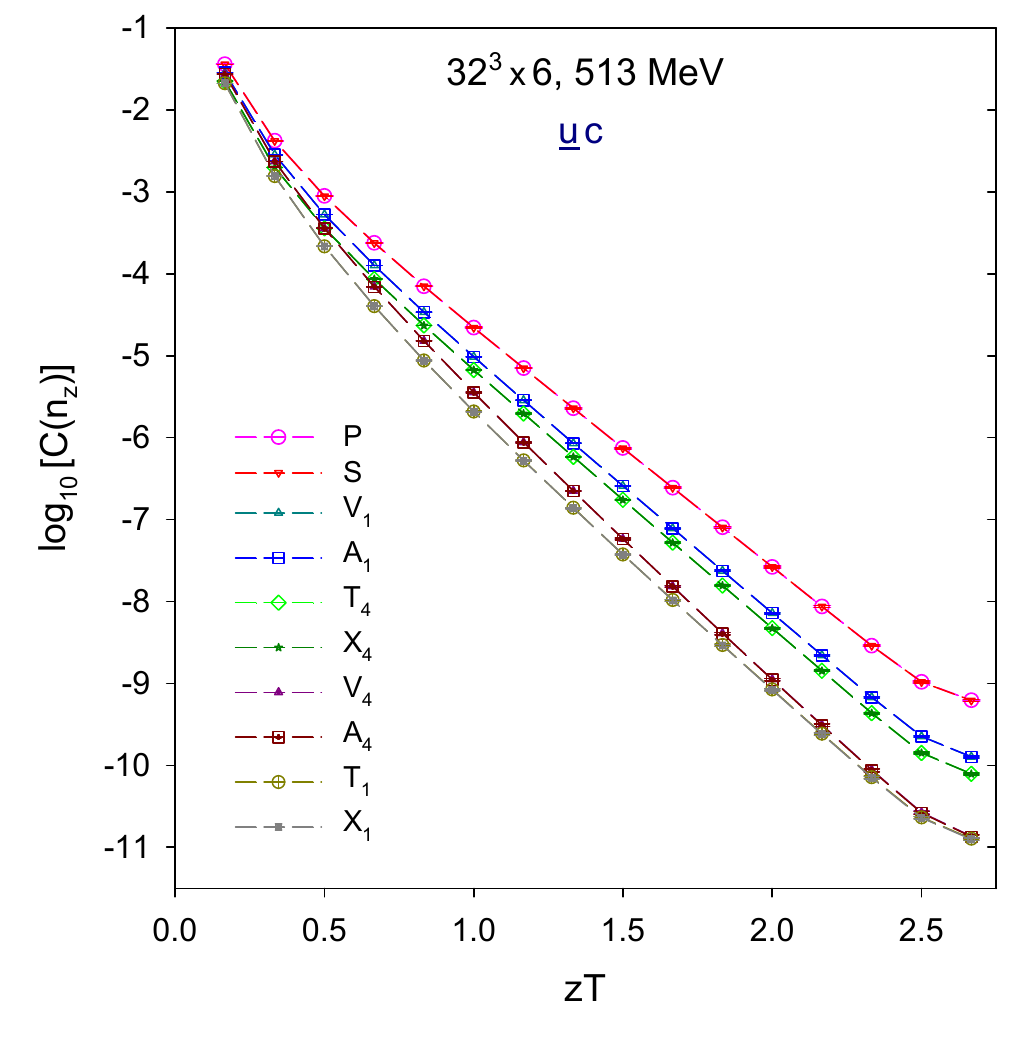}
\\
  \includegraphics[width=7.2cm,clip=true]{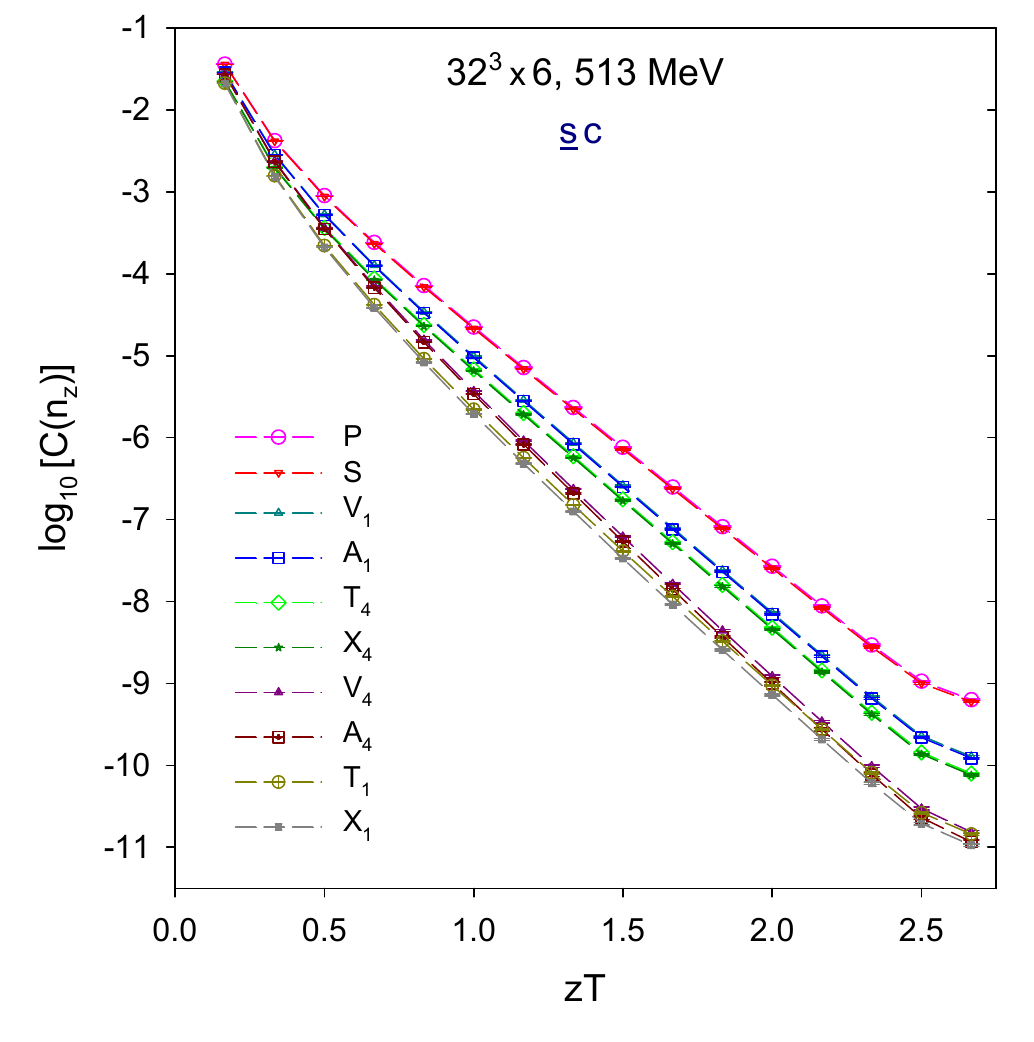}
&
  \includegraphics[width=7.2cm,clip=true]{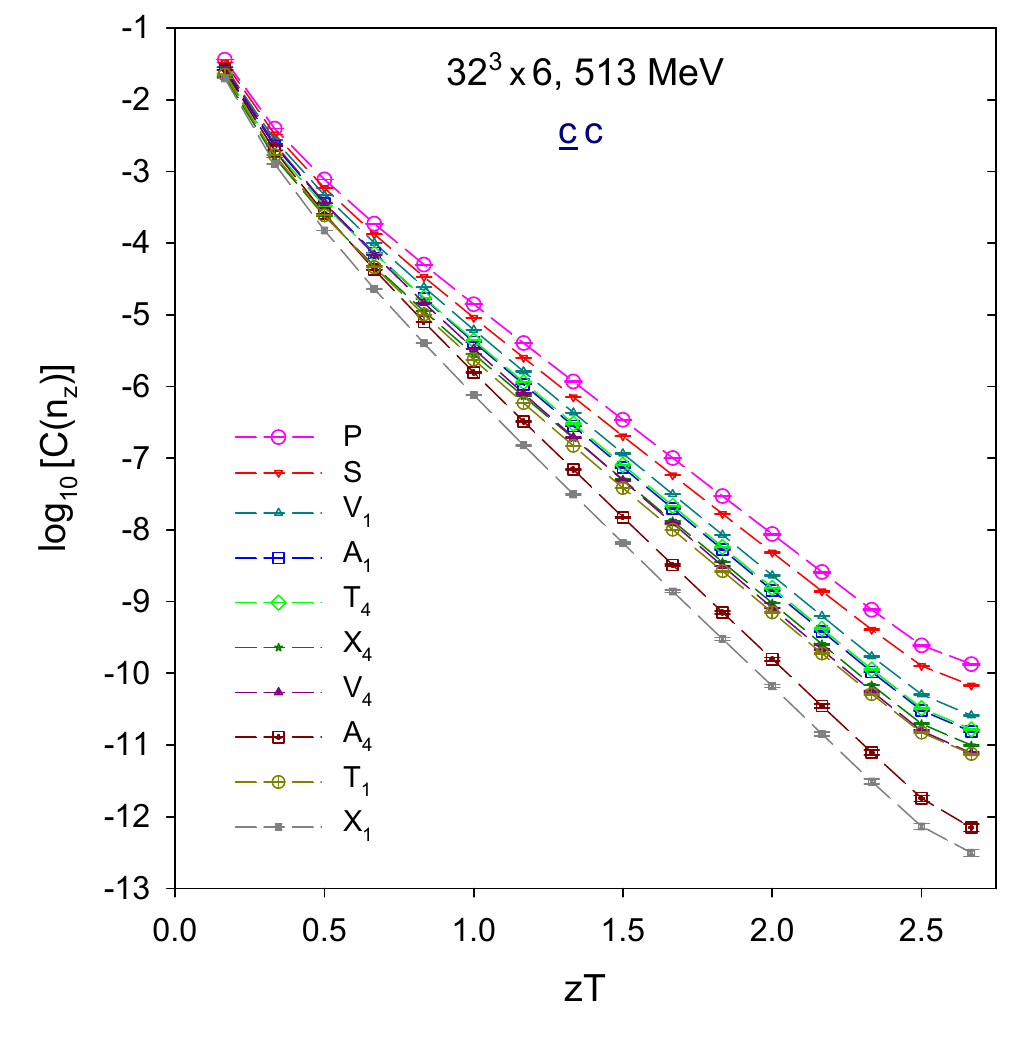}
  \end{tabular}
  \caption{
   The spatial $z$ correlators of meson interpolators for six flavor combinations
($\bar u d$, $\bar u s$, $\bar s s$, $\bar u c$, $\bar s c$, and $\bar c c $)
in $N_f=2+1+1$ lattice QCD at $T \simeq 513$~MeV.
  }
\label{fig:Cz_T513}
\end{figure*}

\begin{figure*}[tbp]
  \centering
  \begin{tabular}{@{}c@{}c@{}}
  \includegraphics[width=7.2cm,clip=true]{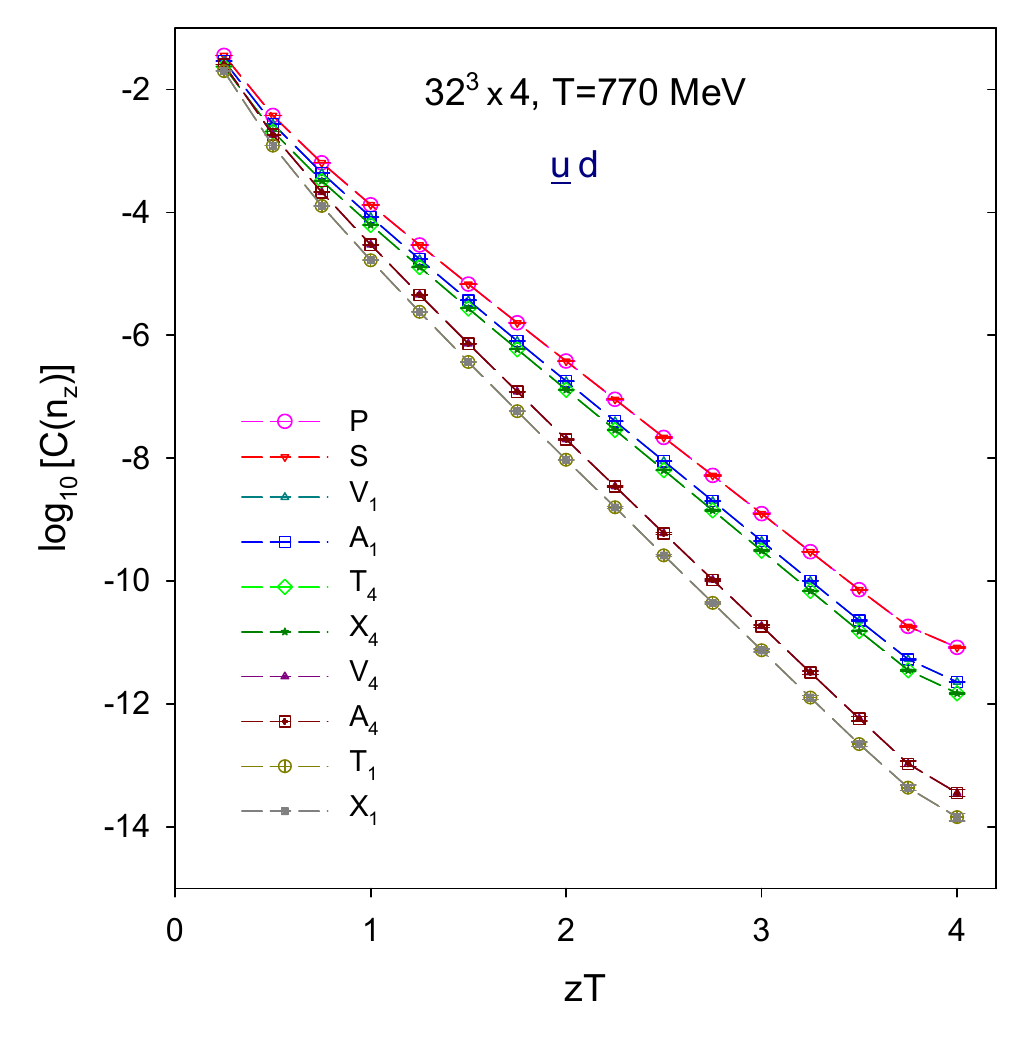}
&
  \includegraphics[width=7.2cm,clip=true]{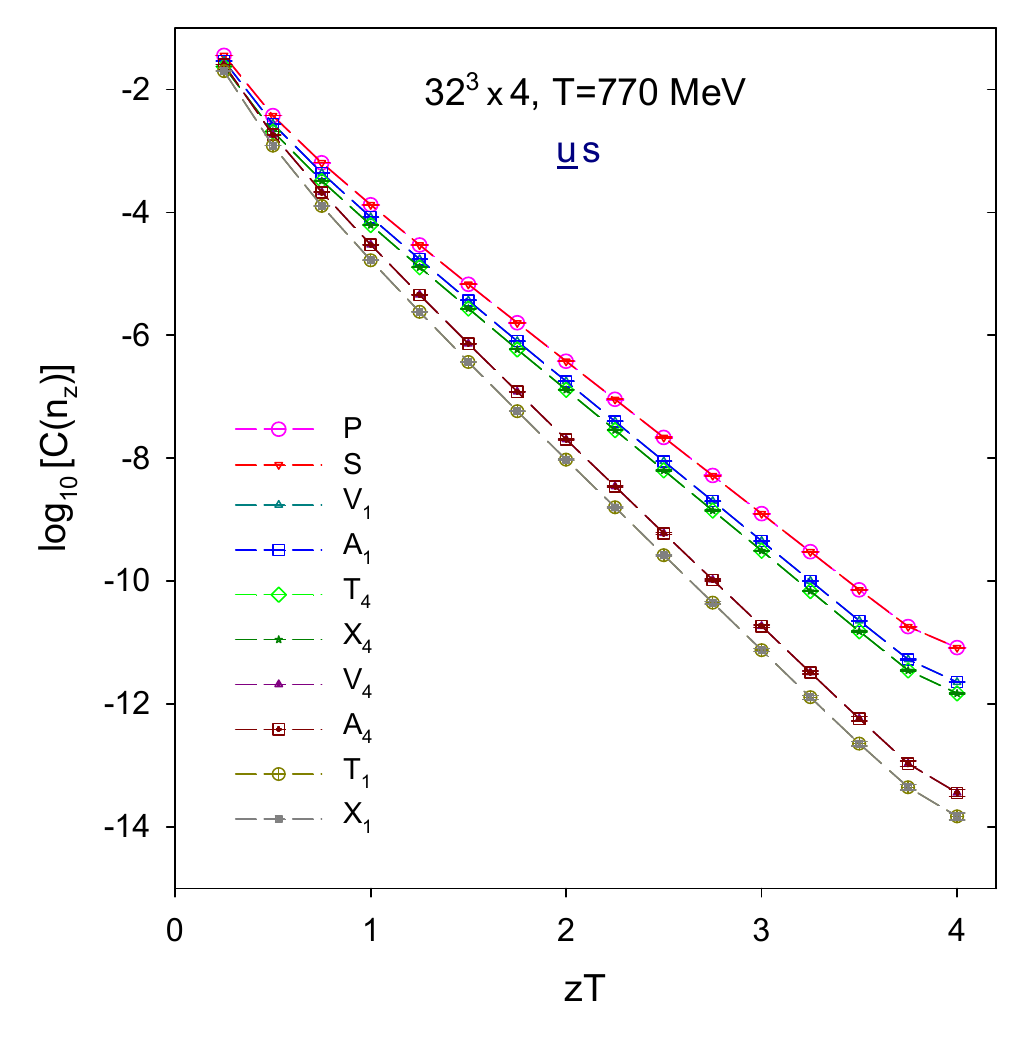}
\\
  \includegraphics[width=7.2cm,clip=true]{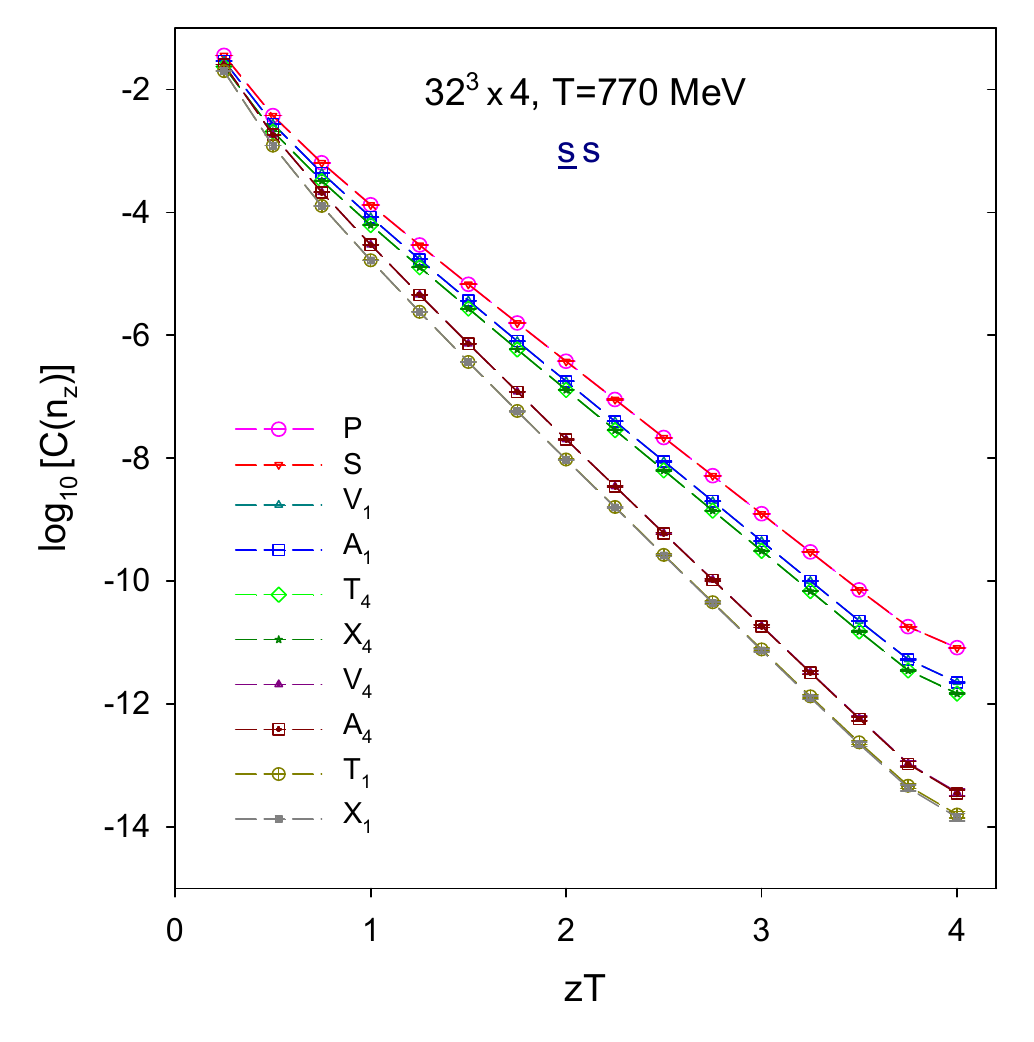}
&
  \includegraphics[width=7.2cm,clip=true]{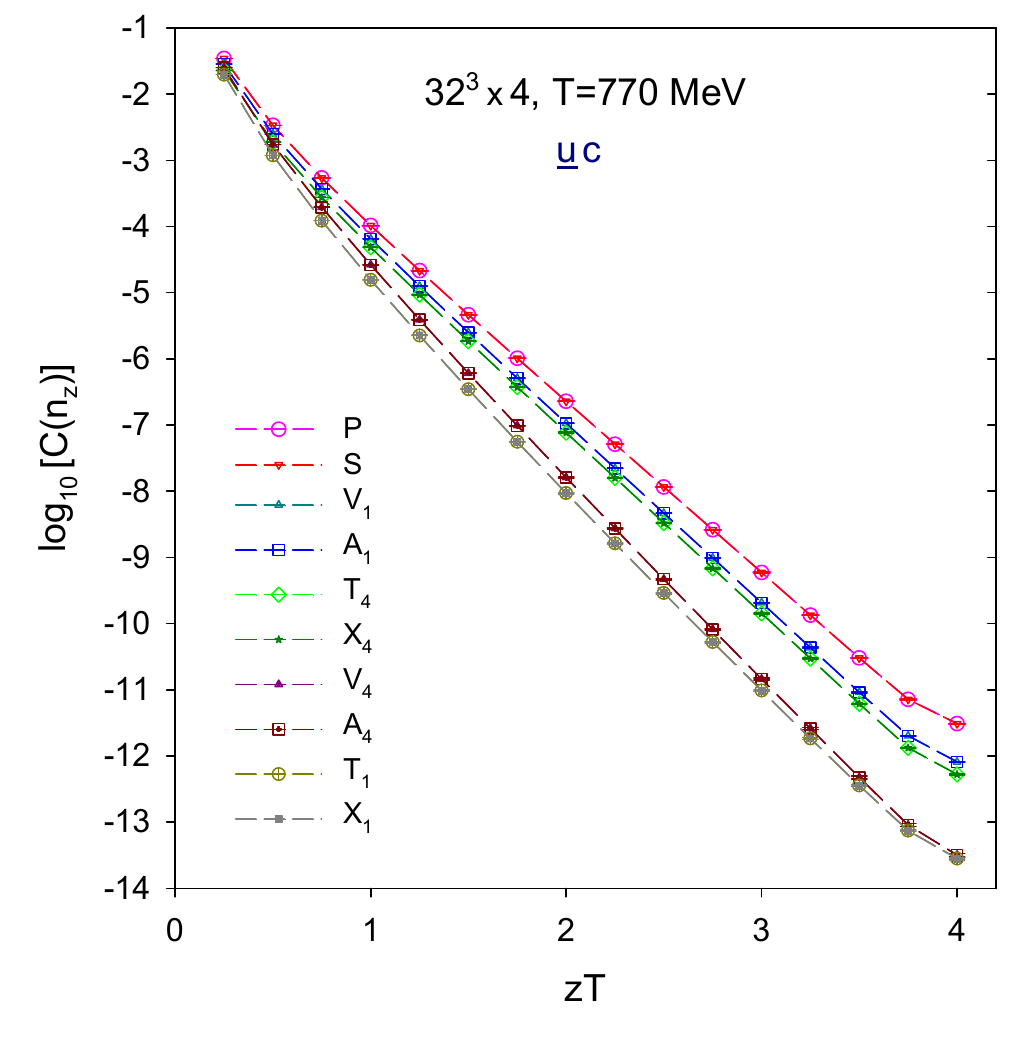}
\\
  \includegraphics[width=7.2cm,clip=true]{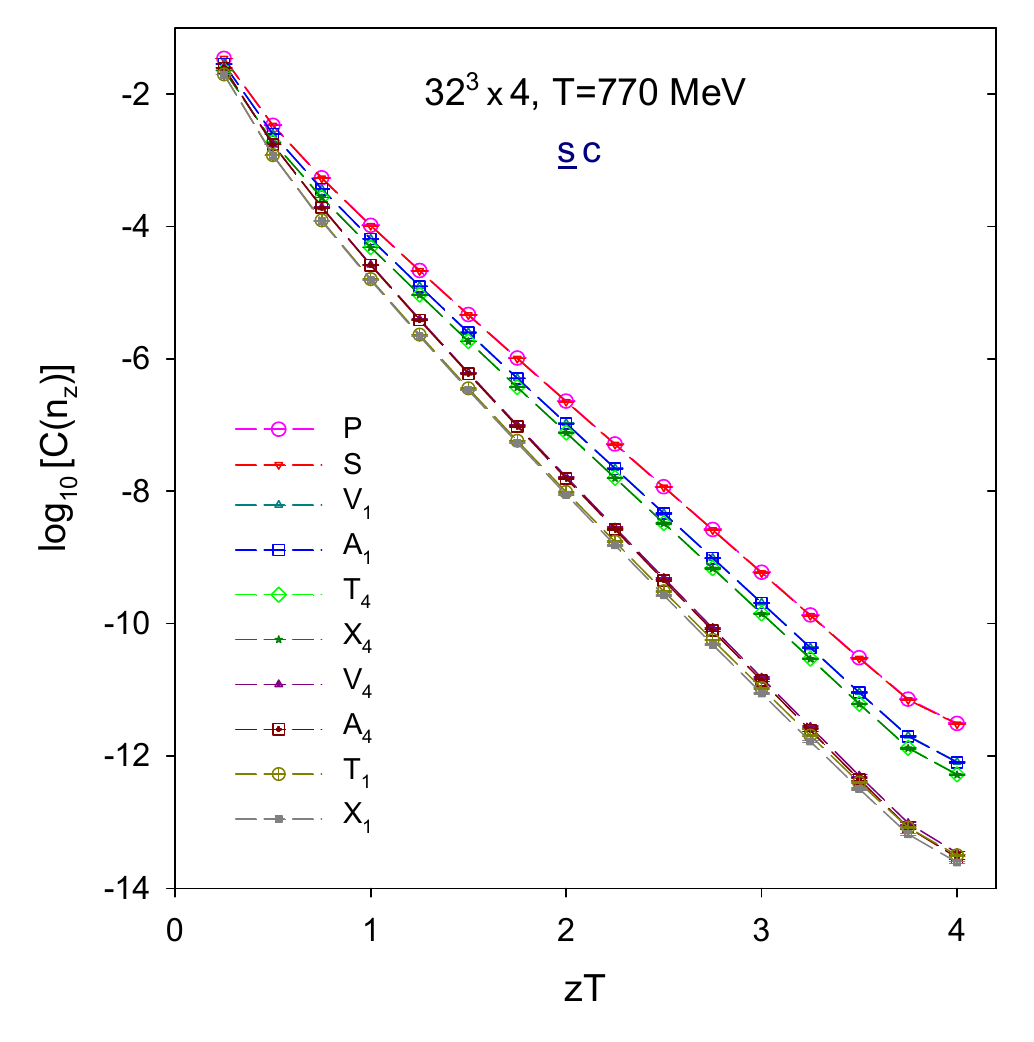}
&
  \includegraphics[width=7.2cm,clip=true]{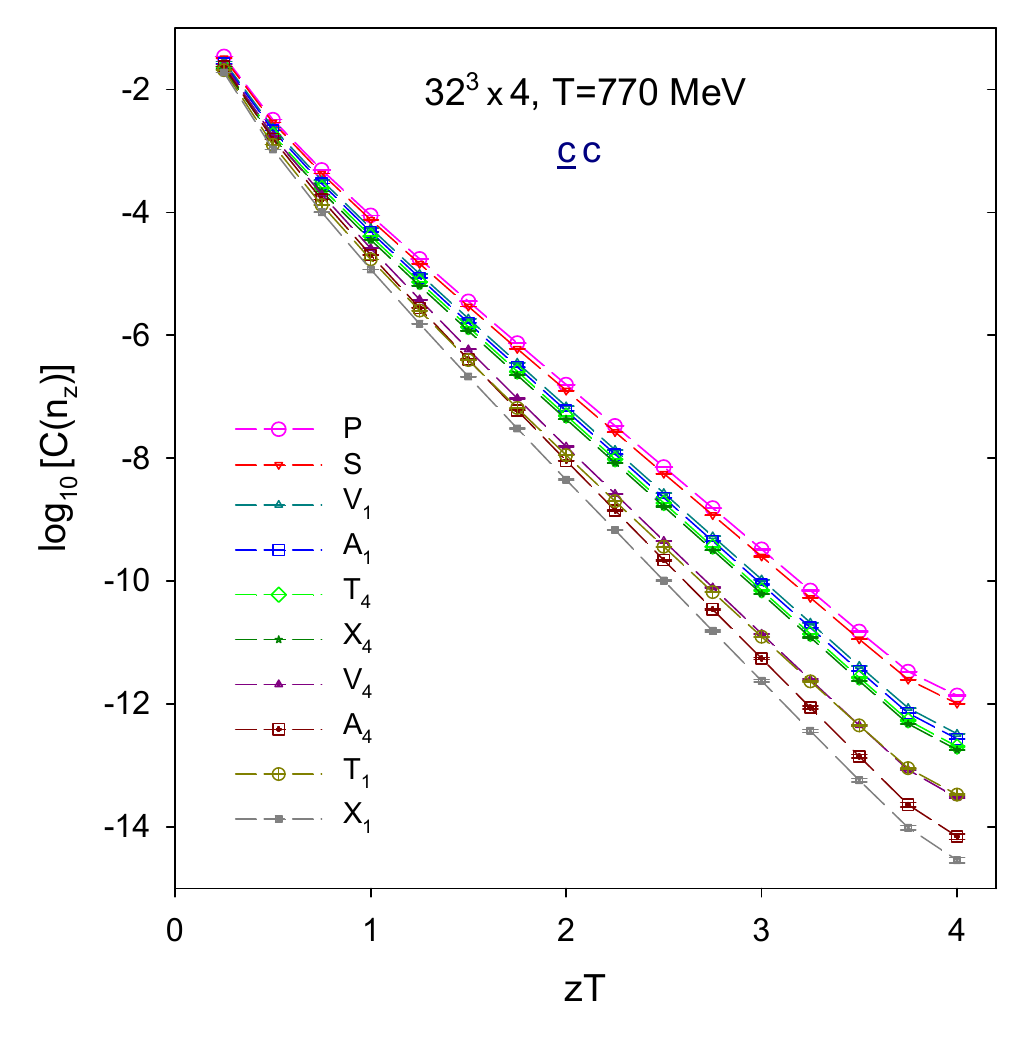}
  \end{tabular}
  \caption{
   The spatial $z$ correlators of meson interpolators for six flavor combinations
($\bar u d$, $\bar u s$, $\bar s s$, $\bar u c$, $\bar s c$, and $\bar c c $)
in $N_f=2+1+1$ lattice QCD at $T \simeq 770$~MeV.
  }
\label{fig:Cz_T770}
\end{figure*}

\begin{figure*}[tbp]
  \centering
  \begin{tabular}{@{}c@{}c@{}}
  \includegraphics[width=7.2cm,clip=true]{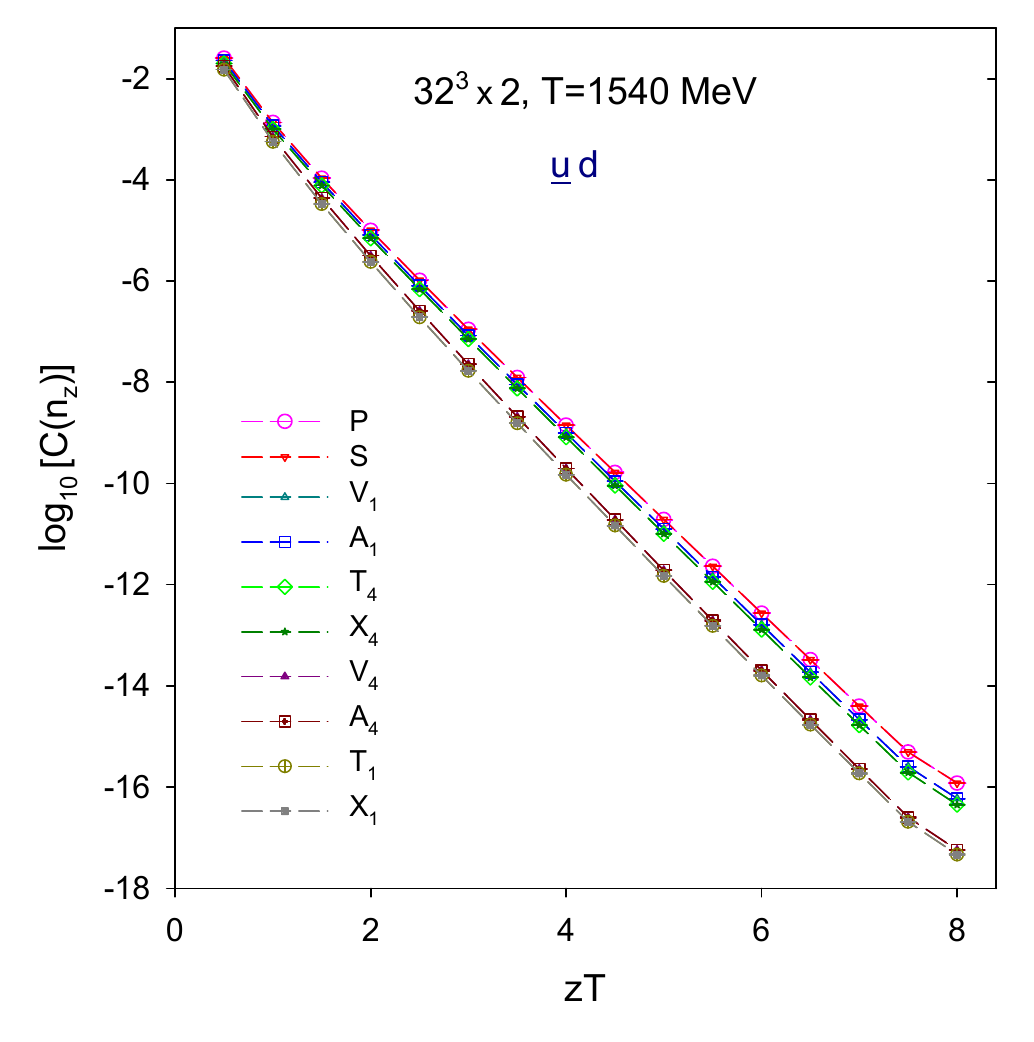}
&
  \includegraphics[width=7.2cm,clip=true]{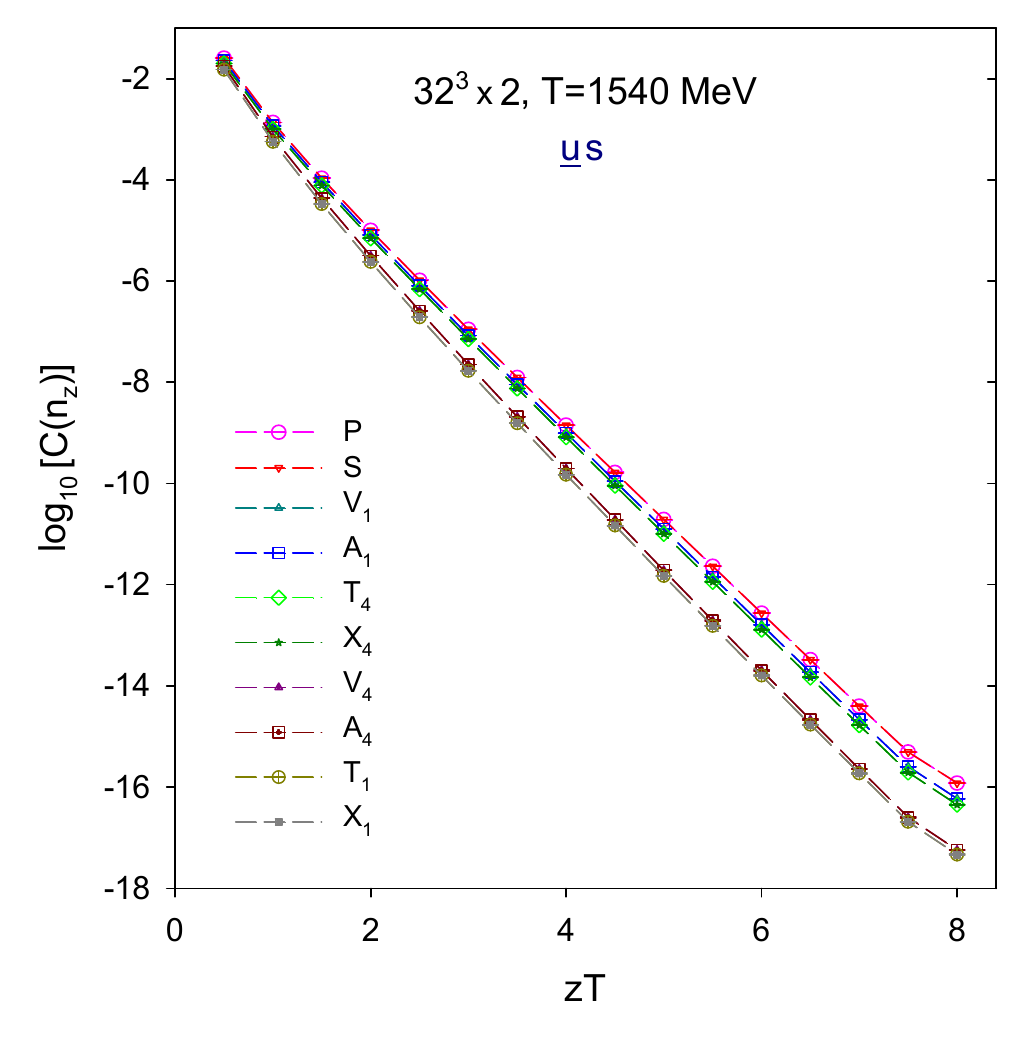}
\\
  \includegraphics[width=7.2cm,clip=true]{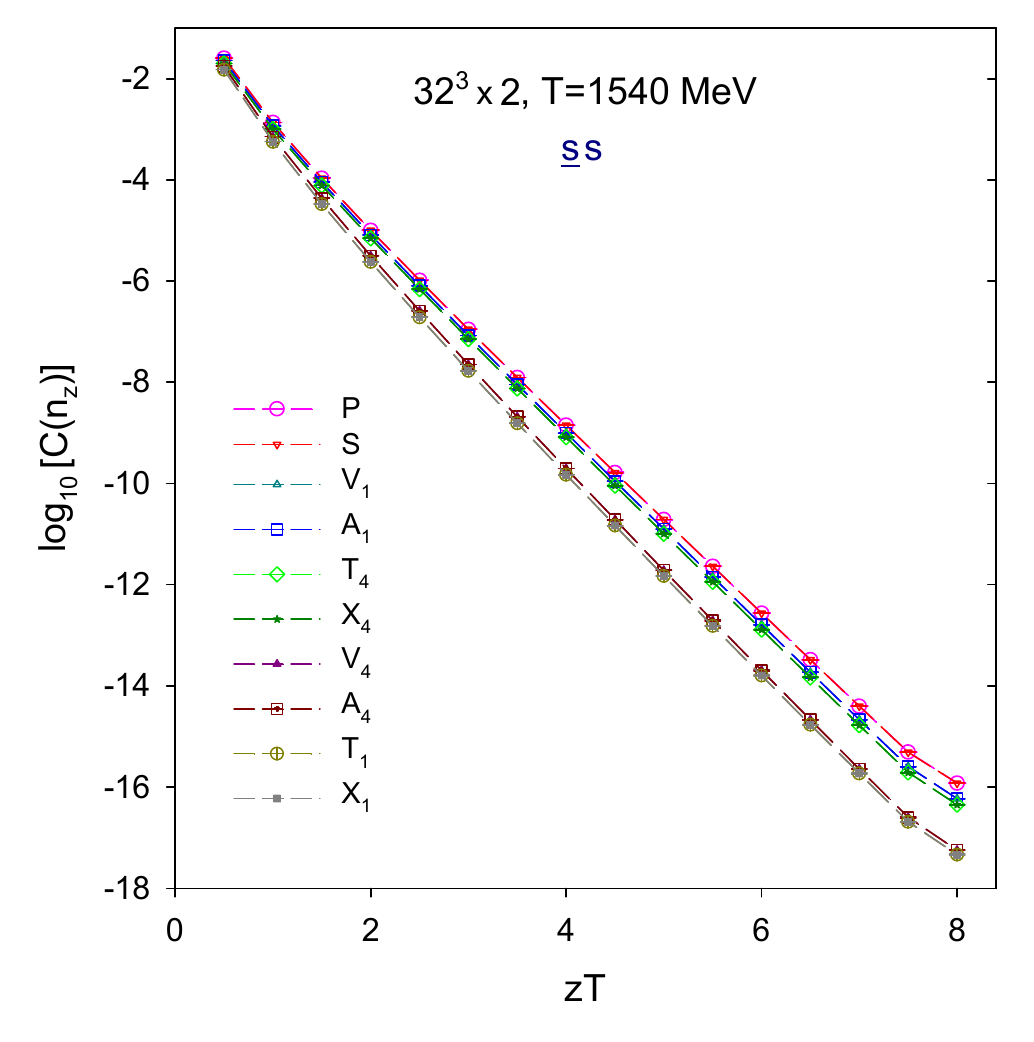}
&
  \includegraphics[width=7.2cm,clip=true]{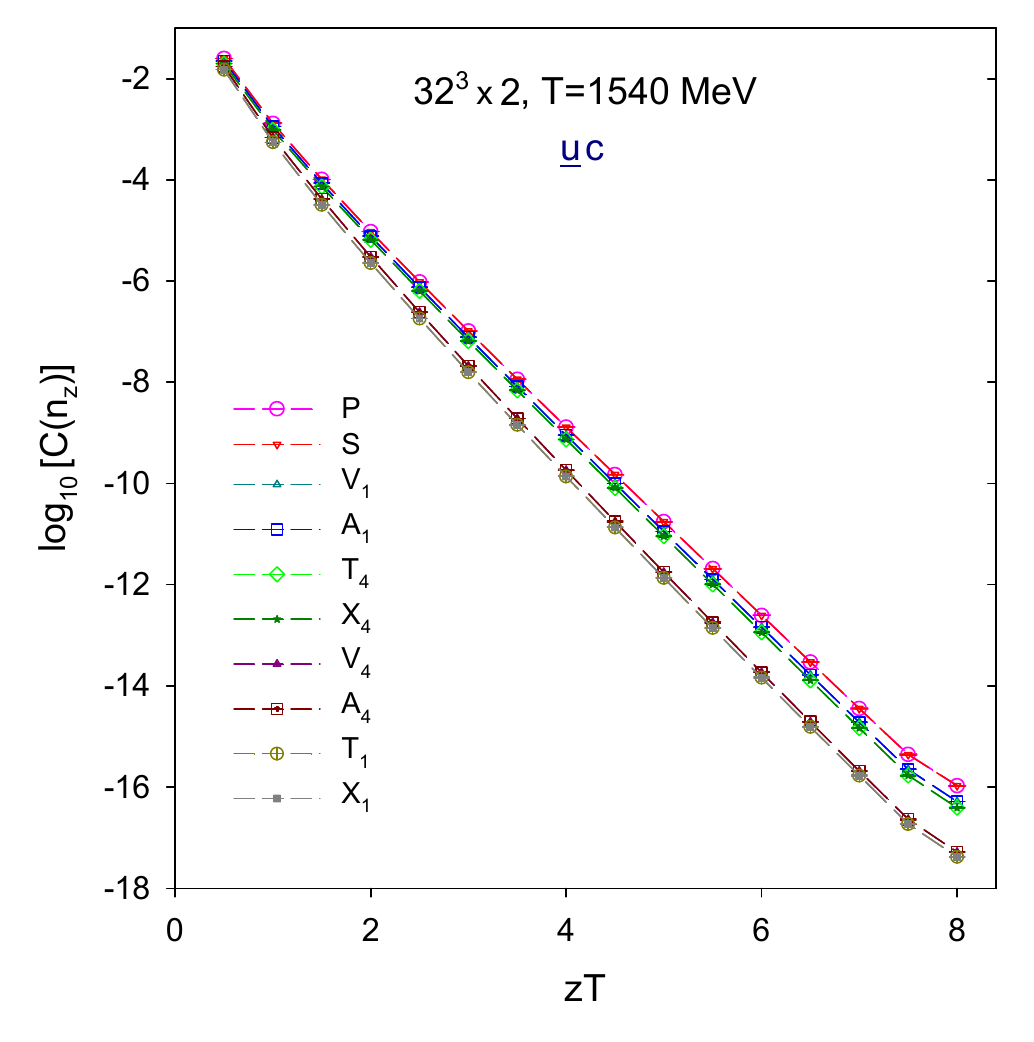}
\\
  \includegraphics[width=7.2cm,clip=true]{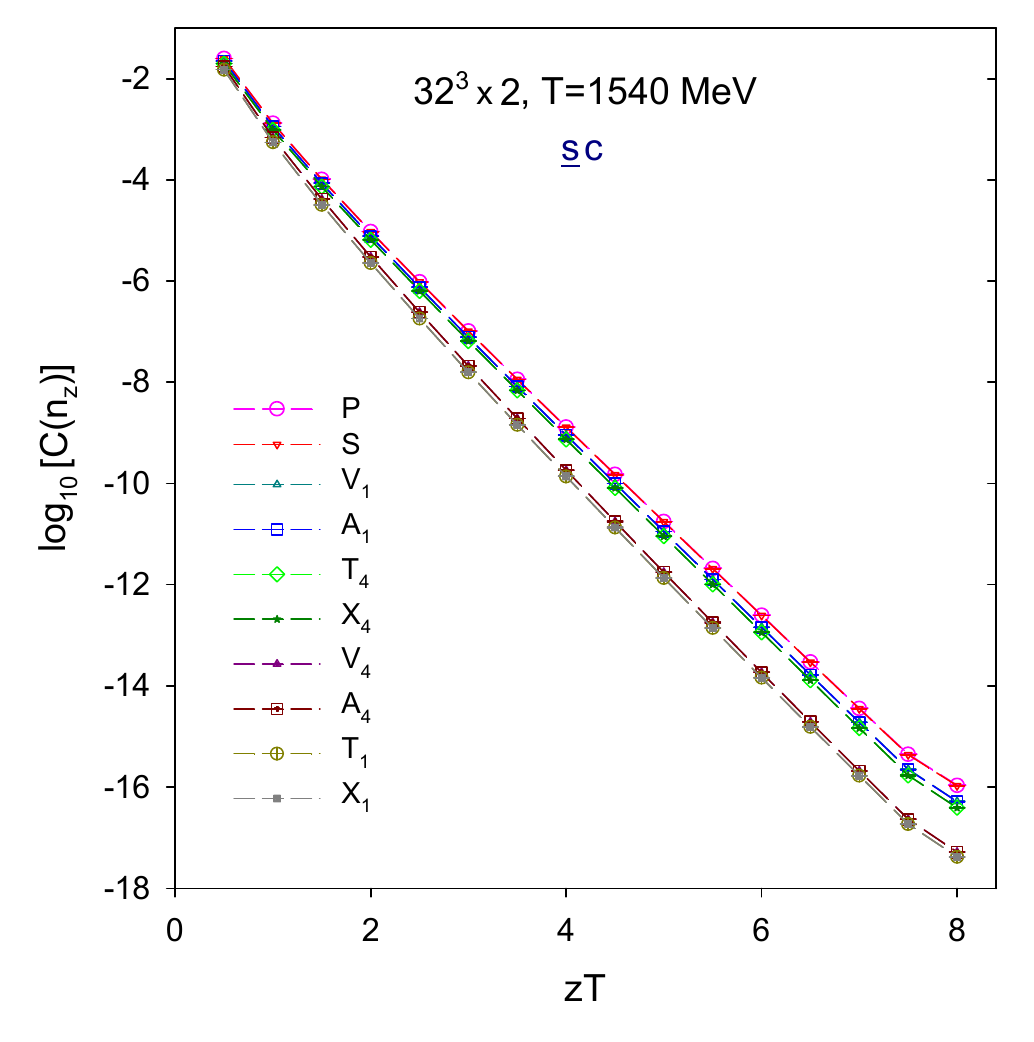}
&
  \includegraphics[width=7.2cm,clip=true]{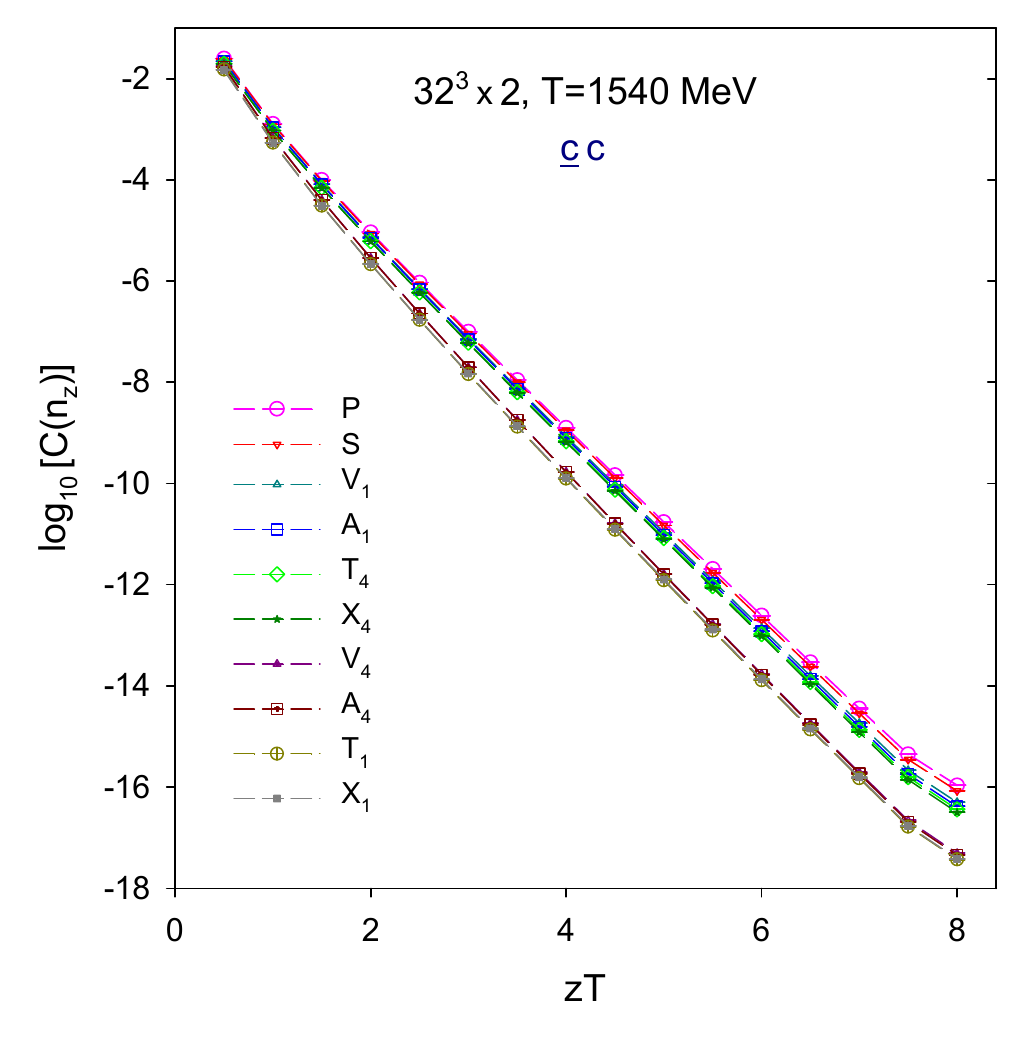}
  \end{tabular}
  \caption{
   The spatial $z$ correlators of meson interpolators for six flavor combinations
($\bar u d$, $\bar u s$, $\bar s s$, $\bar u c$, $\bar s c$, and $\bar c c $)
in $N_f=2+1+1$ lattice QCD at $T \simeq 1540$~MeV.
  }
\label{fig:Cz_T1540}
\end{figure*}

\section{Meson $z$ correlators of $(\bar u d, \bar u s, \bar s s, \bar u c, \bar s c, \bar c c)$} 
\label{meson_Cz}

Following the prescription proposed in Ref. \cite{Chiu:2023hnm} 
for the cancellation of the contribution of unphysical meson states to the $z$ correlators, 
we compute two sets of quark propagators with periodic and antiperiodic
boundary conditions in the $z$ direction, while their boundary conditions in $(x,y,t)$ directions 
are the same, i.e., periodic in the $(x,y)$ directions, and antiperiodic in the $t$ direction.  
Each set of quark propagators is used to construct the $z$ correlators independently, 
and finally taking the average of these two $z$ correlators. 
Then, the contribution of unphysical meson states to the $z$ correlators 
can be canceled configuration by configuration, up to the numerical precision of the quark propagators. 

In each of Figs. \ref{fig:Cz_T192}--\ref{fig:Cz_T1540}, 
the $z$ correlators for six flavor contents 
$(\bar u d, \bar u s, \bar s s, \bar u c, \bar s c, \bar c c)$
at the same $T$ are plotted as a function of the dimensionless variable $zT$ (\ref{eq:zT}). 
Because of the degeneracy (the $S_2$ symmetry) of the ``1" and ``2" components 
in the $z$ correlators of vector mesons, only the ``1" components are plotted. 
In general, each panel plots ten $C_\Gamma(zT)$ for 
$\Gamma = \{ P, S, V_1, A_1, T_4, X_4, V_4, A_4, T_1, X_1 \} $. 
For the classification and notations of meson interpolators, 
see Table \ref{tab:bilinear}. 

\begin{table}[tbp]
\centering
\caption{The classification of meson interpolators $\bar q_1 \Gamma q_2 $,
and their names and notations.}
\setlength{\tabcolsep}{4pt}
\vspace{2mm}
\begin{tabular}{c c }
\hline
\hline
Name and notation & $\Gamma$ (for $z$ correlators)  \\
\hline
\multicolumn{1}{l}{Scalar ($S$)}
&  $\Id$                         \\
\multicolumn{1}{l}{Pseudoscalar ($P$)}
&  $\gamma_5$                    \\
\multicolumn{1}{l}{Vector ($V_k$)}
&  $\gamma_k \ (k=1,2,4)$                    \\
\multicolumn{1}{l}{Axial vector ($A_k$)}
&  $\gamma_5 \gamma_k \ (k=1,2,4)$           \\
\multicolumn{1}{l}{Tensor vector ($T_k$)}
&  $\gamma_3 \gamma_k \ (k=1,2,4)$           \\
\multicolumn{1}{l}{Axial-tensor vector ($X_k$)}
&  $\gamma_5 \gamma_3 \gamma_k \ (k=1,2,4)$  \\
\hline
\hline
\end{tabular}
\label{tab:bilinear}
\end{table}

For any flavor combination, if the $SU(2)_L \times SU(2)_R $ chiral symmetry is restored, then its   
$(V_1, A_1)$ and $(V_4, A_4)$ become degenerate, and the number of distinct $z$ correlators 
appears to be reduced to 8. 
Furthermore, if the $U(1)_A$ symmetry is also restored, then its $(P, S)$,  
$(T_4, X_4)$ and $(T_1, X_1)$ also become degenerate, and the number of distinct $z$ correlators 
is further reduced to 5. Thus, one can visualize the effective restoration  
of $SU(2)_L \times SU(2)_R \times U(1)_A$ chiral symmetry when the number of distinct $z$ correlators 
becomes 5. This provides a simple guideline to look for the restoration 
of chiral symmetry from the panels in Figs. \ref{fig:Cz_T192}--\ref{fig:Cz_T1540}. 

Following the above guideline, we can visually identify the hierarchical restoration of chiral symmetries 
in $N_f=2+1+1$ QCD with $(u, d, s, c)$ quarks.

In Fig. \ref{fig:Cz_T192}, at $T=192$~MeV, we see that both 
$T_c^{\bar u d}$ 
[the temperature for the restoration of $SU(2)_L \times SU(2)_R$ chiral symmetry in the $\bar u d$ sector] 
and $T_1^{\bar u d}$ 
[the temperature for the restoration of $U(1)_A$ symmetry in the $\bar u d$ sector] 
are lower than 190 MeV, i.e., $T_c^{\bar u d} < 190$~MeV and $T_1^{\bar u d} < 190$~MeV.
Thus the $SU(2)_L \times SU(2)_R \times U(1)_A$ chiral symmetry of $\bar u d$ has been restored 
at some temperature lower than 190 MeV, 
i.e., $T_{c1}^{\bar u d} < 190$~MeV.

Moreover, in Fig. \ref{fig:Cz_T192}, 
the panels of $\bar u s$ and $\bar u c$ show that     
$T_c^{\bar u s} < 190$~MeV, $T_1^{\bar u s} < 190$~MeV, $T_c^{\bar u c} < 190$~MeV, and   
$T_1^{\bar u c} < 190$~MeV. Thus, $T_{c1}^{\bar u s} < 190$~MeV, and $T_{c1}^{\bar u c} < 190$~MeV. 

Next, we look at the $\bar s s$ panels in Figs. \ref{fig:Cz_T192}--\ref{fig:Cz_T1540}. 
In Fig. \ref{fig:Cz_T257}, at $T = 257$~MeV, 
it appears to have five distinct $z$ correlators in the channels of  
$(P, S)$, $(V_1, A_1)$, $(T_4, X_4)$, $(V_4, A_4)$ and $(T_1, X_1)$, 
in spite of the small splittings at large $z$ in the channels of $(V_4, A_4)$ and $(T_1, X_1)$.
Thus, the $SU(2)_L \times SU(2)_R \times U(1)_A$ chiral symmetry of $\bar s s$ 
can be regarded to be restored at $T_{c1}^{\bar s s} \sim 257$~MeV.
This implies that the $SU(3)_L \times SU(3)_R \times U(1)_A $ chiral symmetry of $(u, d, s)$ quarks
is restored at $T_{c1}^{\bar s s} \sim 257$~MeV, since the $SU(2)_L \times SU(2)_R \times U(1)_A$ chiral symmetry in both 
$\bar u d$ and $\bar u s$ sectors has been restored at $T < 190$~MeV.
This is the first step of the hierarchical restoration of chiral symmetries 
in $N_f=2+1+1$ lattice QCD at the physical point,   
from the restoration of $SU(2)_L \times SU(2)_R \times U(1)_A$ chiral symmetry of $(u, d)$ quarks 
at $T_{c1}^{\bar u d} < 190$~MeV to 
the restoration of $SU(3)_L \times SU(3)_R \times U(1)_A $ chiral symmetry of $(u, d, s)$ quarks
at $T_{c1}^{\bar s s} \sim 257$~MeV.

Note that, as discussed in Secs. \ref{intro} and \ref{kappa}, 
the restoration of $SU(3)_L \times SU(3)_R \times U(1)_A$ chiral symmetry of $(u, d, s)$ quarks 
requires the $SU(2)_L \times SU(2)_R \times U(1)_A$ chiral symmetry 
for all six flavor combinations  
($\bar u d$, $\bar u s$, $\bar d s$, $\bar u u$, $\bar d d$, $\bar s s$), 
which are reduced to ($\bar u d$, $\bar u s$, $\bar s s$) if $m_u = m_d$. 
Here we have assumed that in high temperature QCD, the contribution 
of the disconnected diagrams to the $z$ correlator of $\bar q \Gamma q$  
is negligible in comparison with that of the connected ones, as discussed in Sec. \ref{intro}. 
Similarly, the restoration of $SU(4)_L \times SU(4)_R \times U(1)_A$ chiral symmetry of 
$(u/d, s, c)$ quarks requires the $SU(2)_L \times SU(2)_R \times U(1)_A$ chiral symmetry 
for all six flavor combinations  
$\bar u d$, $\bar u s$, $\bar s s$, $\bar u c$, $\bar s c$, and $\bar c c$. 

Next, we look at the $\bar s c$ panels in Figs. \ref{fig:Cz_T192}--\ref{fig:Cz_T1540}. 
The $SU(2)_L \times SU(2)_R \times U(1)_A$ 
chiral symmetry seems to manifest at $T=385$~MeV, and it becomes highly pronounced at $T=513$~MeV.
This implies that $T_{c1}^{\bar s c}$ is in the range of 385--512 MeV. 
In general, a more precise estimate of $T_c$ and $T_1$ can be obtained 
by the criteria (\ref{eq:SU2_crit_z}) and (\ref{eq:U1_TX_crit_z}), which will be given 
in the next section.
 
Finally, we look at the $\bar c c$ panels in Figs. \ref{fig:Cz_T192}--\ref{fig:Cz_T1540}. 
The $SU(2)_L \times SU(2)_R \times U(1)_A $ chiral symmetry of $\bar c c$ seems to manifest 
at $T=770$~MeV, and it becomes highly pronounced at $T=1540$~MeV.
This implies that $T_{c1}^{\bar c c}$ is in the range of 770--1540 MeV, 
and also the restoration of the $SU(4)_L \times SU(4)_R \times U(1)_A $ chiral symmetry 
of $(u, d, s, c)$ quarks at $ T_{c1}^{\bar c c} \sim $ 770--1540 MeV, 
since the $SU(2)_L \times SU(2)_R \times U(1)_A $ chiral symmetry in other sectors 
$(\bar u d, \bar u s, \bar s s, \bar u c, \bar s c)$ has already been restored at lower temperatures. 
This gives the second step of the hierarchical restoration of chiral symmetries 
in $N_f=2+1+1$ lattice QCD at the physical point,   
from the restoration of the $SU(3)_L \times SU(3)_R \times U(1)_A$ chiral symmetry of $(u, d, s)$ quarks 
at $T_{c1}^{\bar s s } \sim 257$~MeV to 
the restoration of $SU(4)_L \times SU(4)_R \times U(1)_A $ chiral symmetry of $(u, d, s, c)$ quarks
at $T_{c1}^{\bar c c} \sim 770$--$1540$~MeV.
A more precise estimate of $T_c^{\bar c c}$ and $T_1^{bar c c}$ can be obtained 
by the criteria (\ref{eq:SU2_crit_z}) and (\ref{eq:U1_TX_crit_z}), 
which will be given in the next subsection.

Besides the hierarchical restoration of chiral symmetries, we are also interested in 
visually identifying the emergence of the approximate $SU(2)_{CS}$ chiral spin symmetry 
in each of the six flavor sectors. 
To this end, we look for the appearance of three approximately distinct multiplets 
\BAN
\label{eq:M0}
&& M_0 = ( P, S ), \\
\label{eq:M2}
&& M_2 = ( V_1, A_1, T_4, X_4 ), \\
\label{eq:M4}
&& M_4 = ( V_4, A_4, T_1, X_1 ),
\EAN
which become more pronounced at higher temperatures, and they are in the order
\BAN
C_{M_0} > C_{M_2} > C_{M_4}.
\EAN
The emergence of $M_2$ and $M_4$ is in agreement with the $SU(2)_{CS}$ multiplets 
of (\ref{eq:SU2CS_z_a}) and (\ref{eq:SU2CS_z_b}),
and the $SU(2)_{CS} \times SU(2)_L \times SU(2)_R$ multiplets of 
(\ref{eq:SU2CS_SU2A_z_a}) and (\ref{eq:SU2CS_SU2A_z_b}). 
This suggests the emergence of the approximate $SU(2)_{CS}$ and $SU(4)$ symmetries. 
Moreover, the separation between the multiplets $M_2$ and $M_0$ is decreased
as the temperature is increased further. Thus, at sufficiently high temperatures, say $T > T_f$, 
$M_2$ and $M_0$ merge together to form a single multiplet,  
then the approximate $SU(2)_{CS}$ symmetry becomes washed out,  
and only the $SU(2)_L \times SU(2)_R \times U(1)_A $ chiral symmetry remains.
In other words, the approximate $SU(2)_{CS}$ symmetry
can only appear in a window of $T$ above $T_{c1}$,
i.e., $T_{c1} < T_{cs} \lesssim T \lesssim T_f $,
where $T_{cs} $ ($T_{f}$) depends on $\epsilon_{cs}$ ($\epsilon_{fcs}$) in the criterion
(\ref{eq:SU2_CS_crit_z}) for the emergence (fading) of the approximate $SU(2)_{CS}$ symmetry. 
Note that the multiplet $M_4$ never merges with the multiplets $M_0$ and $M_2$,
even in the limit $T \to \infty $, as discussed in Ref. \cite{Chiu:2023hnm}. 
Thus $M_4$ is irrelevant to the fading of the approximate $SU(2)_{CS} $ symmetry.  
The above provides a guideline to look for the emergence and the fading 
of the approximate $SU(2)_{CS}$ symmetry in Figs. \ref{fig:Cz_T192}--\ref{fig:Cz_T1540}. 

First, we look at the panels of $\bar u d$ and $\bar u s$ in Figs. \ref{fig:Cz_T192}--\ref{fig:Cz_T1540}. 
We see that their $z$ correlators are almost identical for all seven temperatures.
Furthermore, as $T$ is increased from 192 to 770 MeV,
we see the emergence of three approximately distinct multiplets  
$M_0$, $M_2$, and $M_4$, which become more pronounced at higher temperatures, 
while the separation of $M_0$ and $M_2$ become smaller.
This suggests the emergence of the approximate $SU(2)_{CS}$ and $SU(4)$ symmetries 
in the window $T \sim $ 308--770 MeV, for both $\bar u d$ and $\bar u s$ sectors.
Finally, at $T = 1540$~MeV, $M_0$ and $M_2$ (for any flavor combination) merge together 
to form a single multiplet, and the approximate $SU(2)_{CS}$ symmetry has become completely washed out, 
and only the chiral symmetry remains. 

Next, from the $\bar s s$ panels in Figs. \ref{fig:Cz_T192}--\ref{fig:Cz_T1540}, 
we see that its window for the approximate $SU(2)_{CS}$ symmetry 
is almost the same as that of $\bar u d$ and $\bar u s$, i.e., $T \sim $ 308--770 MeV. 

Finally, we visually estimate the windows of the approximate $SU(2)_{CS}$ symmetry 
for heavy mesons with the $c$ quark, which seem to be similar to that of the light mesons. 
However, if one performs a more precise estimate    
with the criterion (\ref{eq:SU2_CS_crit_z}), one can reveal some 
salient features of the heavy vector mesons which cannot be easily observed by visual estimate, 
as shown in the next section.


\begin{figure*}[tbp]
  \centering
  \begin{tabular}{@{}c@{}c@{}}
  \includegraphics[width=7.2cm,clip=true]{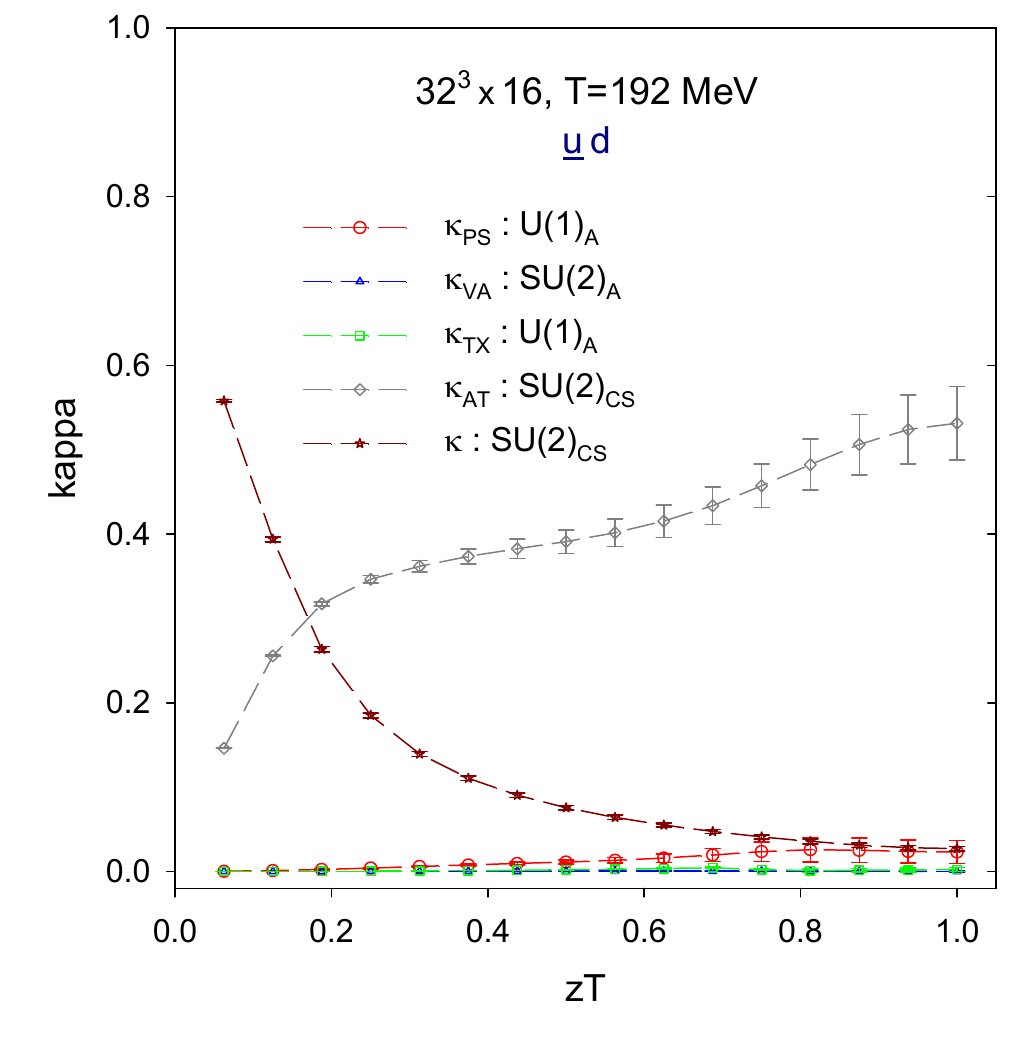}
&
  \includegraphics[width=7.2cm,clip=true]{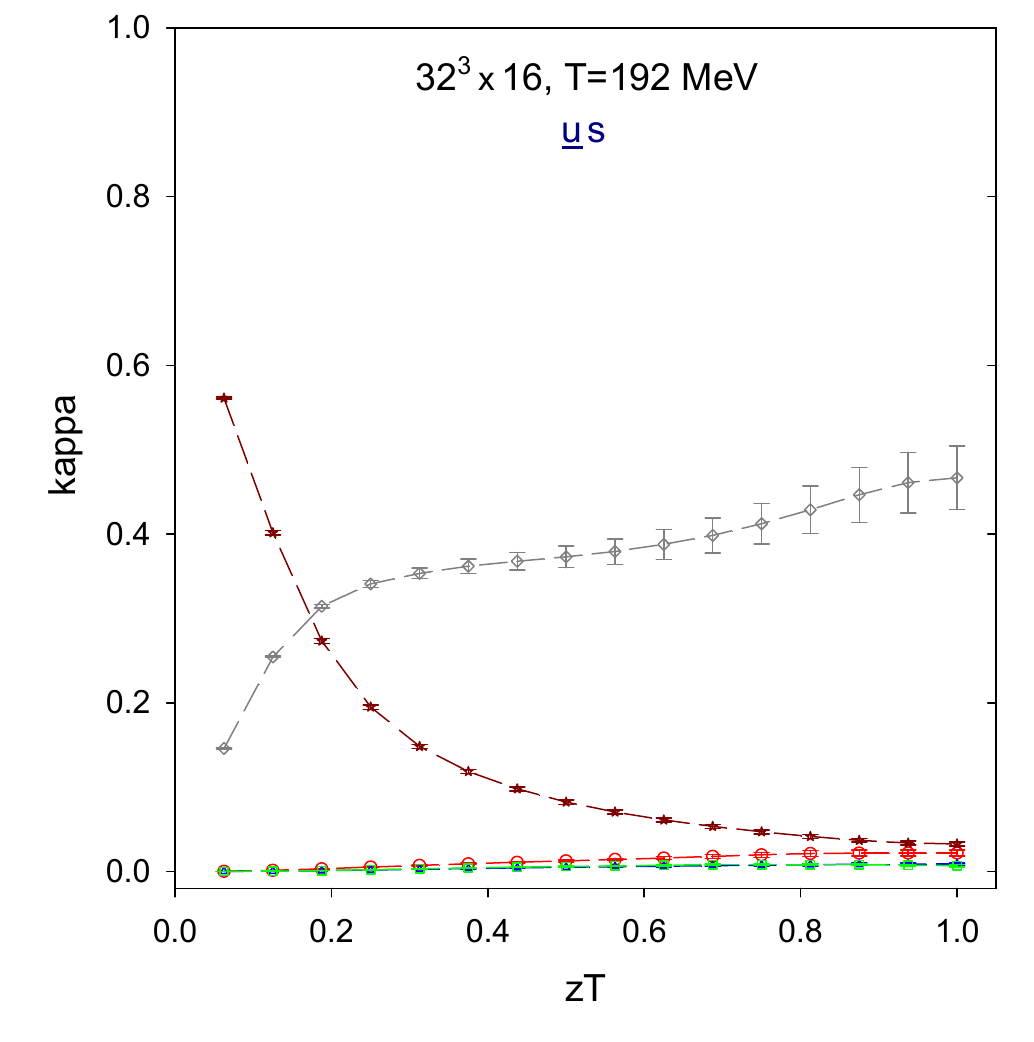} 
\\
  \includegraphics[width=7.2cm,clip=true]{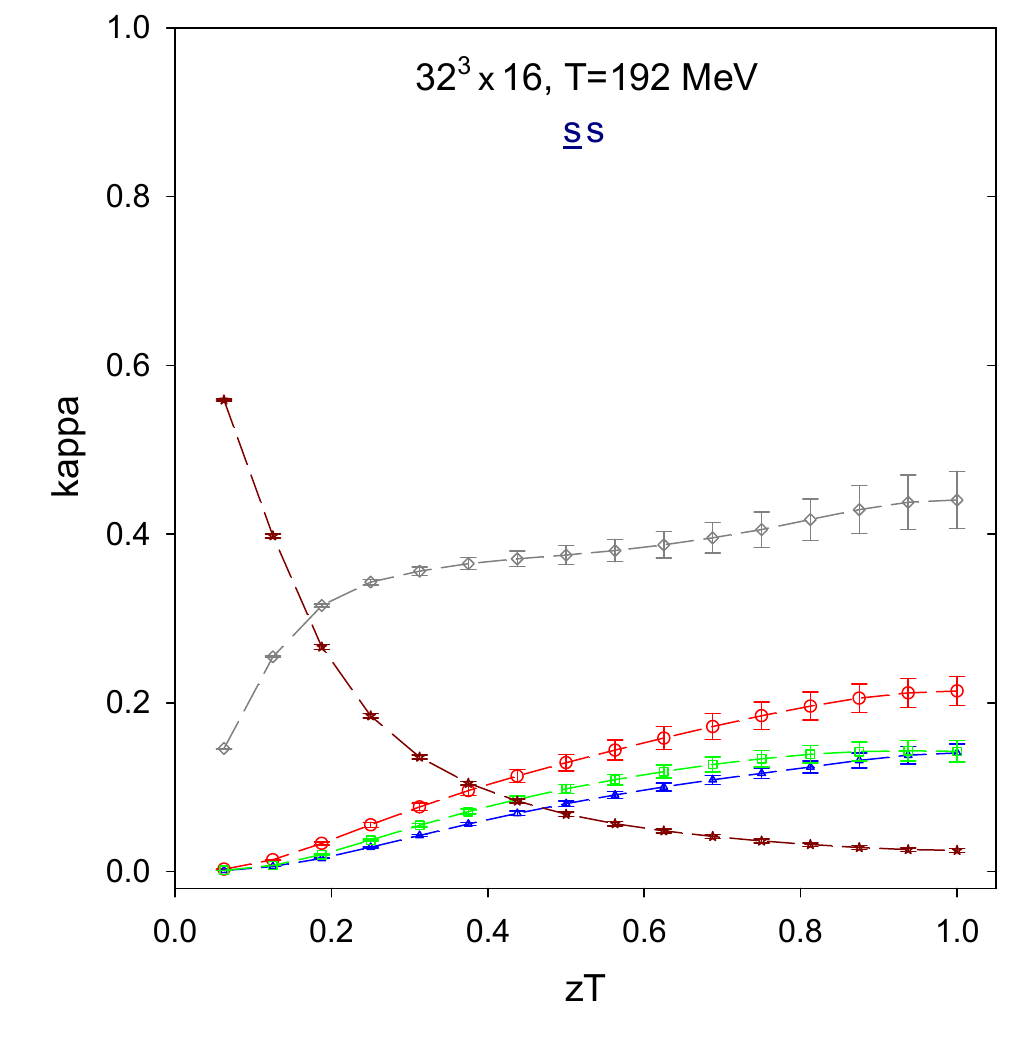}
&
  \includegraphics[width=7.2cm,clip=true]{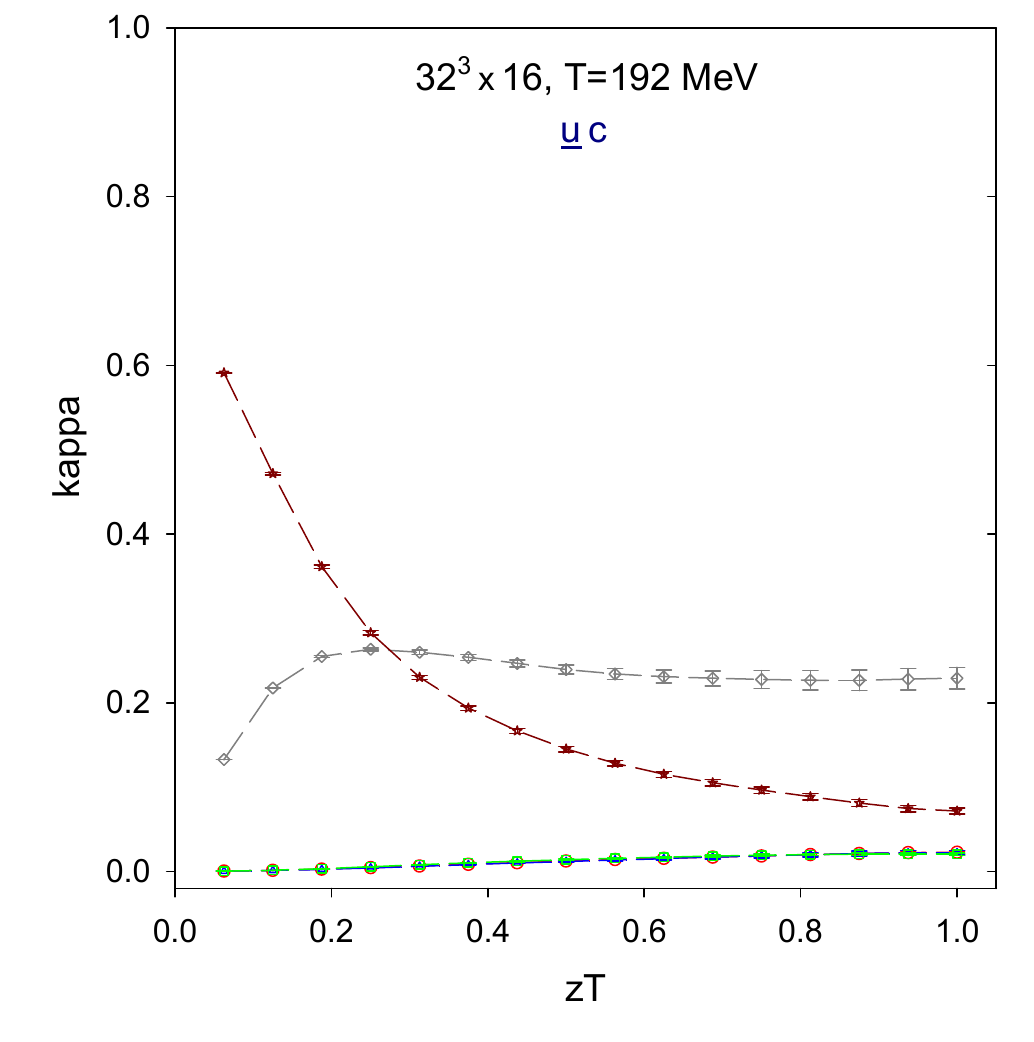} 
\\
  \includegraphics[width=7.2cm,clip=true]{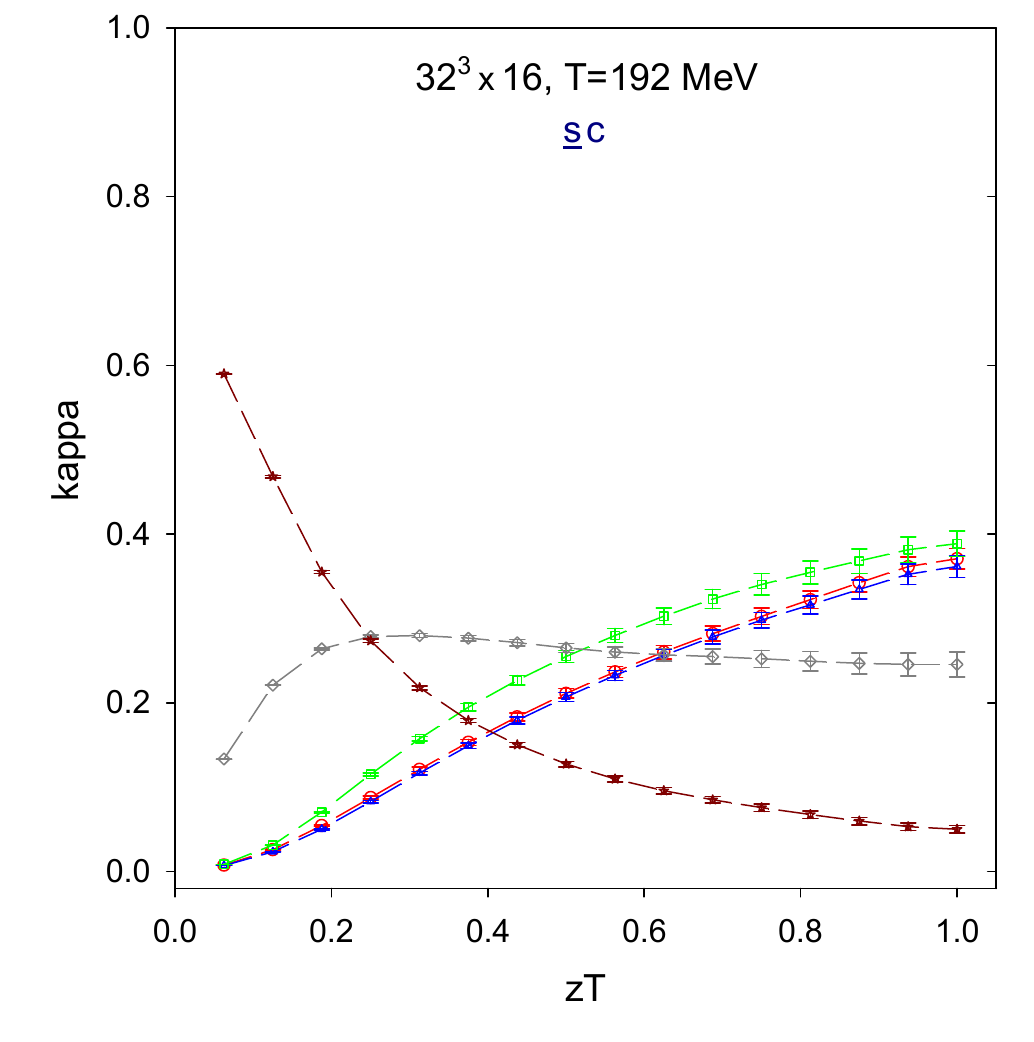}
&
  \includegraphics[width=7.2cm,clip=true]{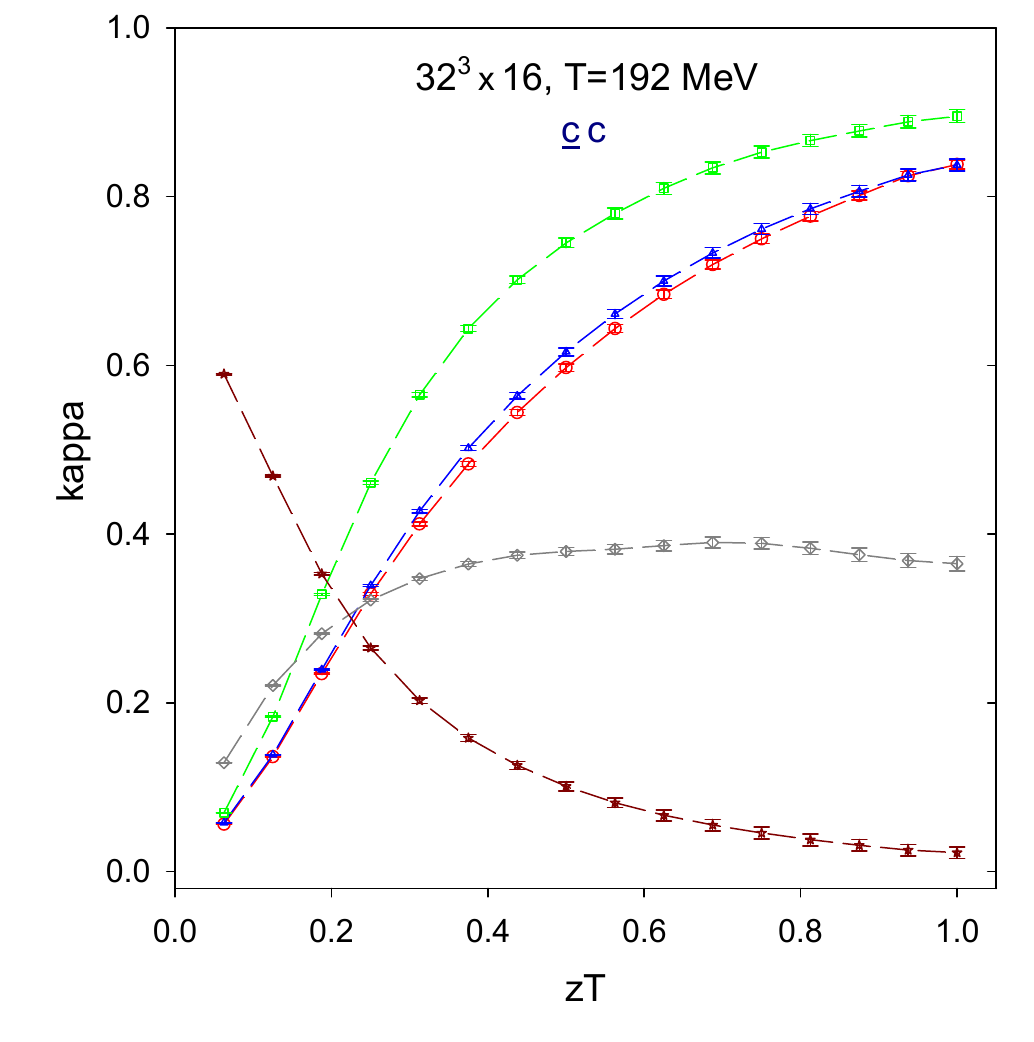}
  \end{tabular}
  \caption{
   The symmetry breaking parameters of spatial $z$ correlators of meson interpolators of six flavor 
   combinations ($\bar u d$, $\bar u s$, $\bar s s$, $\bar u c$, $\bar s c$, and $\bar c c$)
   in $N_f=2+1+1$ lattice QCD at $T \simeq 192$~MeV.
  }
\label{fig:Kz_T192}
\end{figure*}

\begin{figure*}[tbp]
  \centering
  \begin{tabular}{@{}c@{}c@{}}
  \includegraphics[width=7.2cm,clip=true]{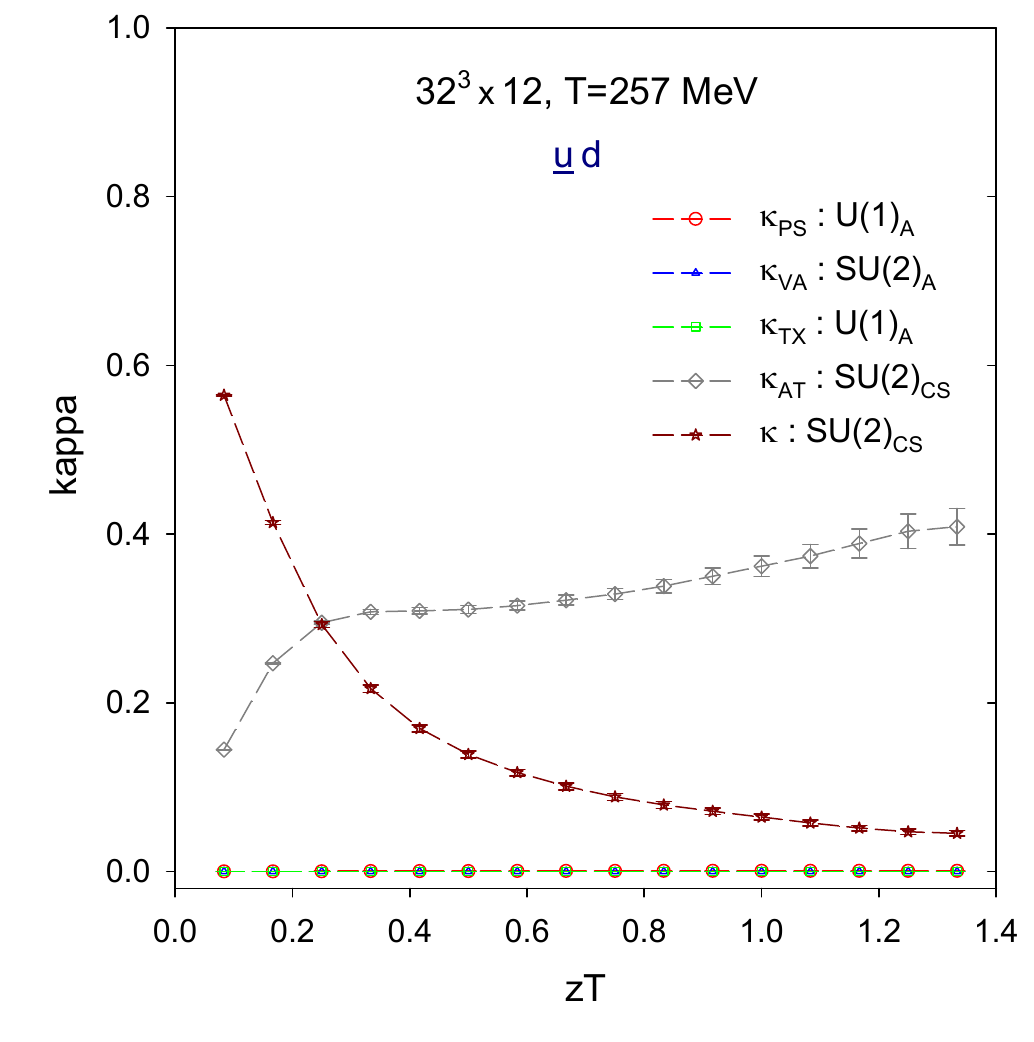}
&
  \includegraphics[width=7.2cm,clip=true]{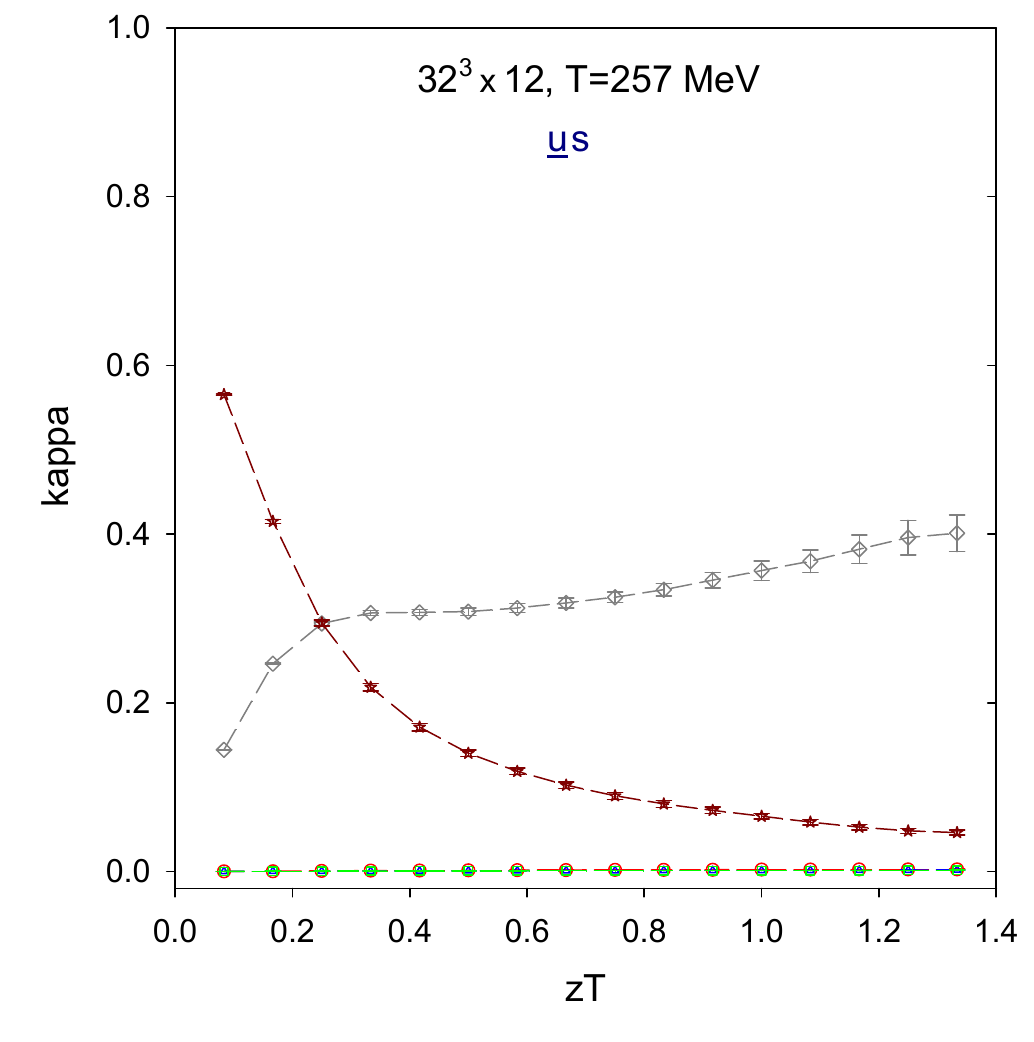}
\\
  \includegraphics[width=7.2cm,clip=true]{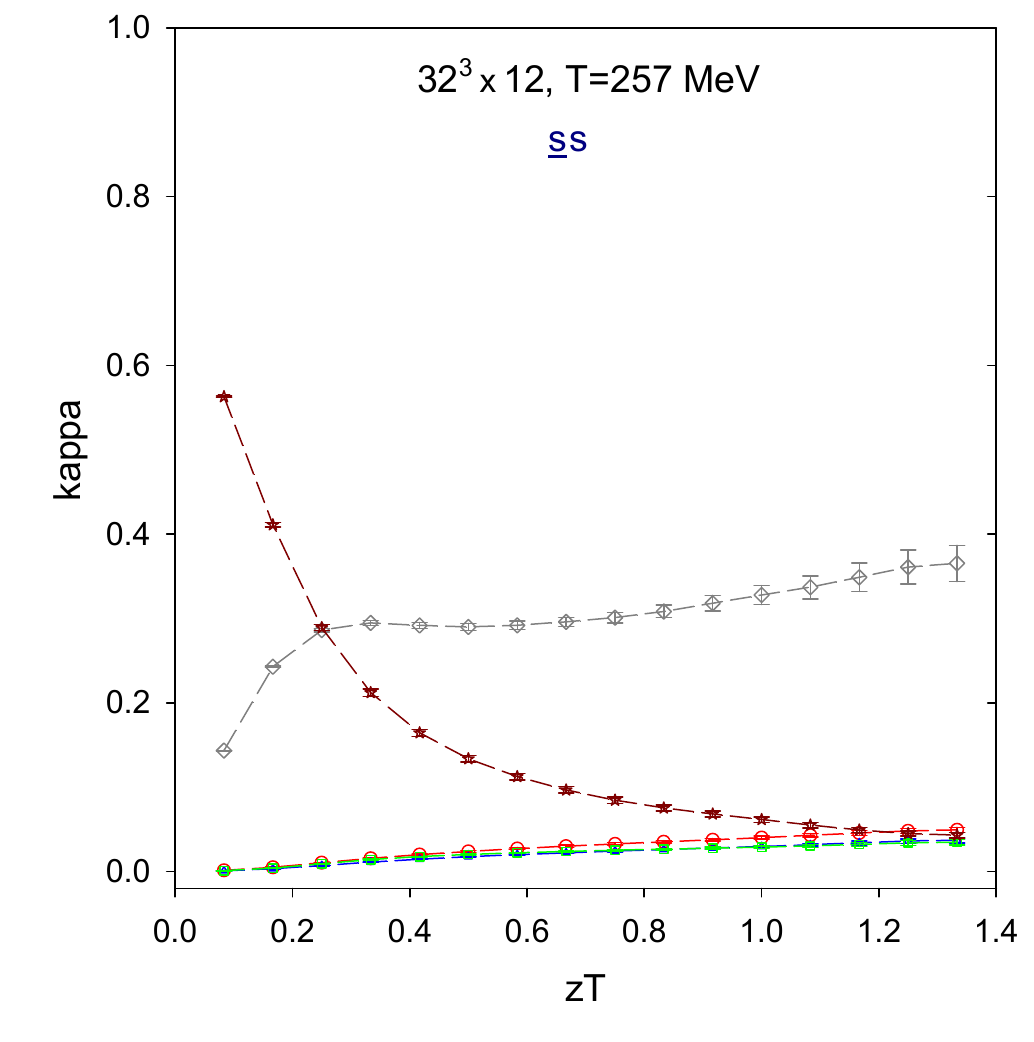}
&
  \includegraphics[width=7.2cm,clip=true]{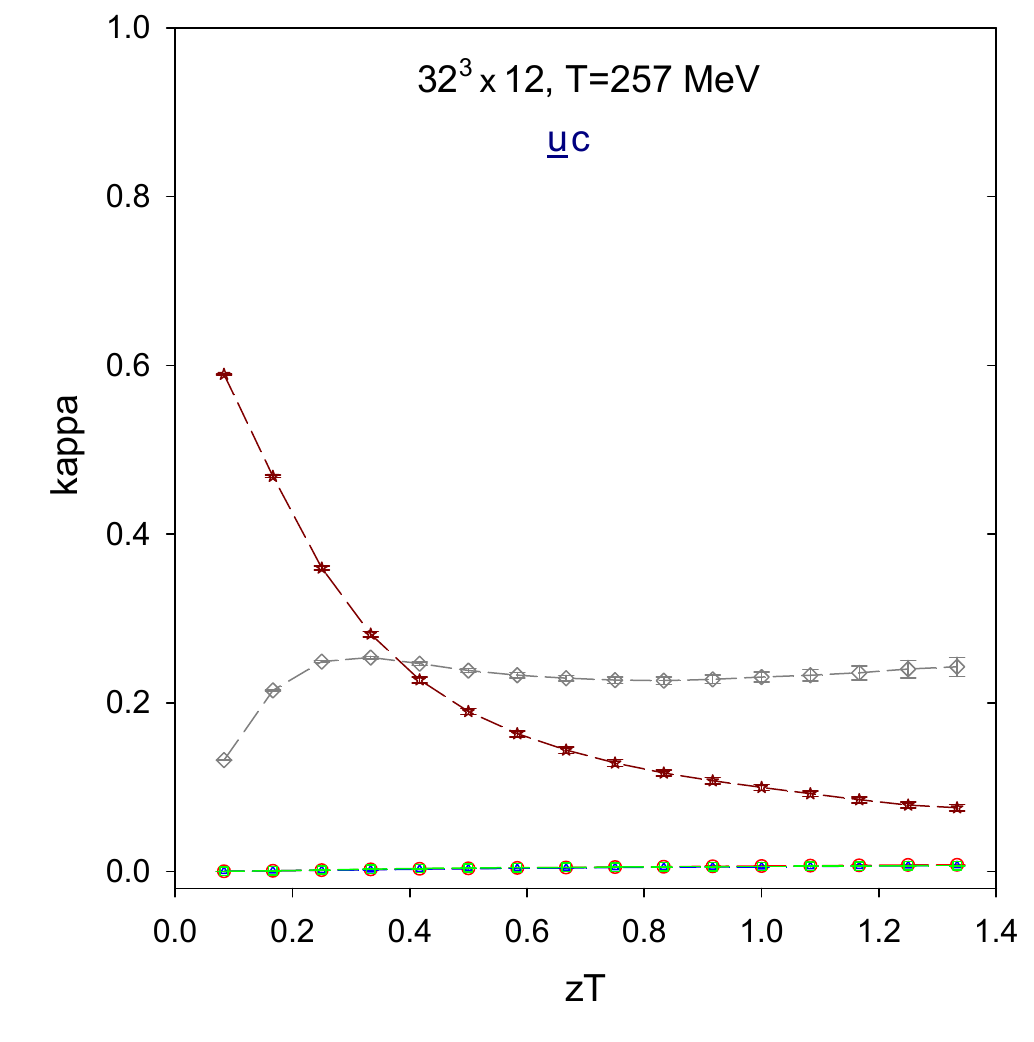}
\\
  \includegraphics[width=7.2cm,clip=true]{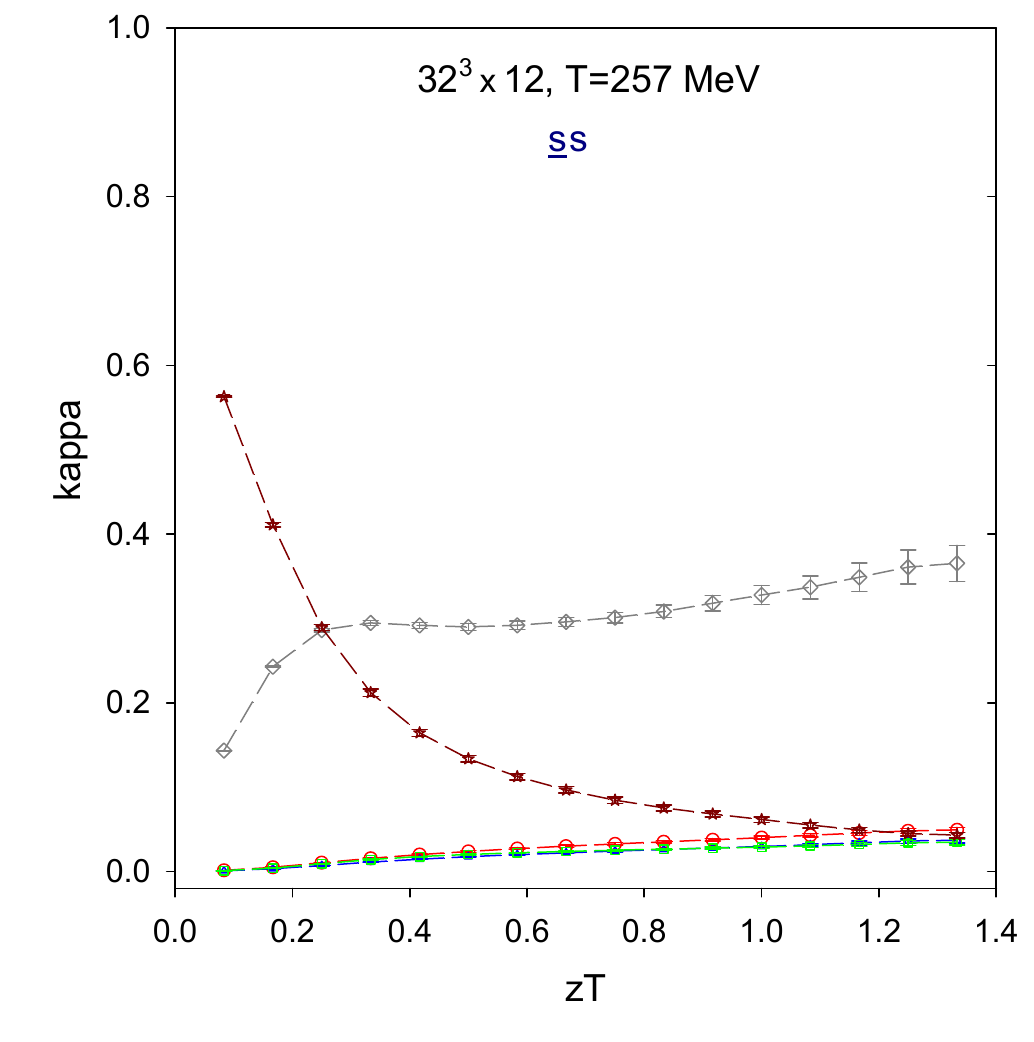}
&
  \includegraphics[width=7.2cm,clip=true]{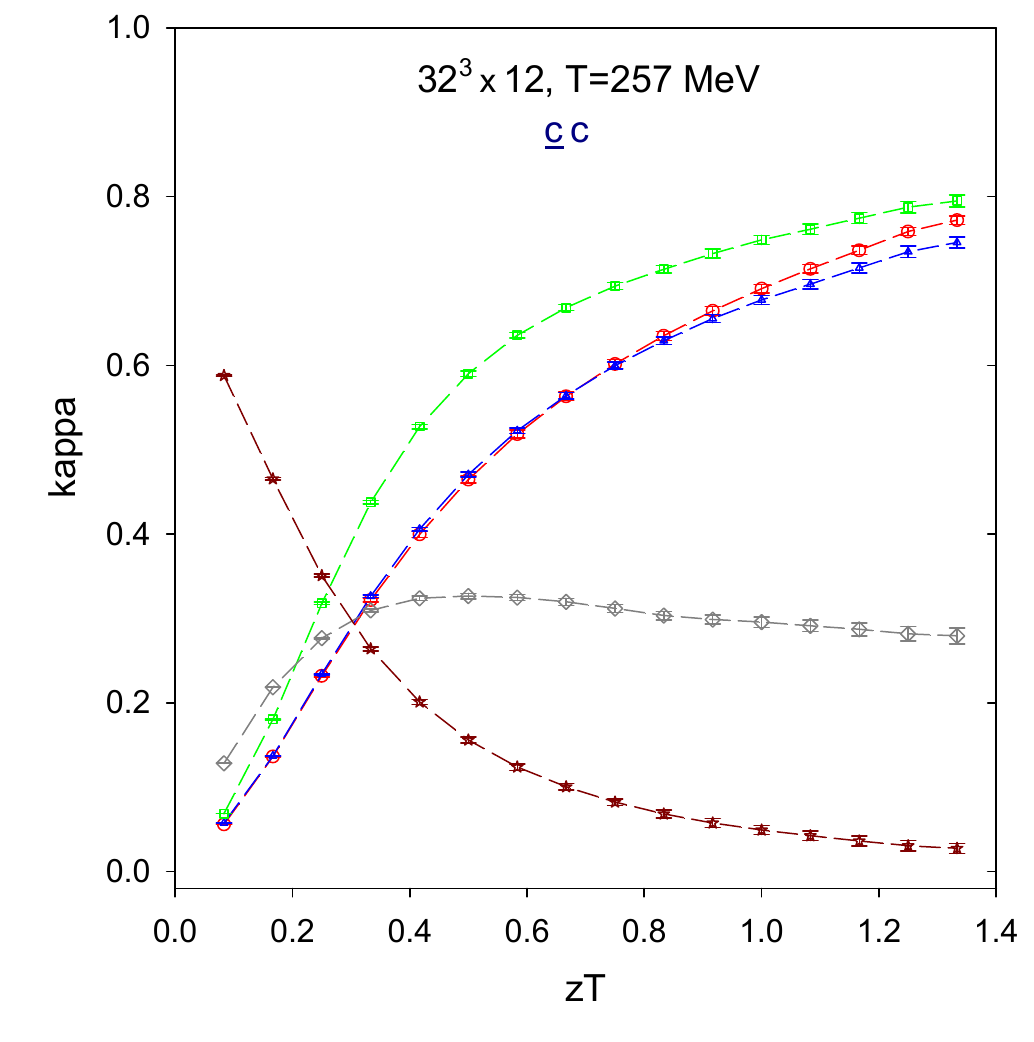}
  \end{tabular}
  \caption{
   The symmetry breaking parameters of spatial $z$ correlators of meson interpolators of six flavor 
   combinations ($\bar u d$, $\bar u s$, $\bar s s$, $\bar u c$, $\bar s c$, and $\bar c c$)
   in $N_f=2+1+1$ lattice QCD at $T \simeq 257$~MeV.
  }
\label{fig:Kz_T257}
\end{figure*}

\begin{figure*}[tbp]
  \centering
  \begin{tabular}{@{}c@{}c@{}}
  \includegraphics[width=7.2cm,clip=true]{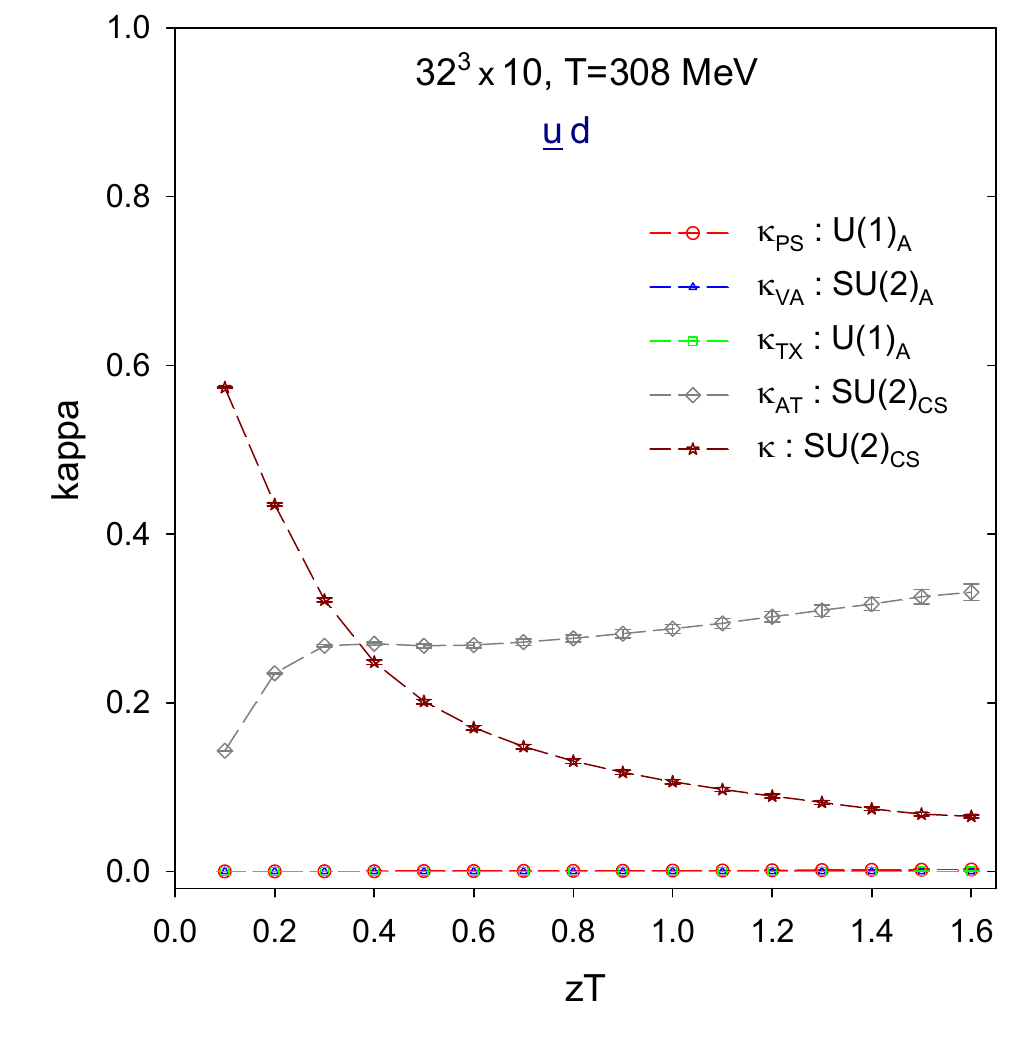}
&
  \includegraphics[width=7.2cm,clip=true]{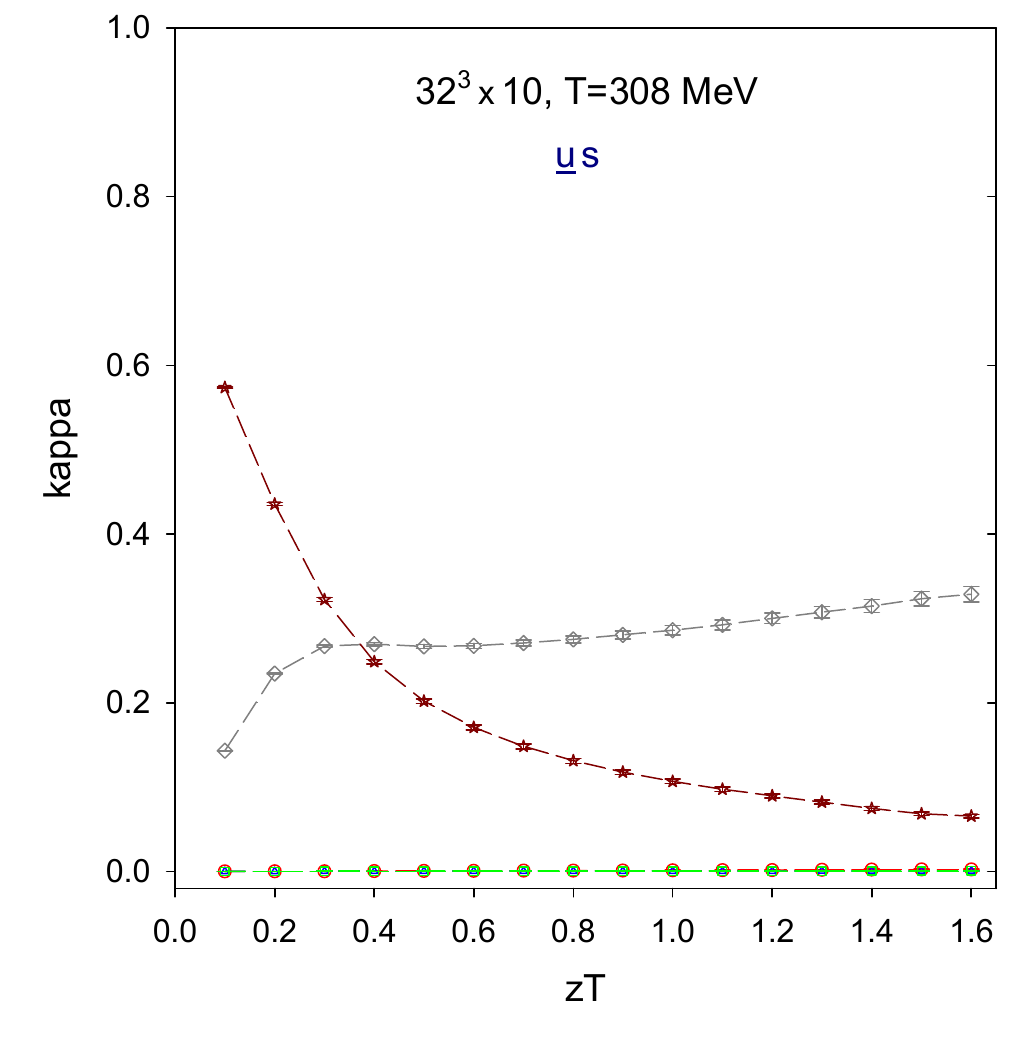}
\\
  \includegraphics[width=7.2cm,clip=true]{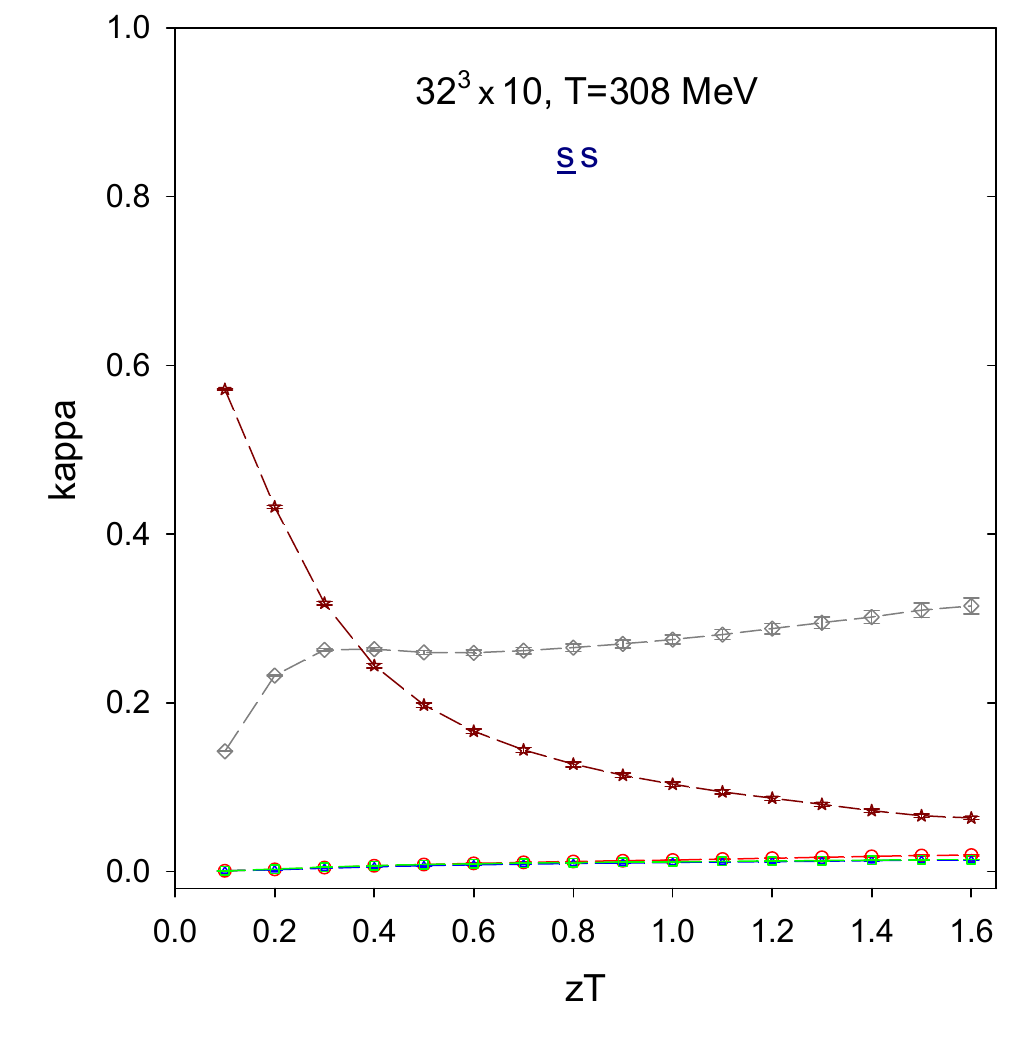}
&
  \includegraphics[width=7.2cm,clip=true]{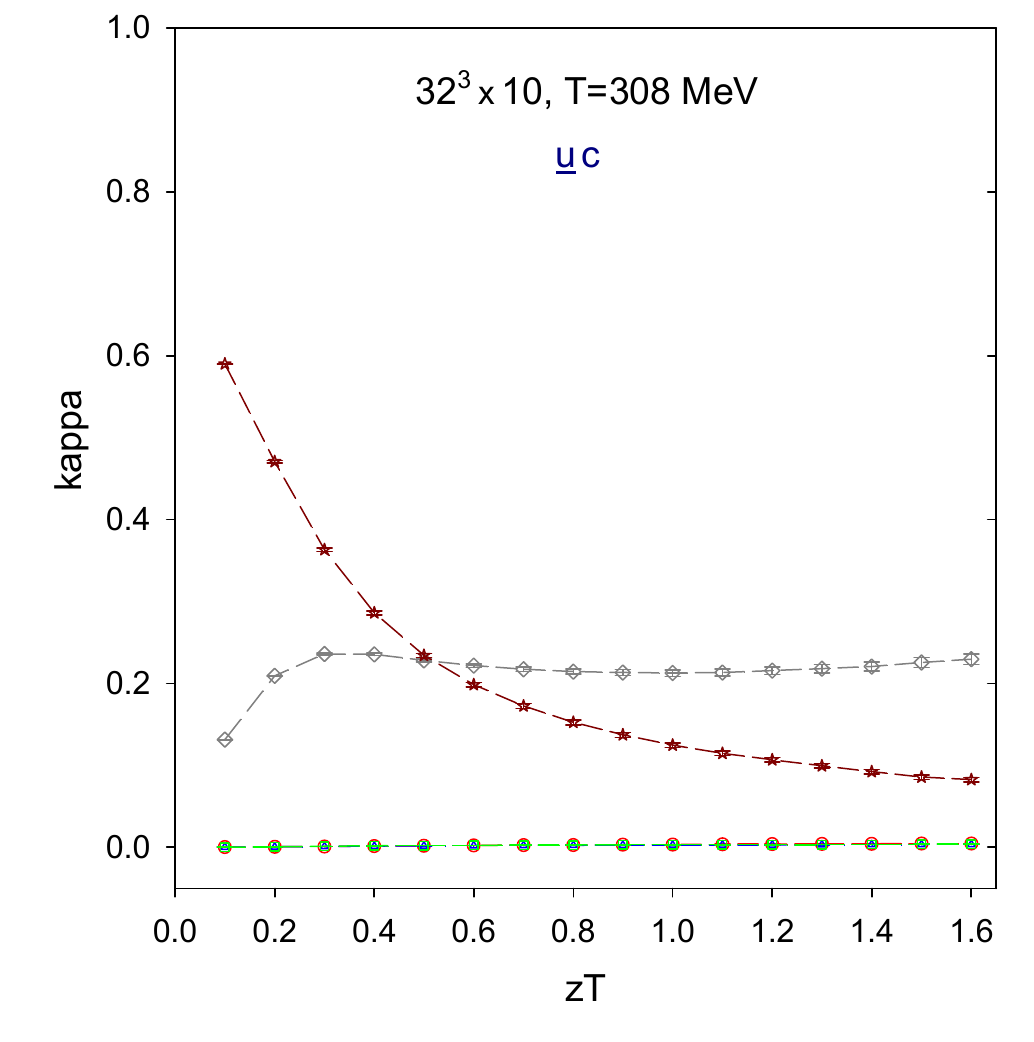}
\\
  \includegraphics[width=7.2cm,clip=true]{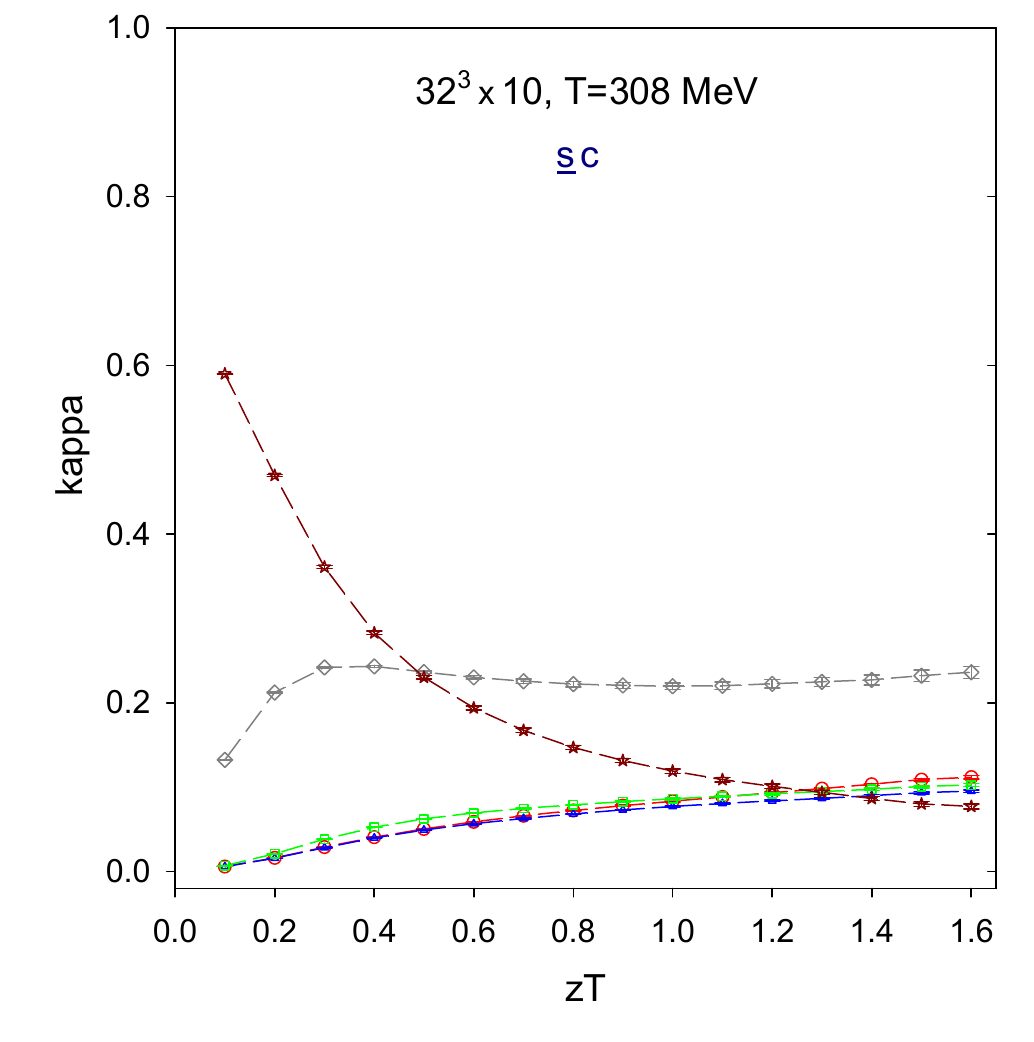}
&
  \includegraphics[width=7.2cm,clip=true]{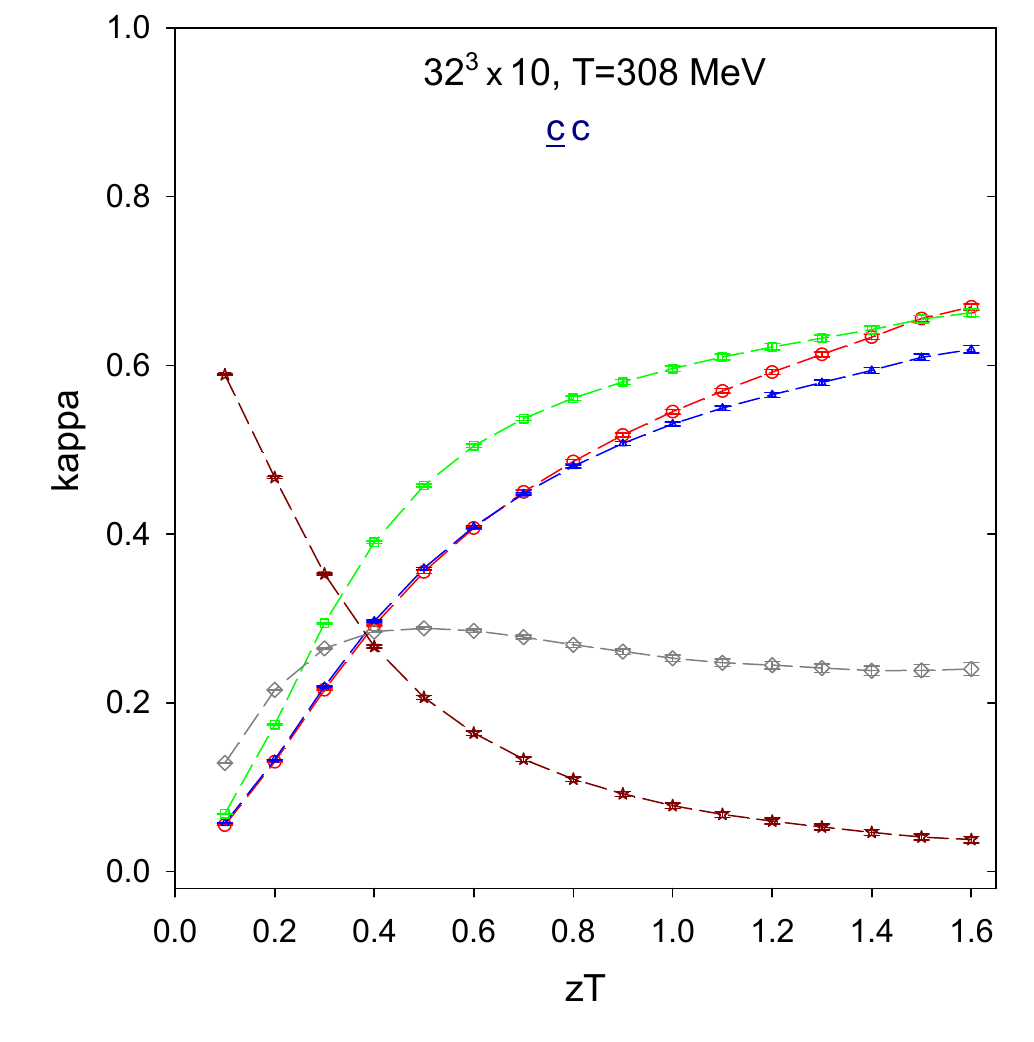}
  \end{tabular}
  \caption{
   The symmetry breaking parameters of spatial $z$ correlators of meson interpolators of six flavor 
   combinations ($\bar u d$, $\bar u s$, $\bar s s$, $\bar u c$, $\bar s c$, and $\bar c c$)
   in $N_f=2+1+1$ lattice QCD at $T \simeq 308$~MeV.
  }
\label{fig:Kz_T308}
\end{figure*}

\begin{figure*}[tbp]
  \centering
  \begin{tabular}{@{}c@{}c@{}}
  \includegraphics[width=7.2cm,clip=true]{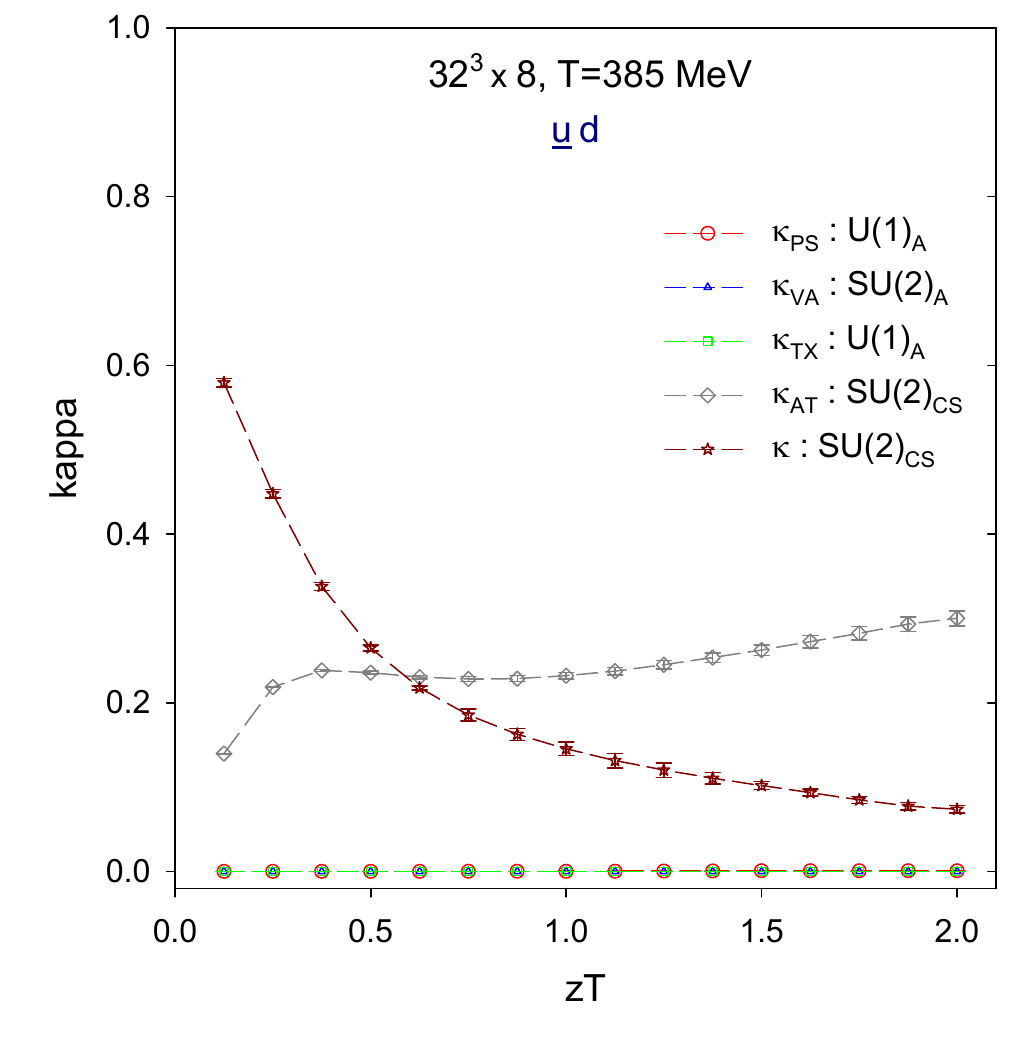}
&
  \includegraphics[width=7.2cm,clip=true]{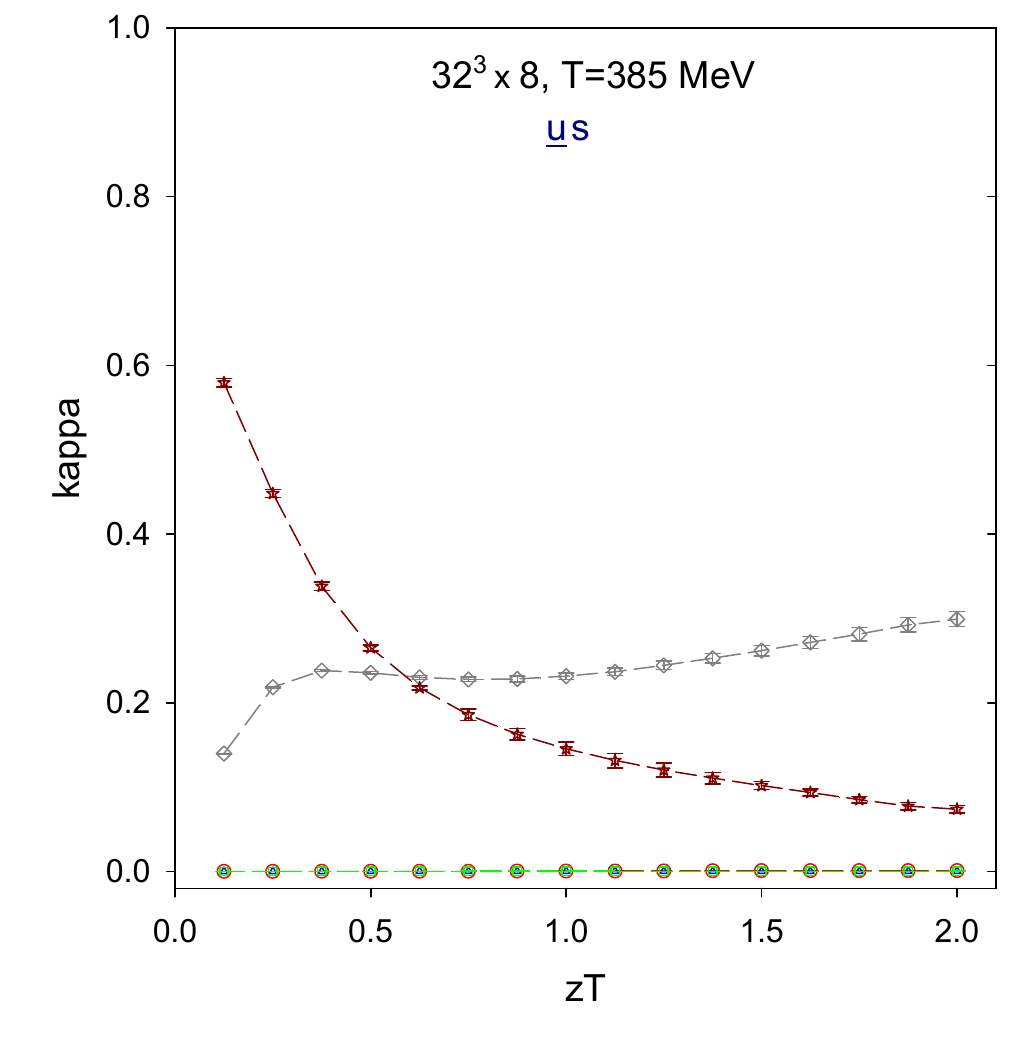}
\\
  \includegraphics[width=7.2cm,clip=true]{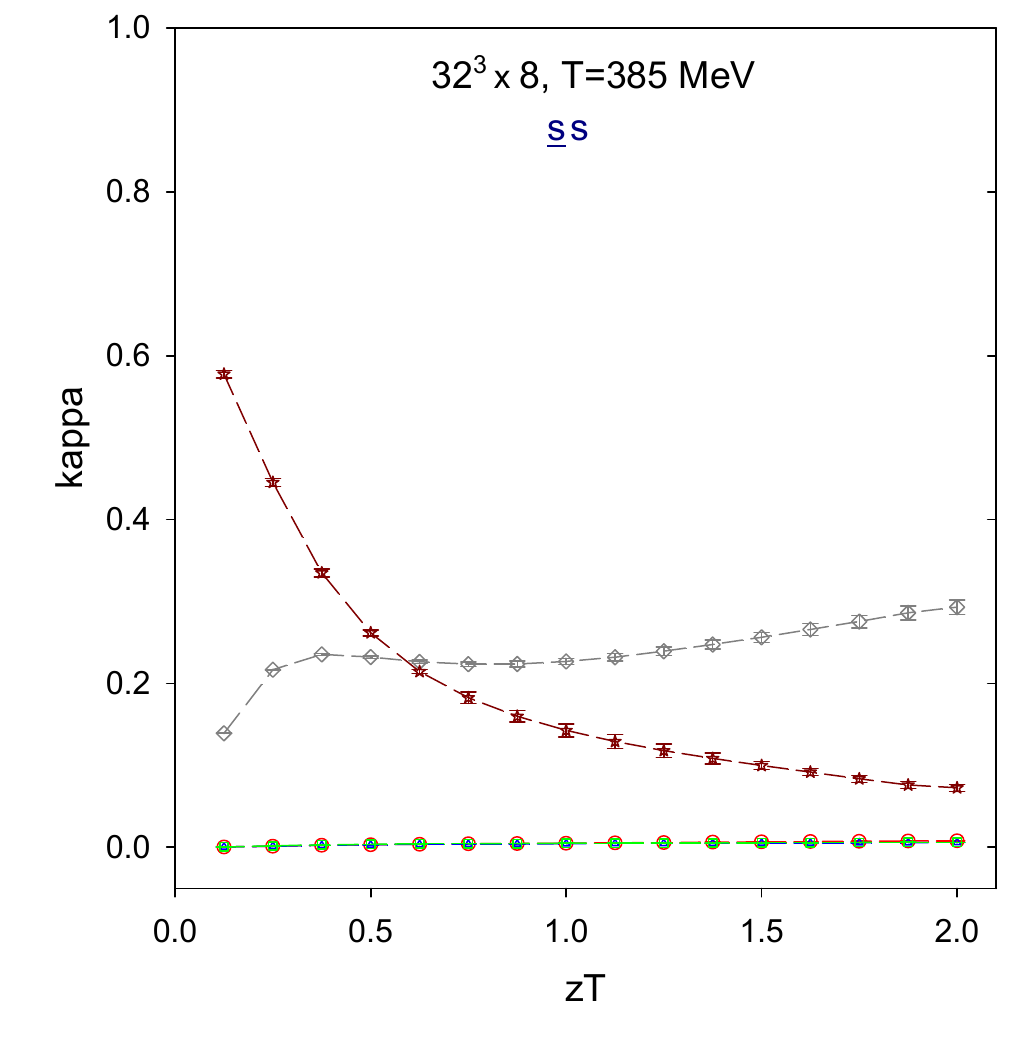}
&
  \includegraphics[width=7.2cm,clip=true]{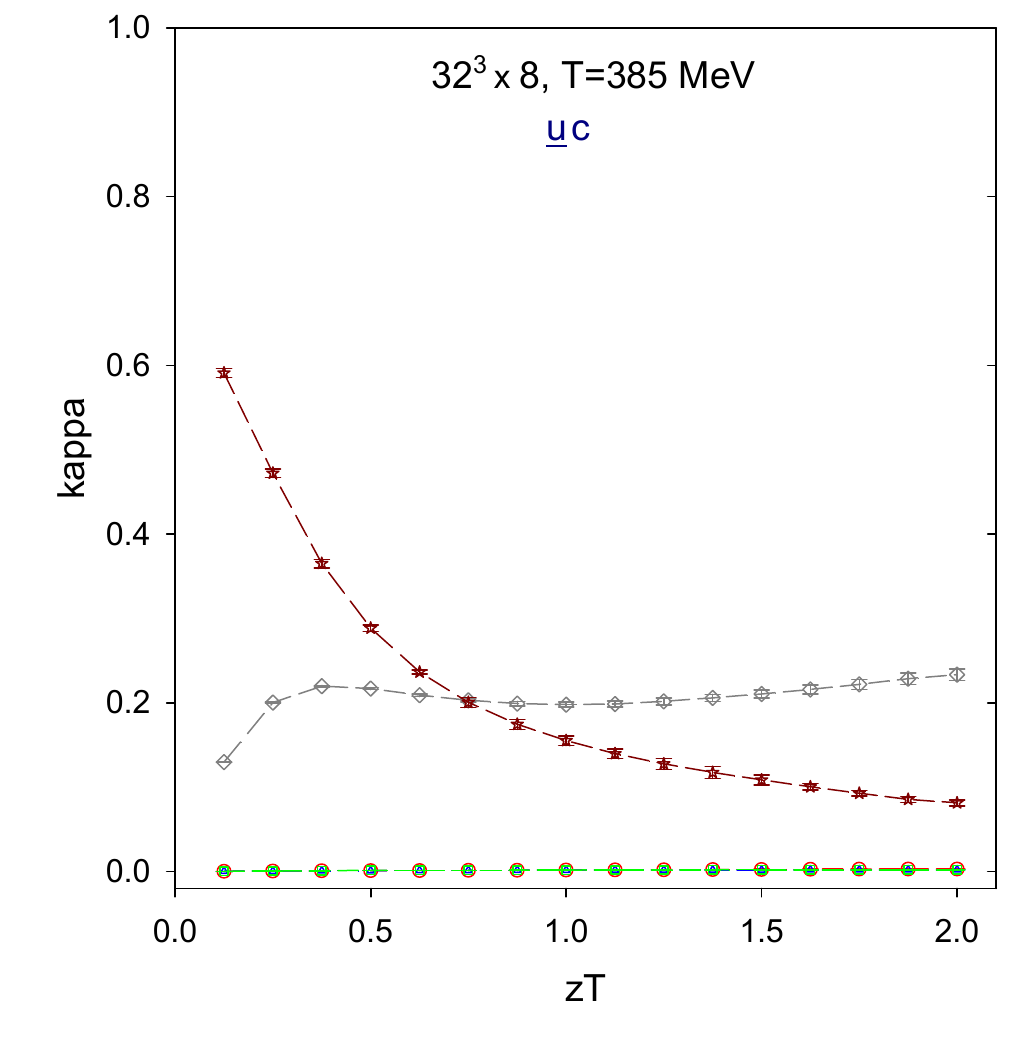}
\\
  \includegraphics[width=7.2cm,clip=true]{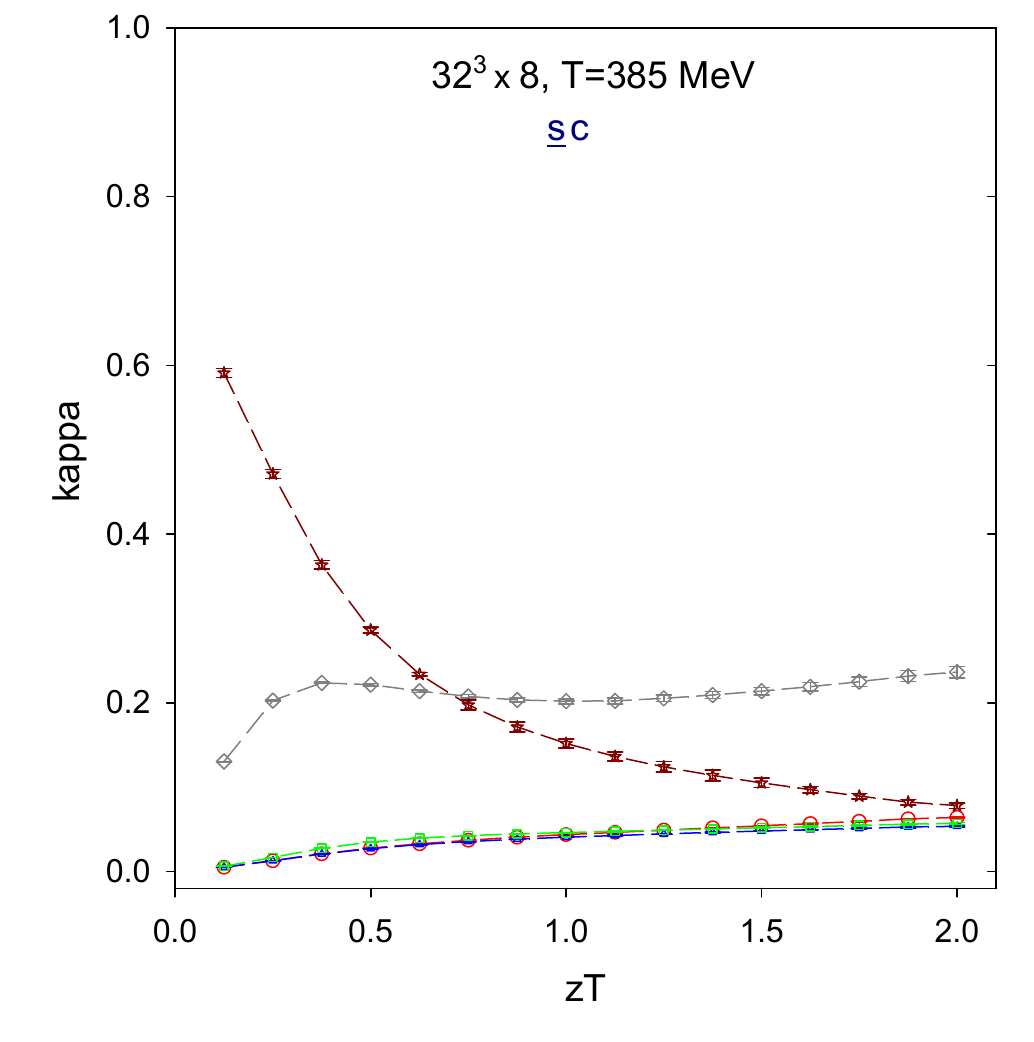}
&
  \includegraphics[width=7.2cm,clip=true]{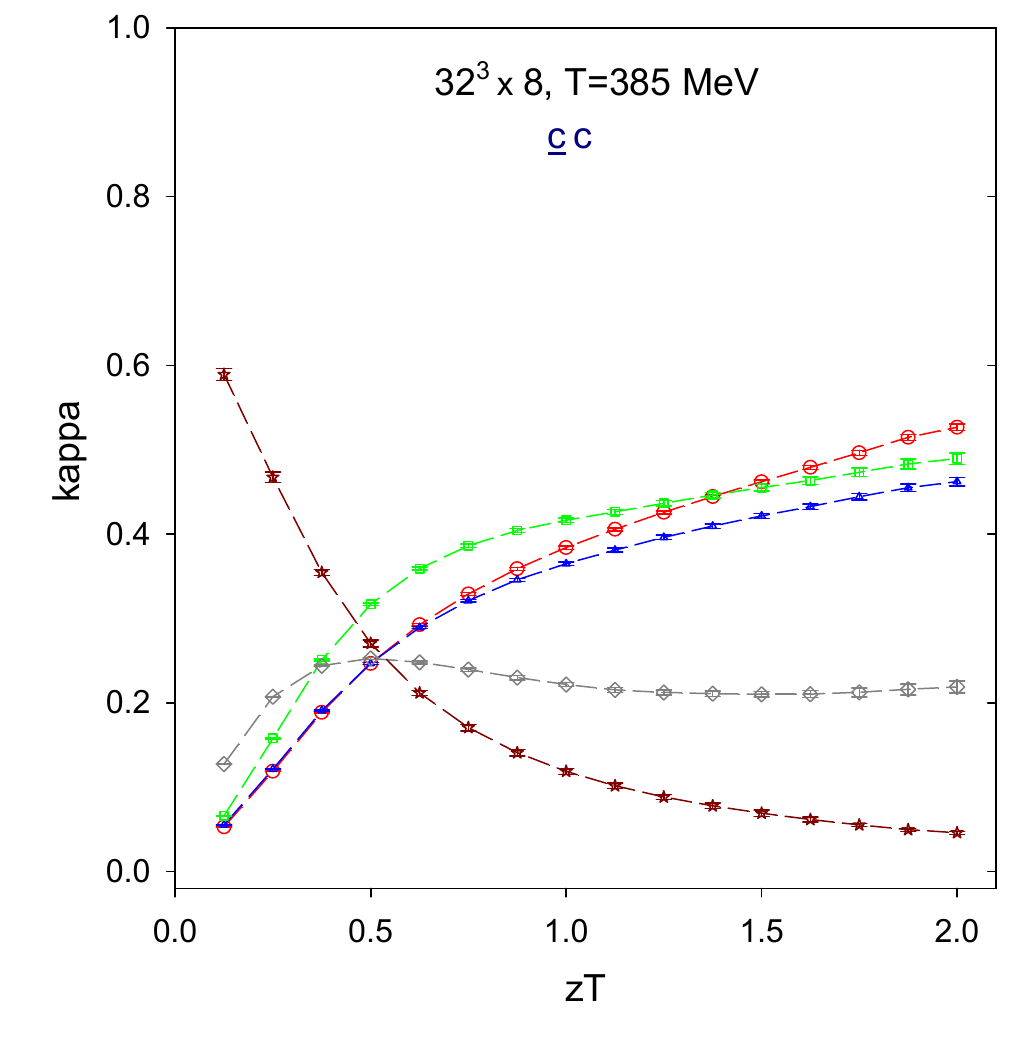}
  \end{tabular}
  \caption{
   The symmetry breaking parameters of spatial $z$ correlators of meson interpolators of six flavor 
   combinations ($\bar u d$, $\bar u s$, $\bar s s$, $\bar u c$, $\bar s c$, and $\bar c c$)
   in $N_f=2+1+1$ lattice QCD at $T \simeq 385$~MeV.
  }
\label{fig:Kz_T385}
\end{figure*}

\begin{figure*}[tbp]
  \centering
  \begin{tabular}{@{}c@{}c@{}}
  \includegraphics[width=7.2cm,clip=true]{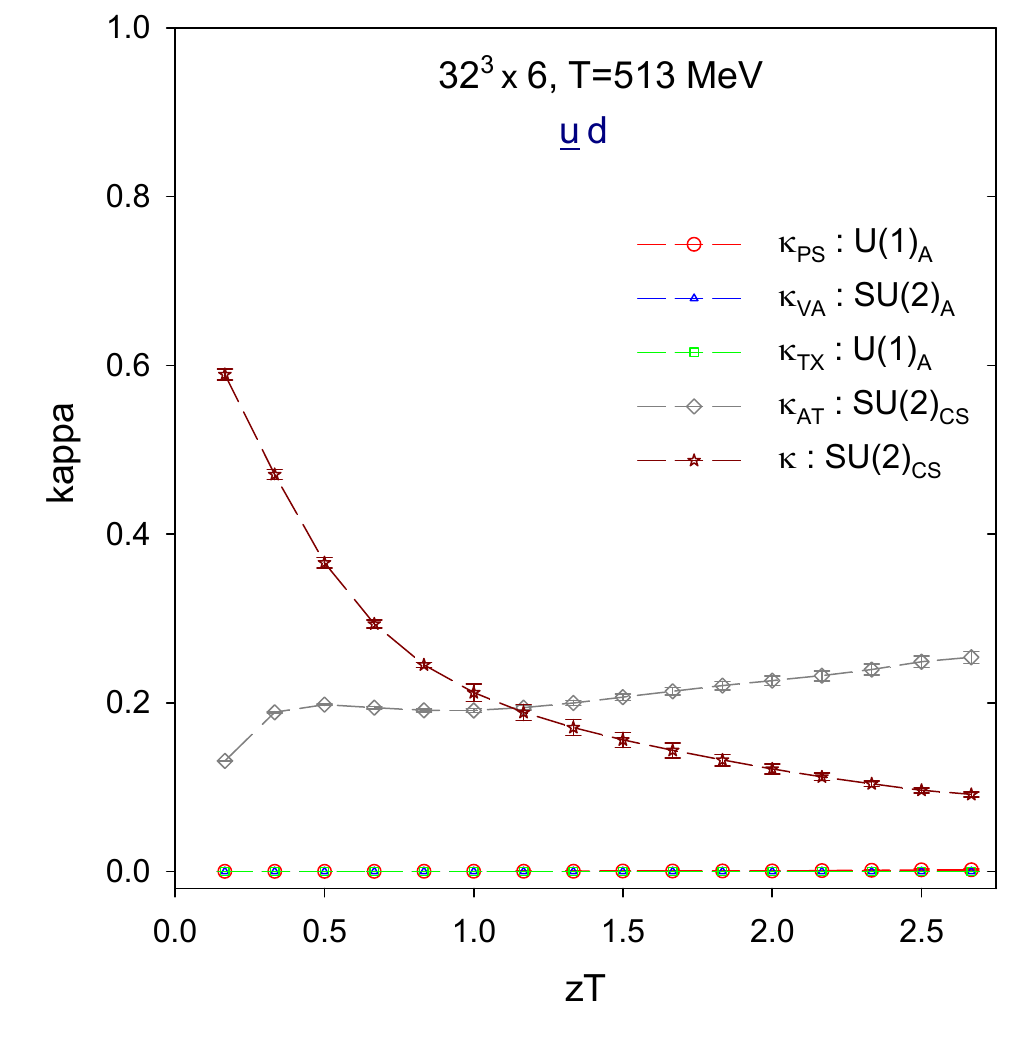}
&
  \includegraphics[width=7.2cm,clip=true]{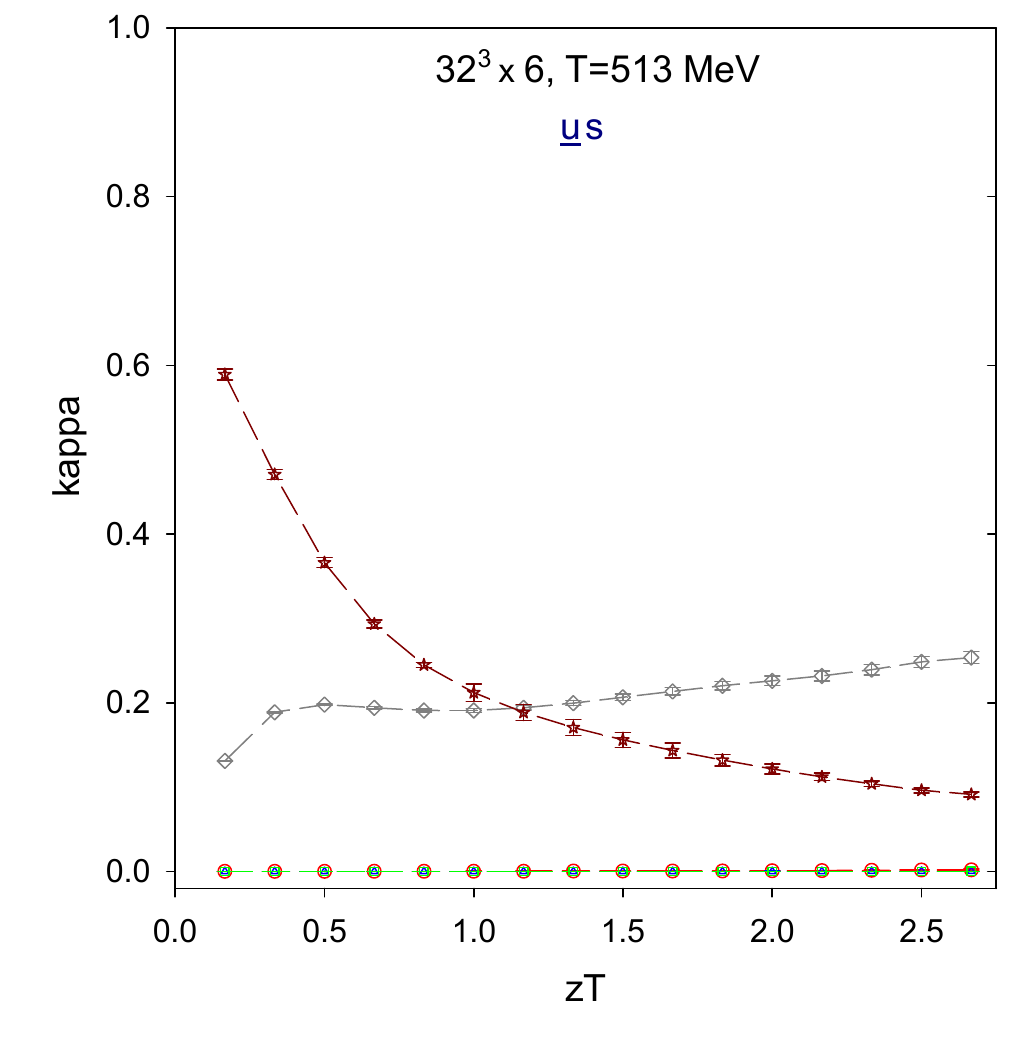}
\\
  \includegraphics[width=7.2cm,clip=true]{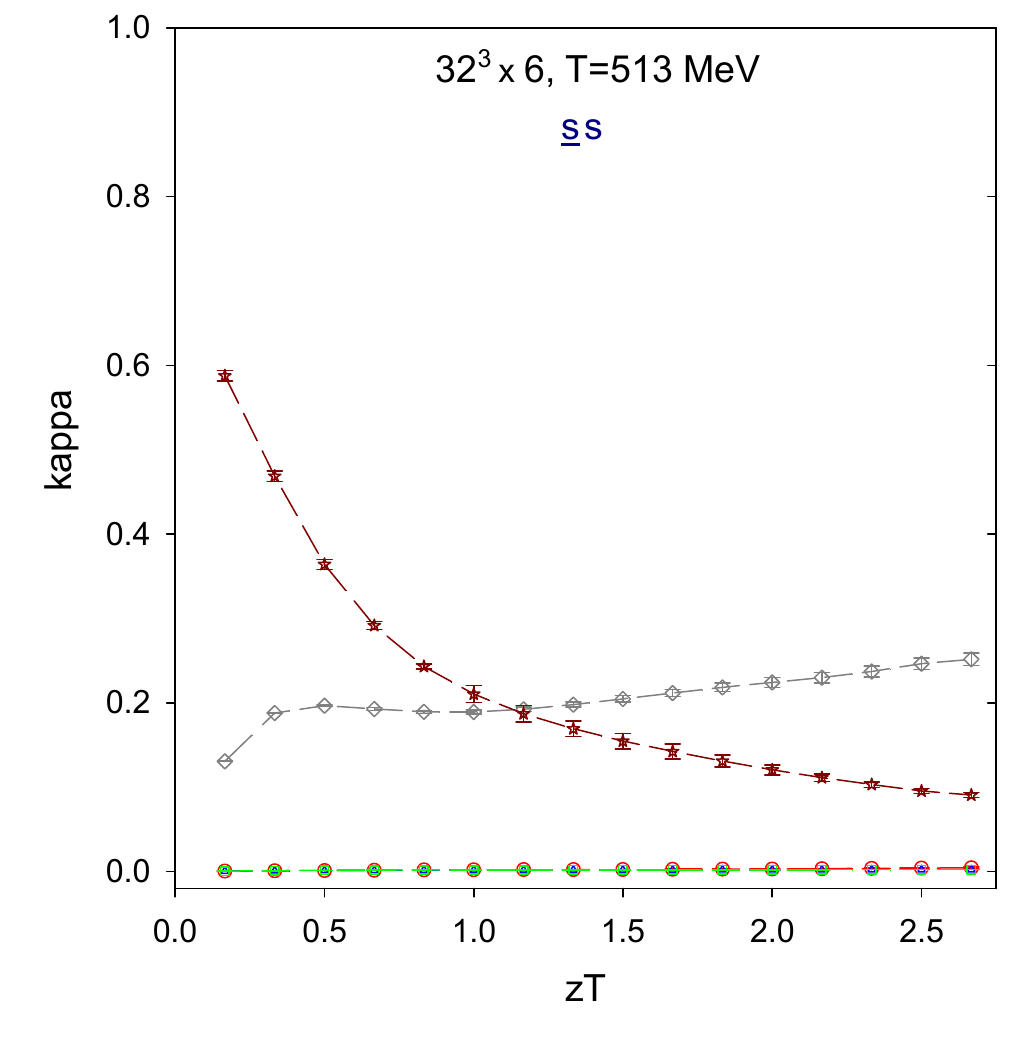}
&
  \includegraphics[width=7.2cm,clip=true]{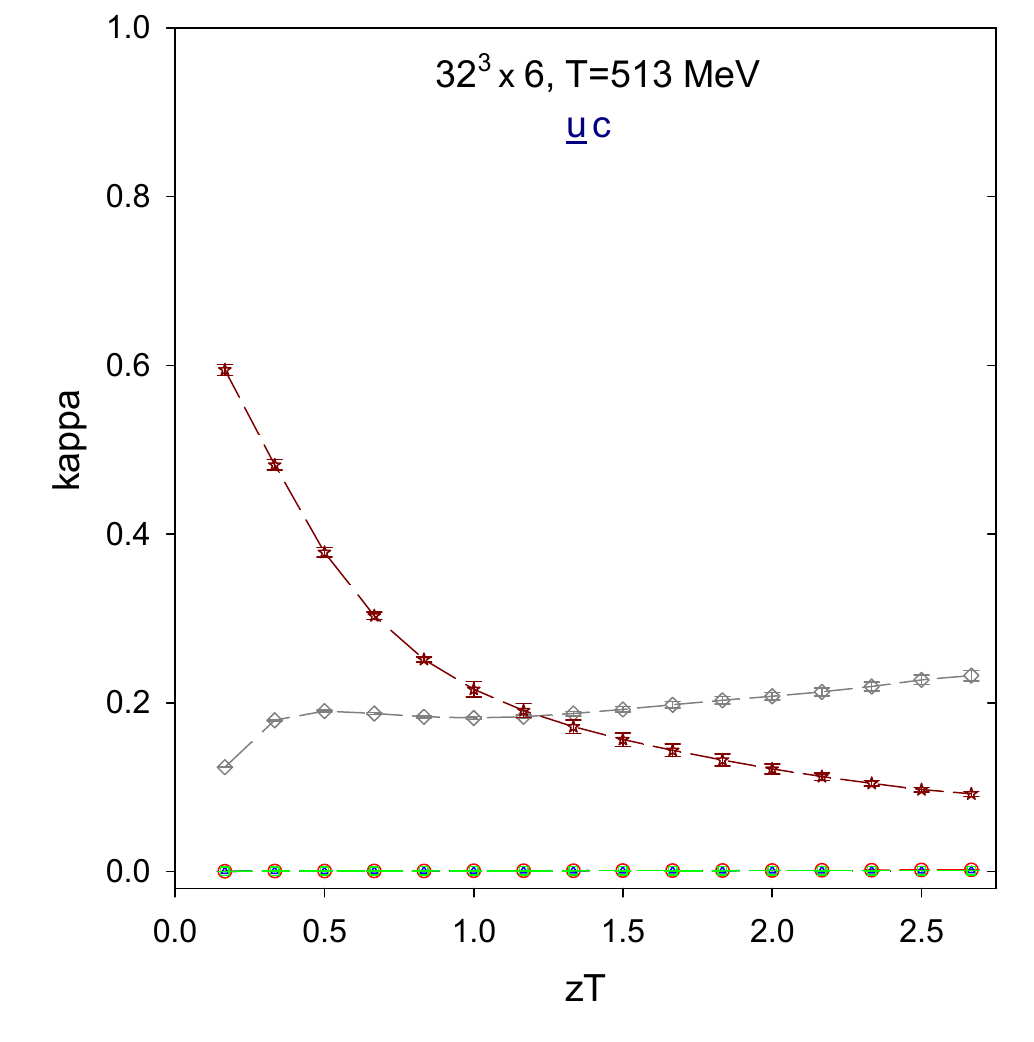}
\\
  \includegraphics[width=7.2cm,clip=true]{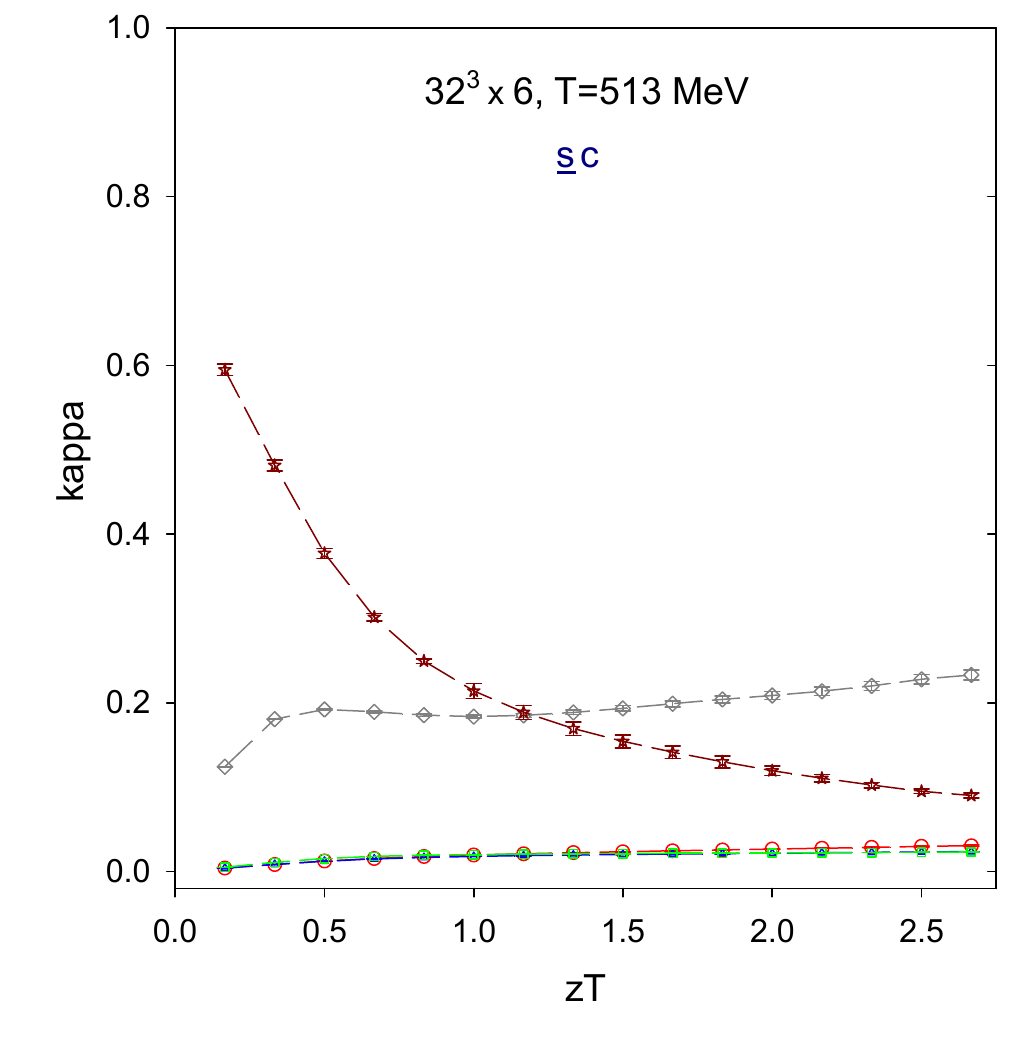}
&
  \includegraphics[width=7.2cm,clip=true]{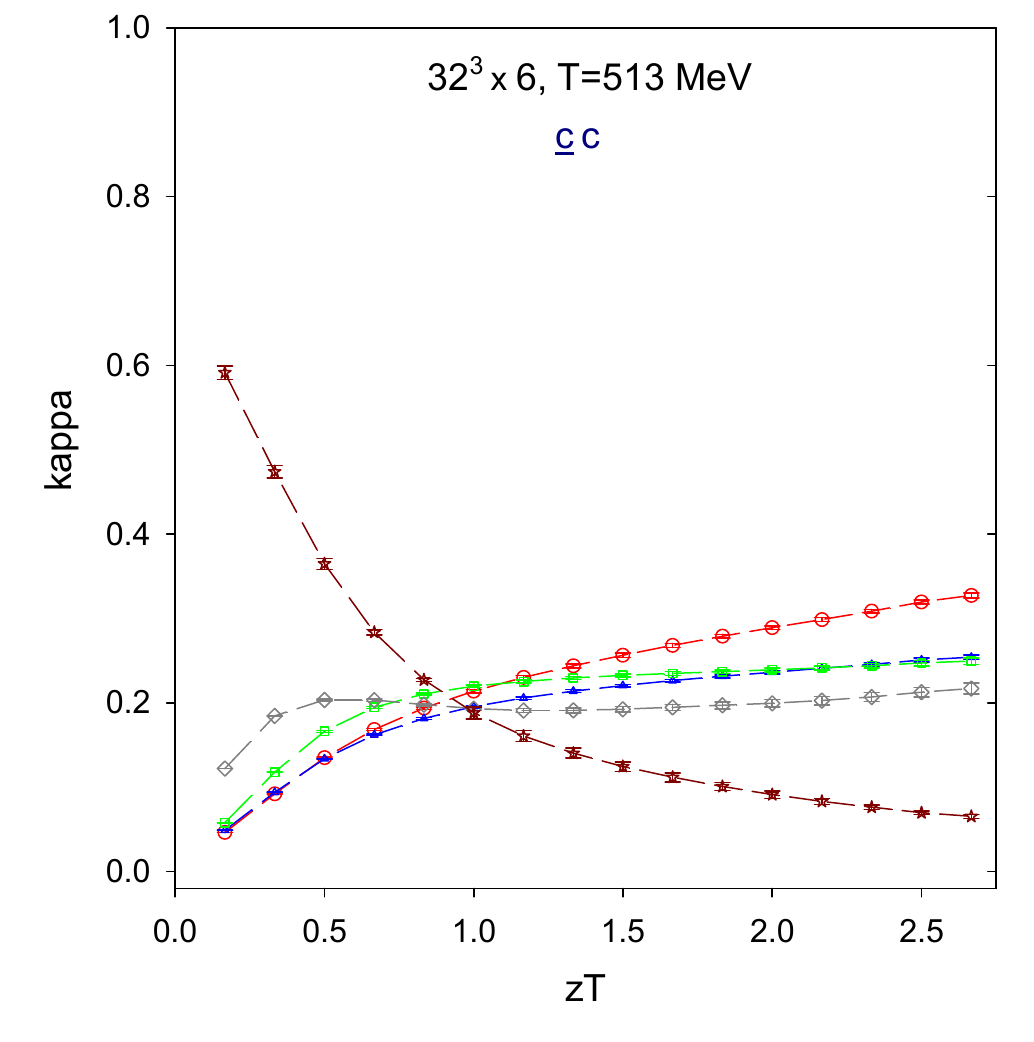}
  \end{tabular}
  \caption{
   The symmetry breaking parameters of spatial $z$ correlators of meson interpolators of six flavor 
   combinations ($\bar u d$, $\bar u s$, $\bar s s$, $\bar u c$, $\bar s c$, and $\bar c c$)
   in $N_f=2+1+1$ lattice QCD at $T \simeq 513$~MeV.
  }
\label{fig:Kz_T513}
\end{figure*}

\begin{figure*}[tbp]
  \centering
  \begin{tabular}{@{}c@{}c@{}}
  \includegraphics[width=7.2cm,clip=true]{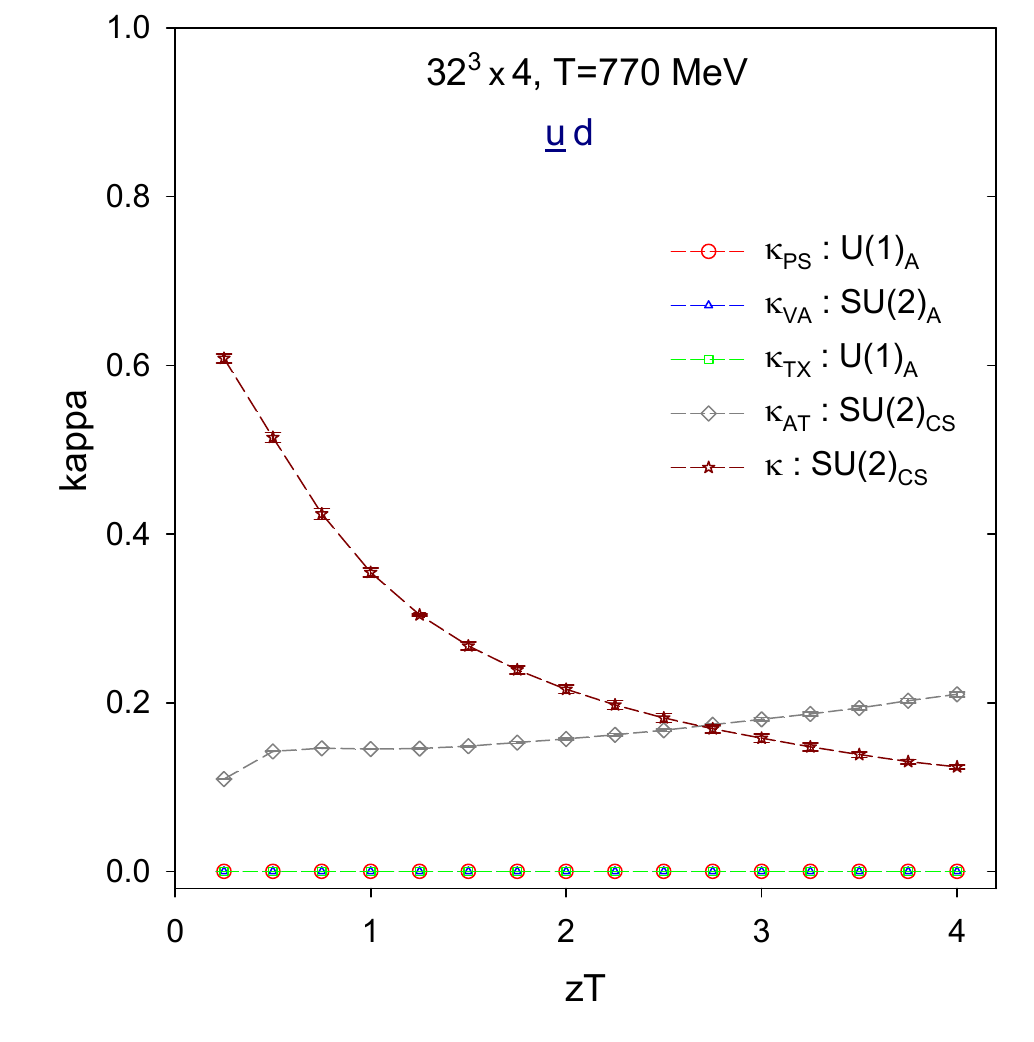}
&
  \includegraphics[width=7.2cm,clip=true]{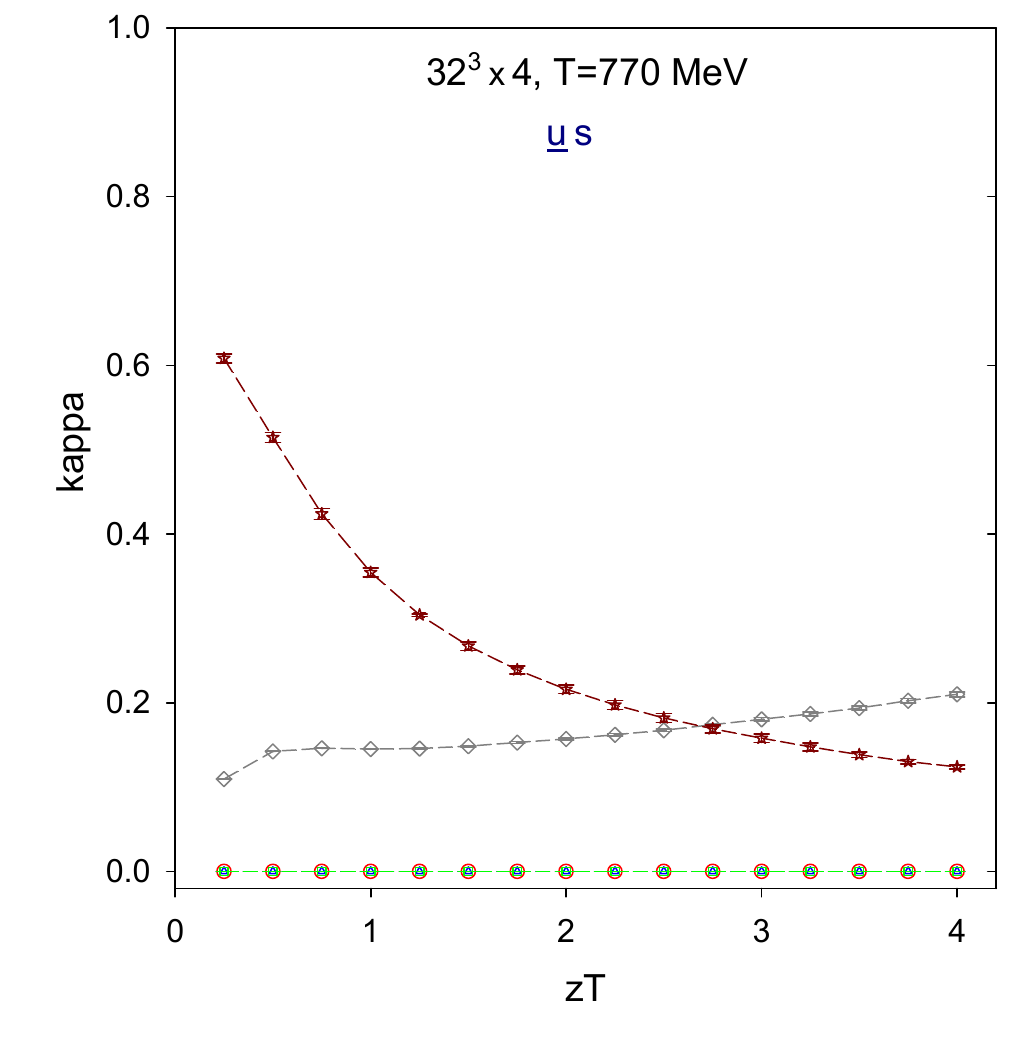}
\\
  \includegraphics[width=7.2cm,clip=true]{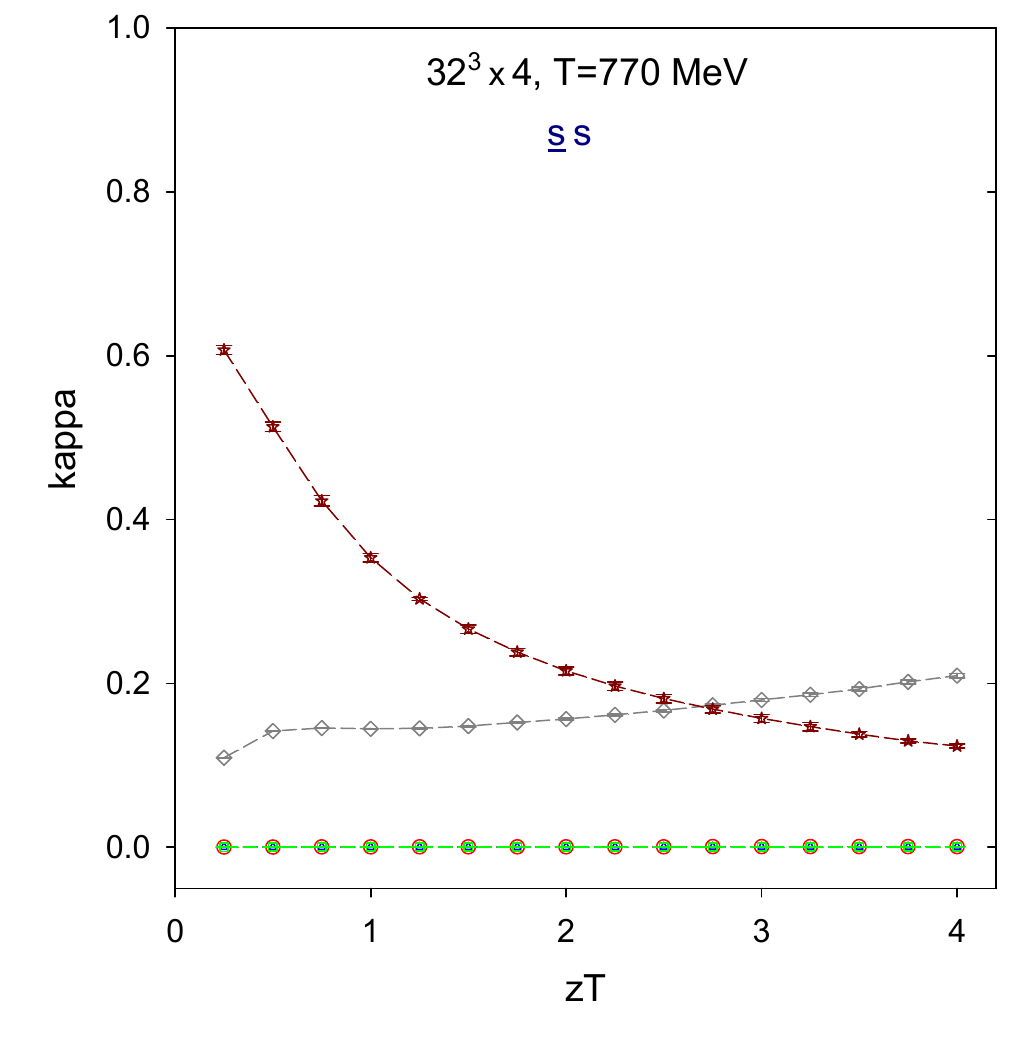}
&
  \includegraphics[width=7.2cm,clip=true]{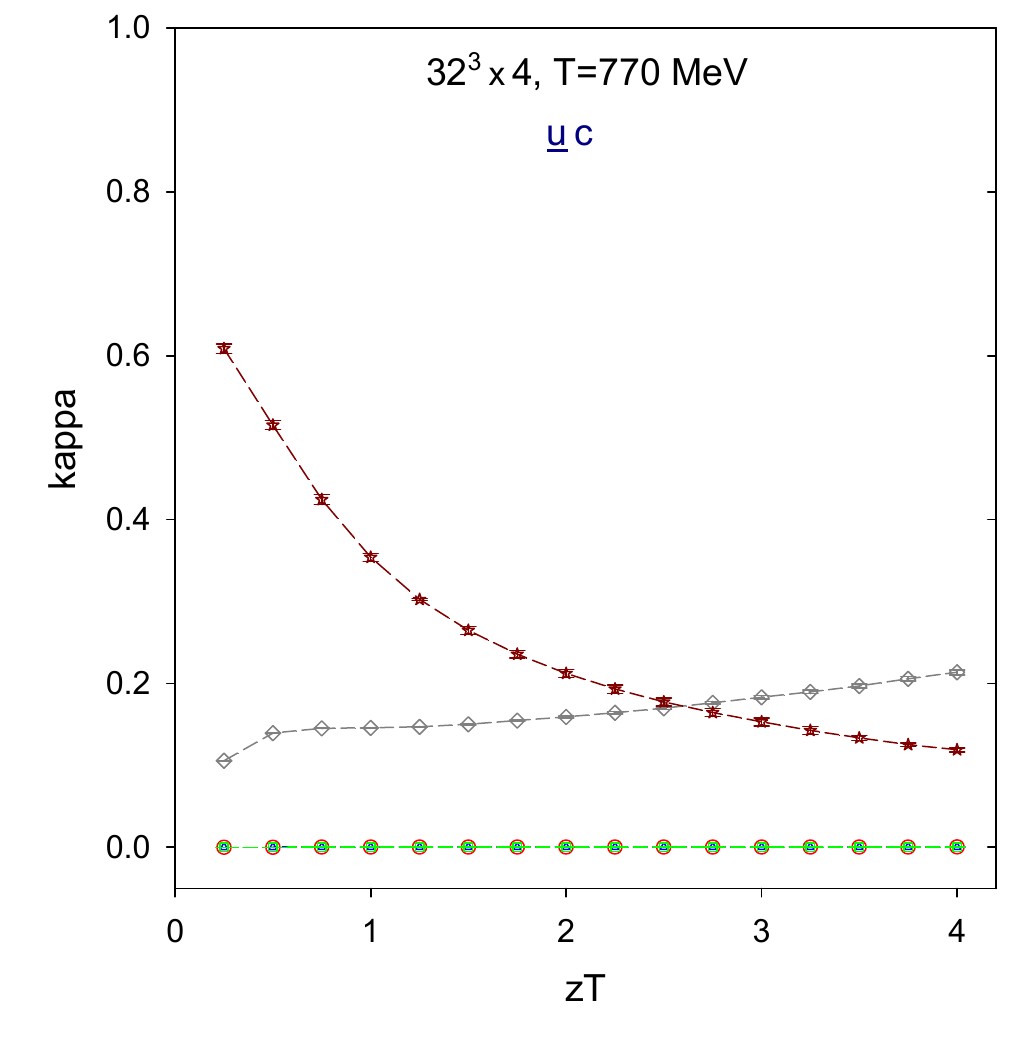}
\\
  \includegraphics[width=7.2cm,clip=true]{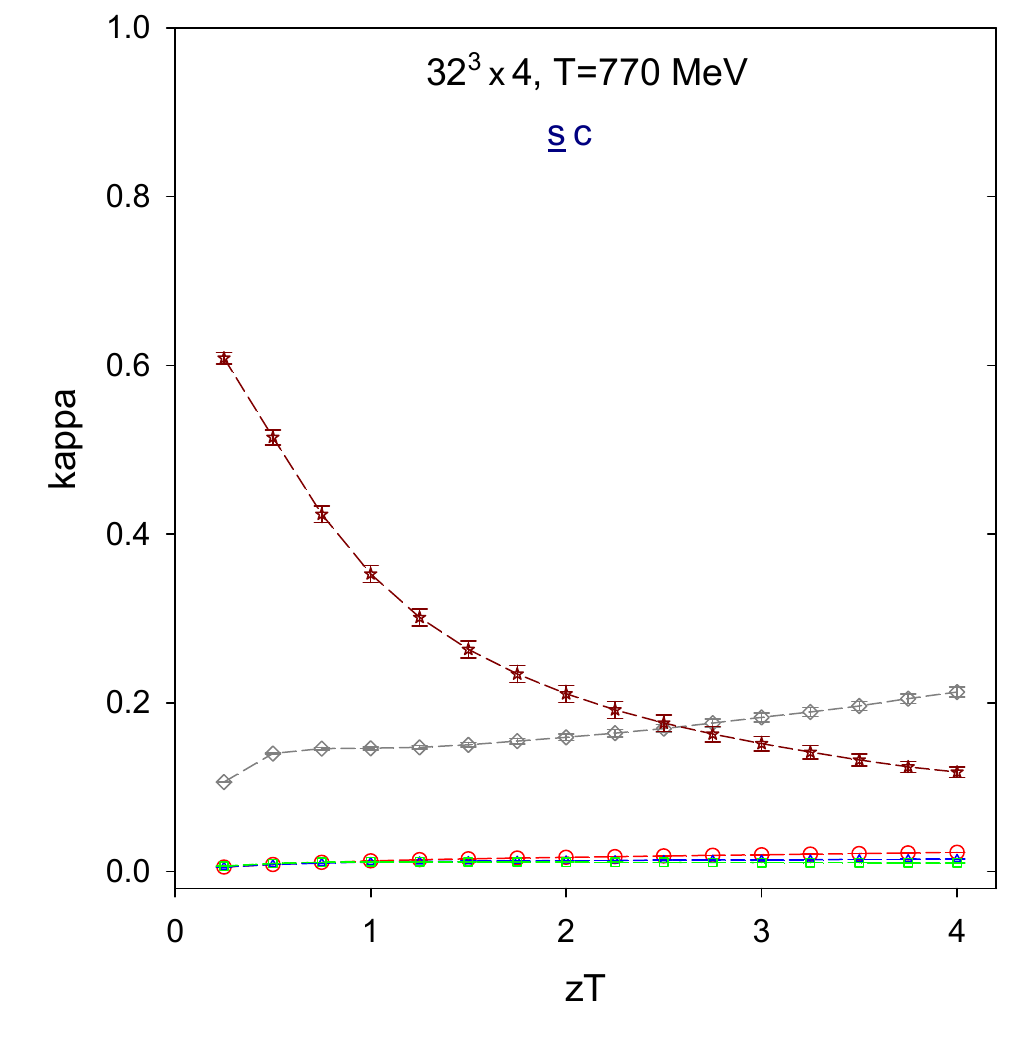}
&
  \includegraphics[width=7.2cm,clip=true]{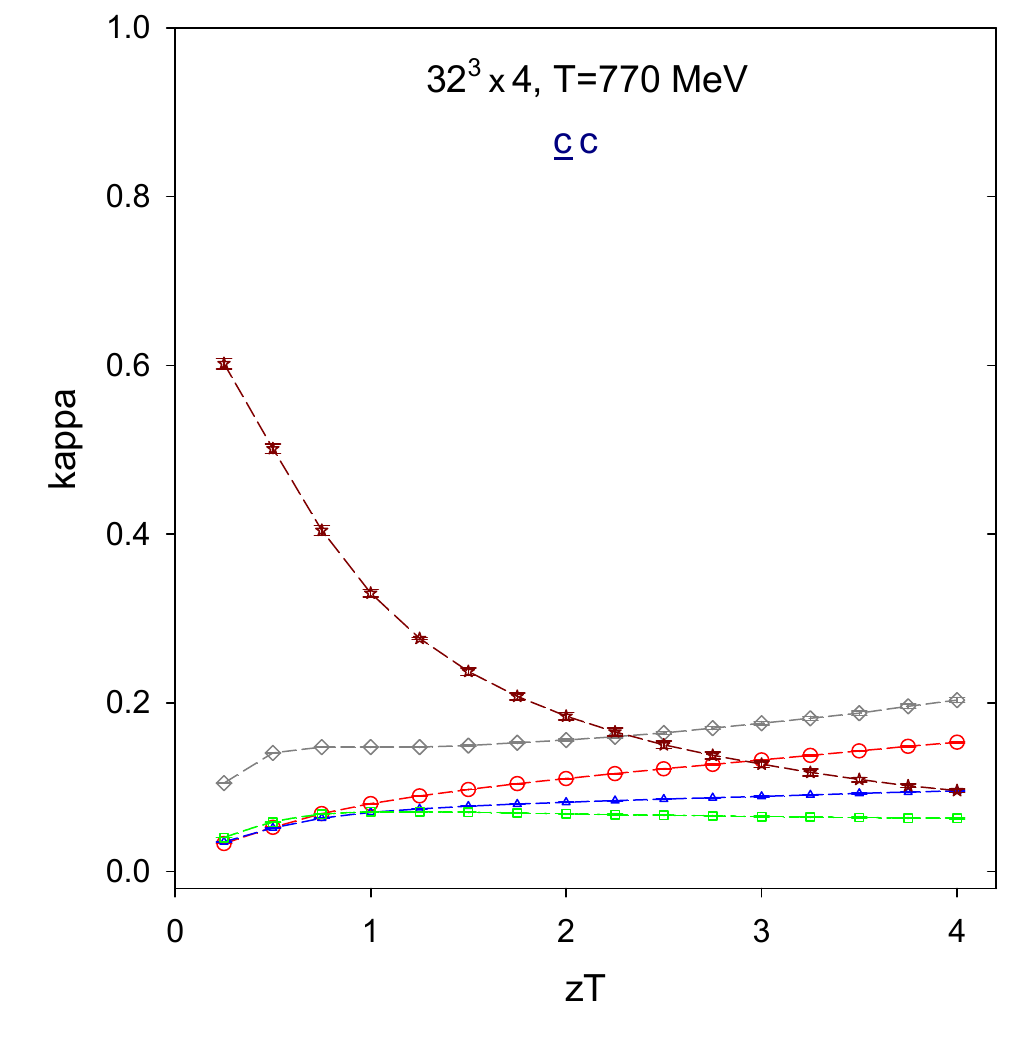}
  \end{tabular}
  \caption{
   The symmetry breaking parameters of spatial $z$ correlators of meson interpolators of six flavor 
   combinations ($\bar u d$, $\bar u s$, $\bar s s$, $\bar u c$, $\bar s c$, and $\bar c c$)
   in $N_f=2+1+1$ lattice QCD at $T \simeq 770$~MeV.
  }
\label{fig:Kz_T770}
\end{figure*}

\begin{figure*}[tbp]
  \centering
  \begin{tabular}{@{}c@{}c@{}}
  \includegraphics[width=7.2cm,clip=true]{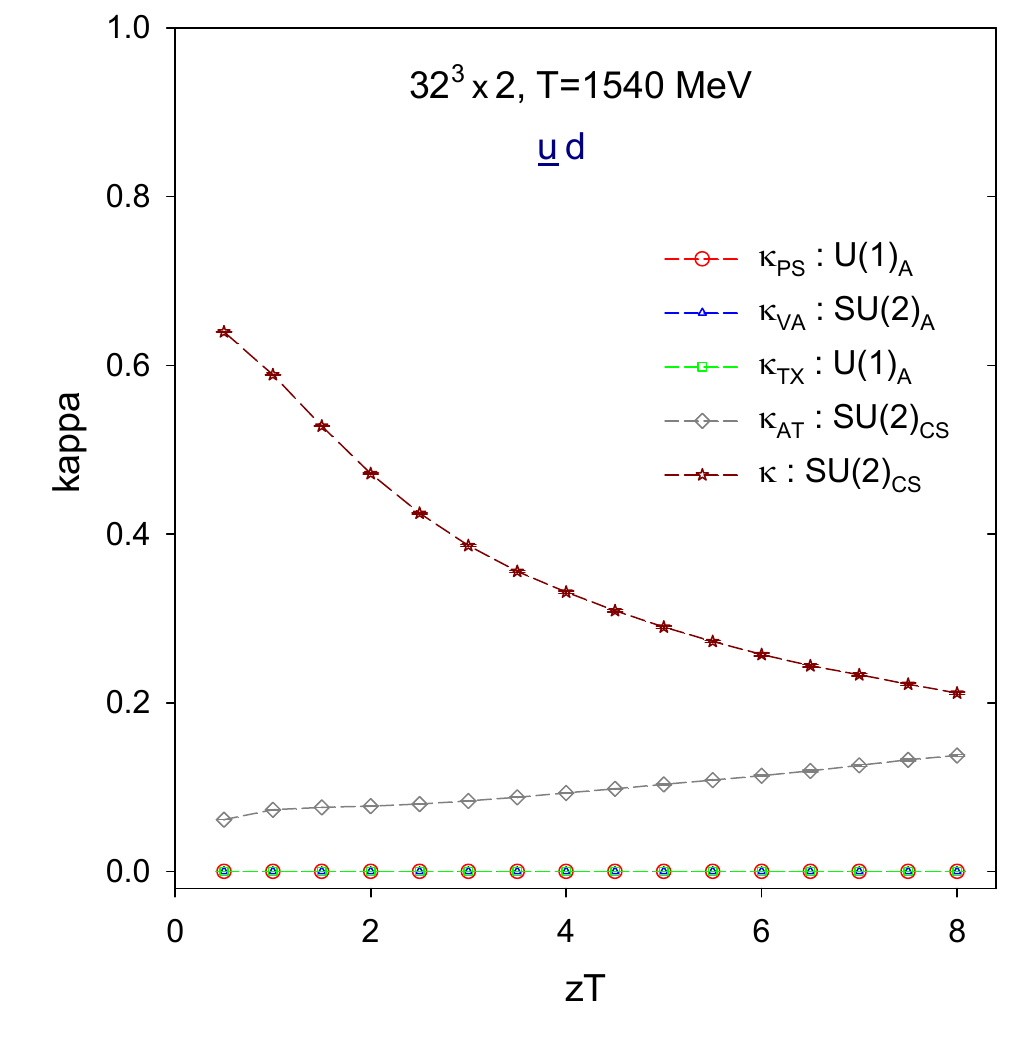}
&
  \includegraphics[width=7.2cm,clip=true]{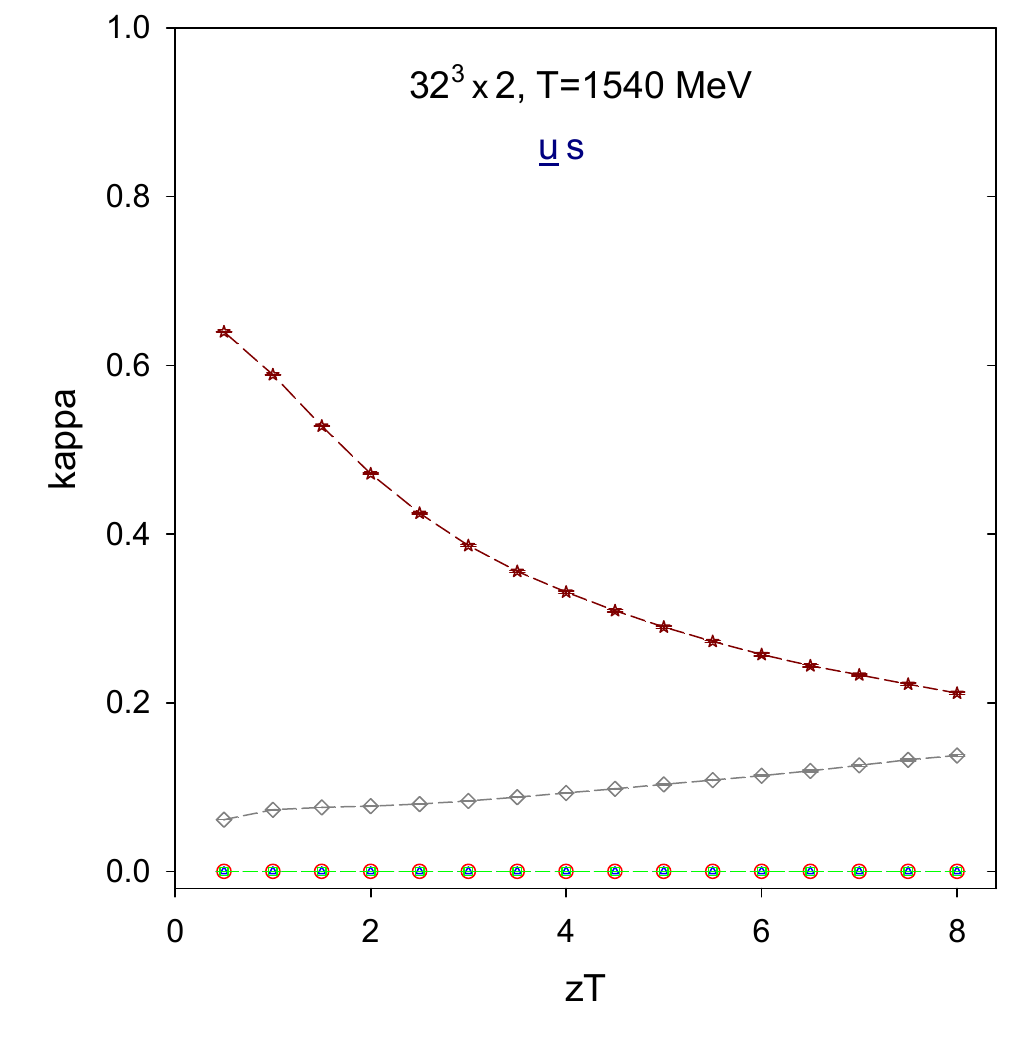}
\\
  \includegraphics[width=7.2cm,clip=true]{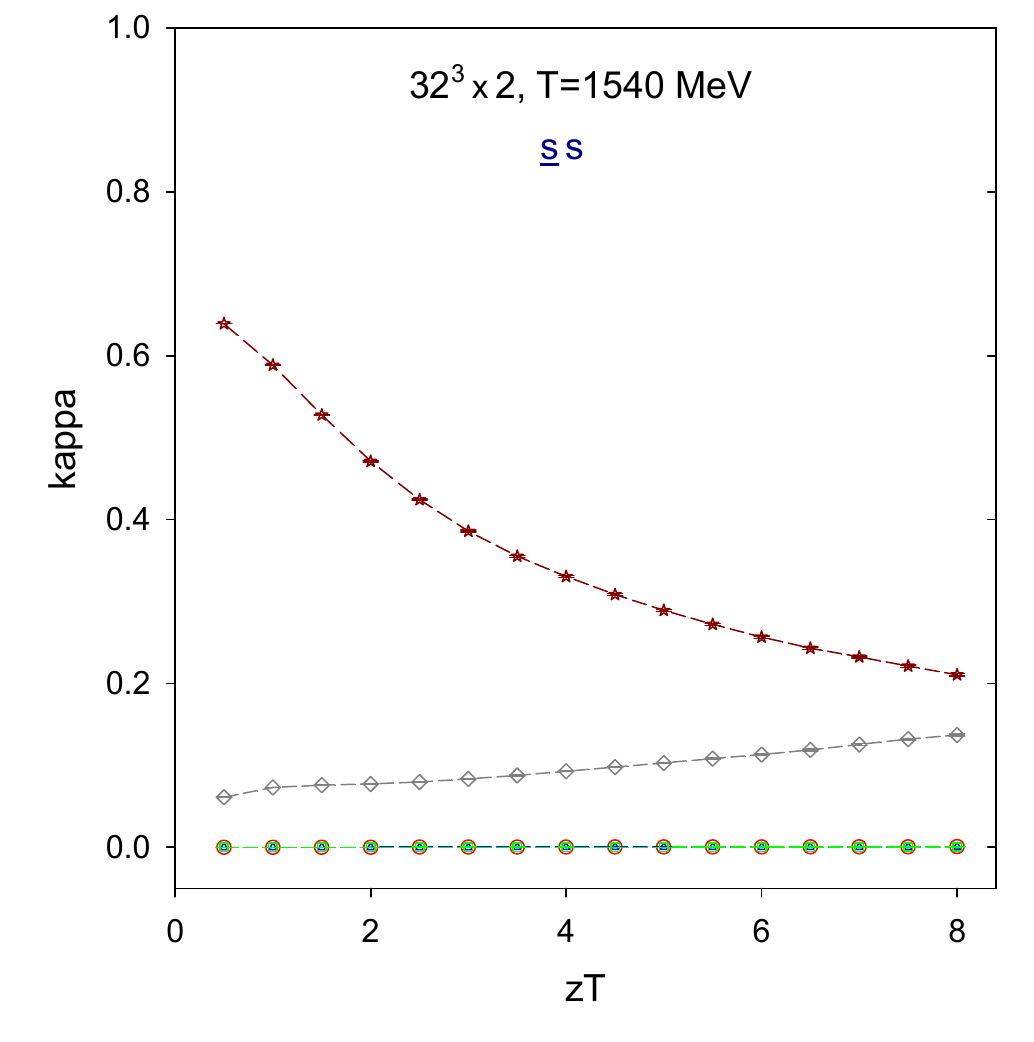}
&
  \includegraphics[width=7.2cm,clip=true]{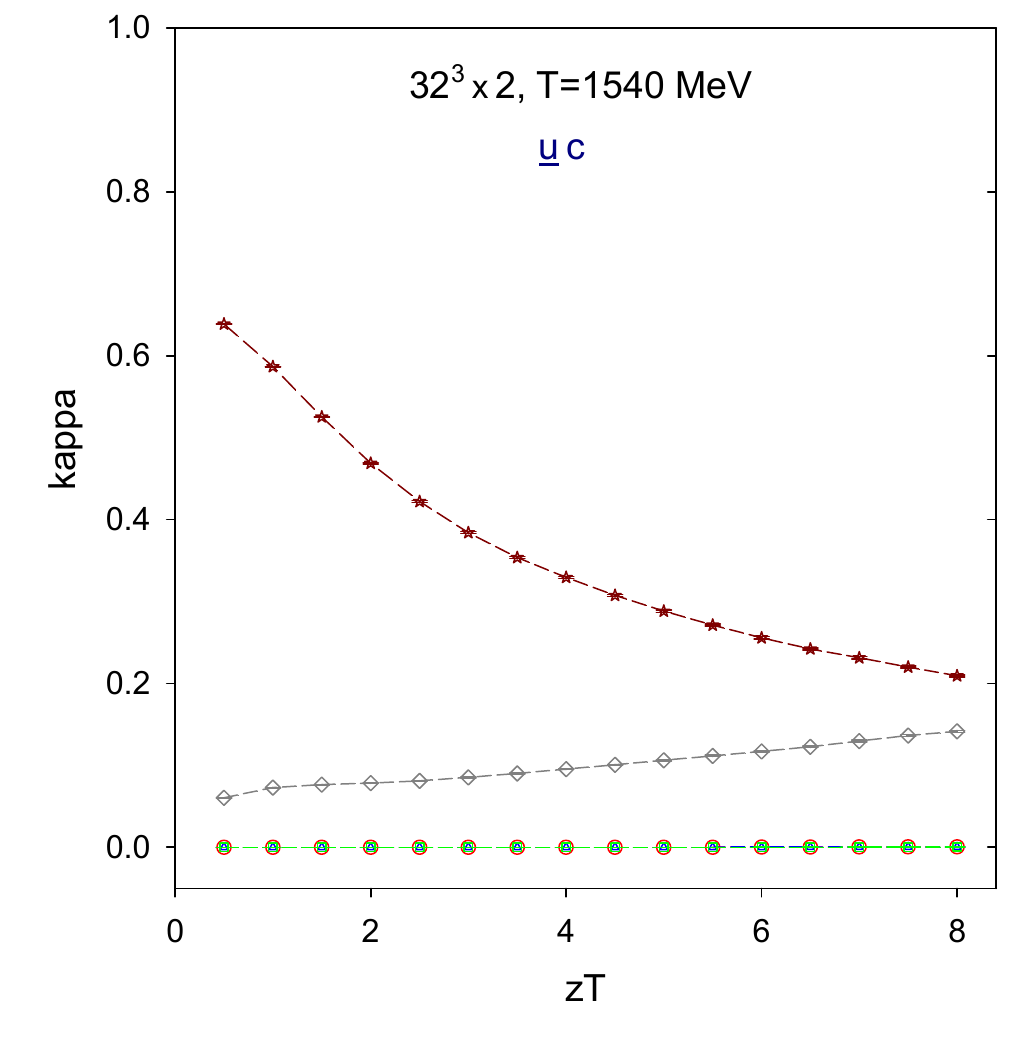}
\\
  \includegraphics[width=7.2cm,clip=true]{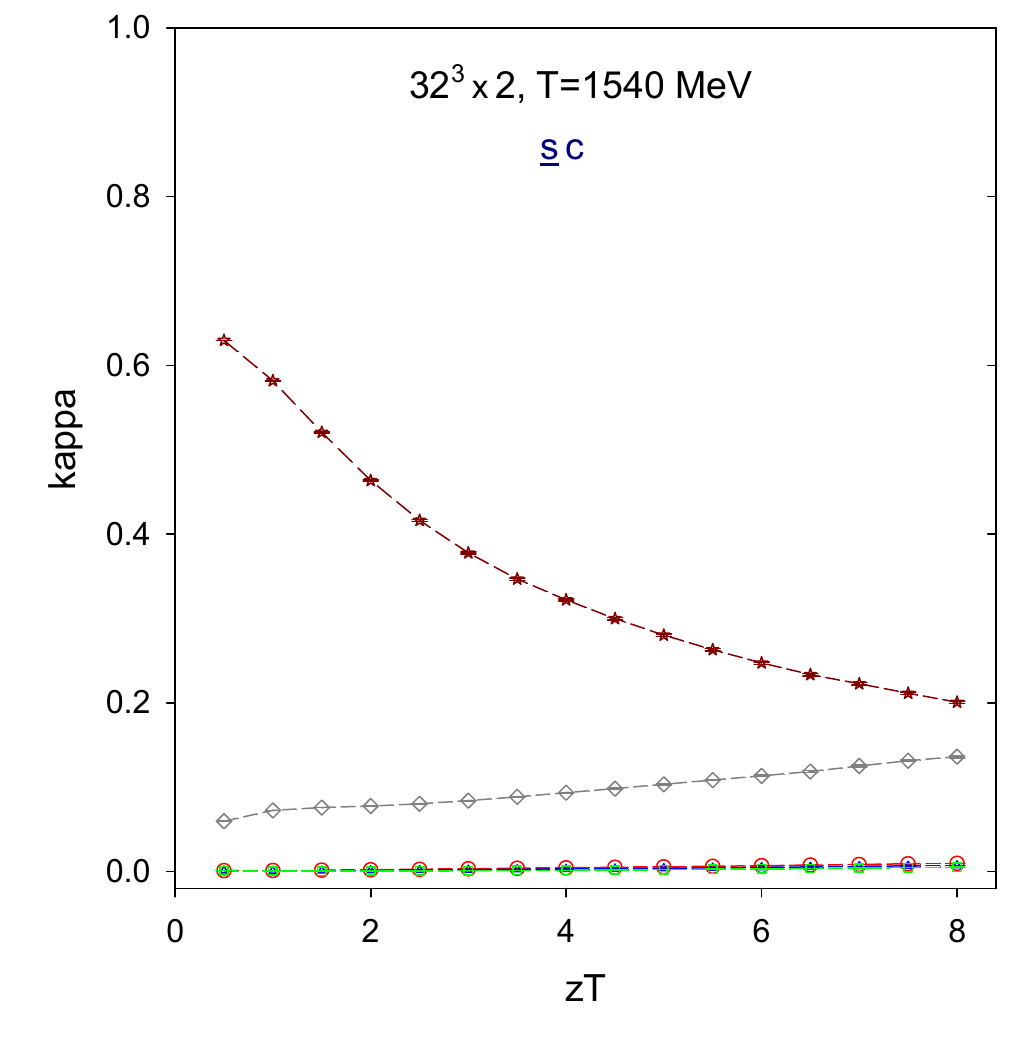}
&
  \includegraphics[width=7.2cm,clip=true]{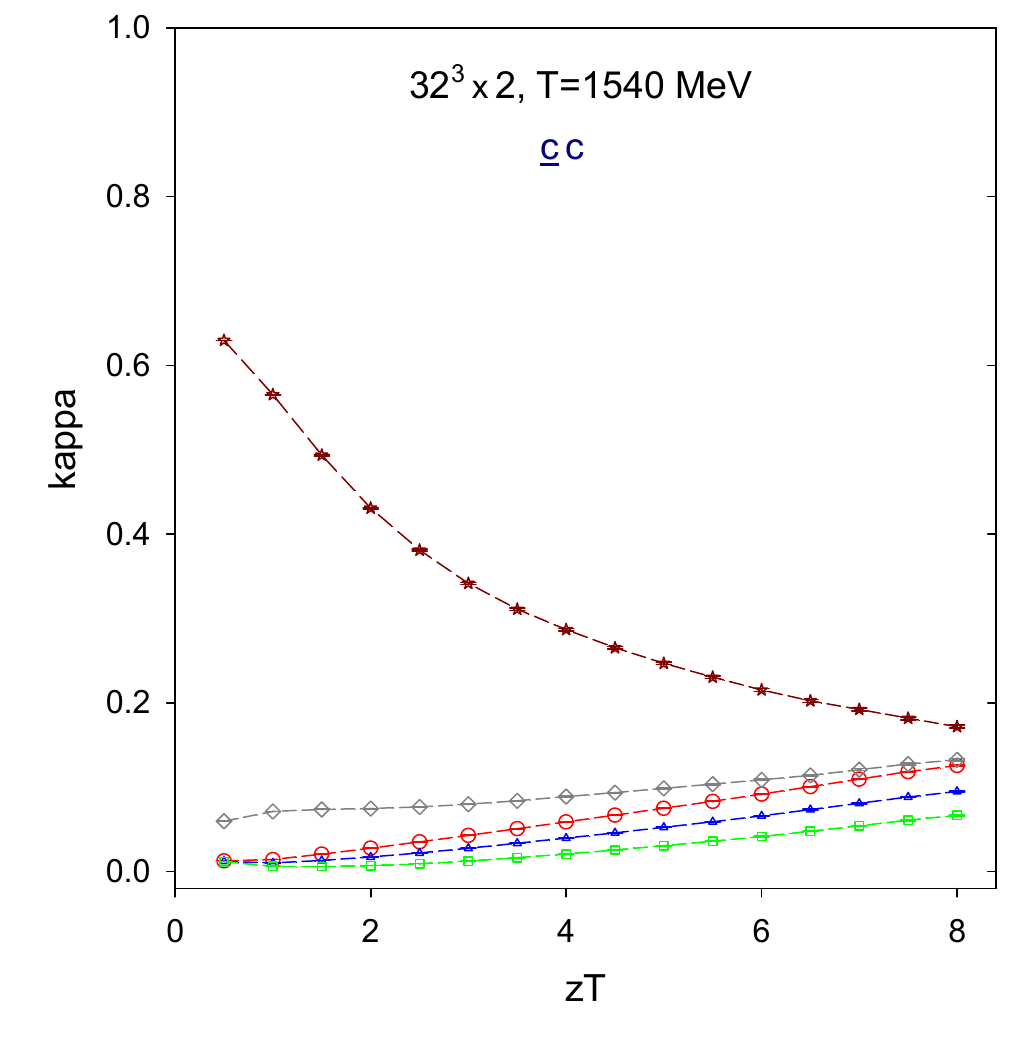}
\\
  \end{tabular}
  \caption{
   The symmetry breaking parameters of spatial $z$ correlators of meson interpolators of six flavor 
   combinations ($\bar u d$, $\bar u s$, $\bar s s$, $\bar u c$, $\bar s c$, and $\bar c c$)
   in $N_f=2+1+1$ lattice QCD at $T \simeq 1540$~MeV.
  }
\label{fig:Kz_T1540}
\end{figure*}

\section{Symmetry breaking parameters of $(\bar u d, \bar u s, \bar s s, \bar u c, \bar s c, \bar c c)$} 
\label{kappa_all}

In this section, we use the criteria (\ref{eq:SU2_crit_z}), (\ref{eq:U1_TX_crit_z}), 
and (\ref{eq:SU2_CS_crit_z}) to obtain $T_c$ 
(the temperature for the restoration of $SU(2)_L \times SU(2)_R$ chiral symmetry),  
$T_1$ (the temperature for the restoration of $U(1)_A$ symmetry), 
and the window of $T$ for the emergence of the approximate $SU(2)_{CS}$ symmetry,  
for six flavor combinations 
$(\bar u d, \bar u s, \bar s s, \bar u c, \bar s c, \bar c c)$, respectively. 
To this end, we use the $z$ correlators in Figs. \ref{fig:Cz_T192}--\ref{fig:Cz_T1540}   
to compute the symmetry-breaking parameters 
$\kappa_{VA}$, $\kappa_{PS}$, $\kappa_{TX}$, 
$\kappa_{AT}$, and $\kappa$, as defined in Sec. \ref{kappa}. 

In Figs. \ref{fig:Kz_T192}--\ref{fig:Kz_T1540}, the symmetry breaking parameters 
of six flavor combinations are plotted 
as a function of the dimensionless variable $zT$, for seven temperatures 
in the range of 190--1540 MeV. 

At each $T$, and for fixed $zT$, the chiral symmetry breakings due to the quark masses of the 
meson operator can be seen clearly from $\kappa_{VA}$, $\kappa_{PS}$, and $\kappa_{TX}$, 
in the order of 
\bea
\label{eq:k_VA_PS_TX}
\kappa_{\alpha}^{\bar u d} < \kappa_{\alpha}^{\bar u s} < \kappa_{\alpha}^{\bar u c}
< \kappa_{\alpha}^{\bar s s} < \kappa_{\alpha}^{\bar s c} < \kappa_{\alpha}^{\bar c c} 
\eea  
for each channel of $\alpha = ({\scriptstyle VA,\ PS,\ TX})$. 
Also, for each flavor content,    
$\kappa_{\alpha} (zT)$ at fixed $zT$ is a monotonic decreasing function of $T$.
Note that for the charmonium $\bar c c$, the chiral symmetry breakings at $T=1540$ 
are still not negligible, e.g., at $zT=4$, 
$ 0.02 \lesssim \kappa_{TX} < \kappa_{VA} < \kappa_{PS} \lesssim 0.06$. 

About the $SU(2)_{CS}$ symmetry breaking parameter $\kappa_{CS}=\max(\kappa_{AT}, \kappa_{TX})$, 
for any flavor combination, it is a monotonic decreasing function of $T$ at fixed $zT$, 
since both $\kappa_{AT}(zT)$ and $\kappa_{TX}(zT)$ are monotonic decreasing function of $T$. 
However, the flavor dependence of $\kappa_{CS}$ turns out to be rather nontrivial, 
and it is temperature dependent. Similarly, the flavor dependence of the $SU(2)_{CS}$ 
symmetry fading parameter $\kappa$ is also temperature dependent. 
Nevertheless, it is interesting to point out that $\kappa_{CS}$ of the $\bar u c$ sector 
is the smallest among all flavor sectors, 
while $\kappa$ is almost the same for all flavor sectors,  
for all seven temperatures in the range of 190--1540 MeV. 
This suggests that the most attractive vector meson channels to detect
the emergence of approximate $SU(2)_{CS}$ symmetry are in the $\bar u c$ sector. 
This will be addressed more quantitatively in Sec.~\ref{SU2_CS},
in terms of the window of $T$ for the approximate $SU(2)_{CS}$ symmetry.

%
%

\begin{figure*}[tbp]
  \centering
  \includegraphics[width=0.9\linewidth,clip=true]{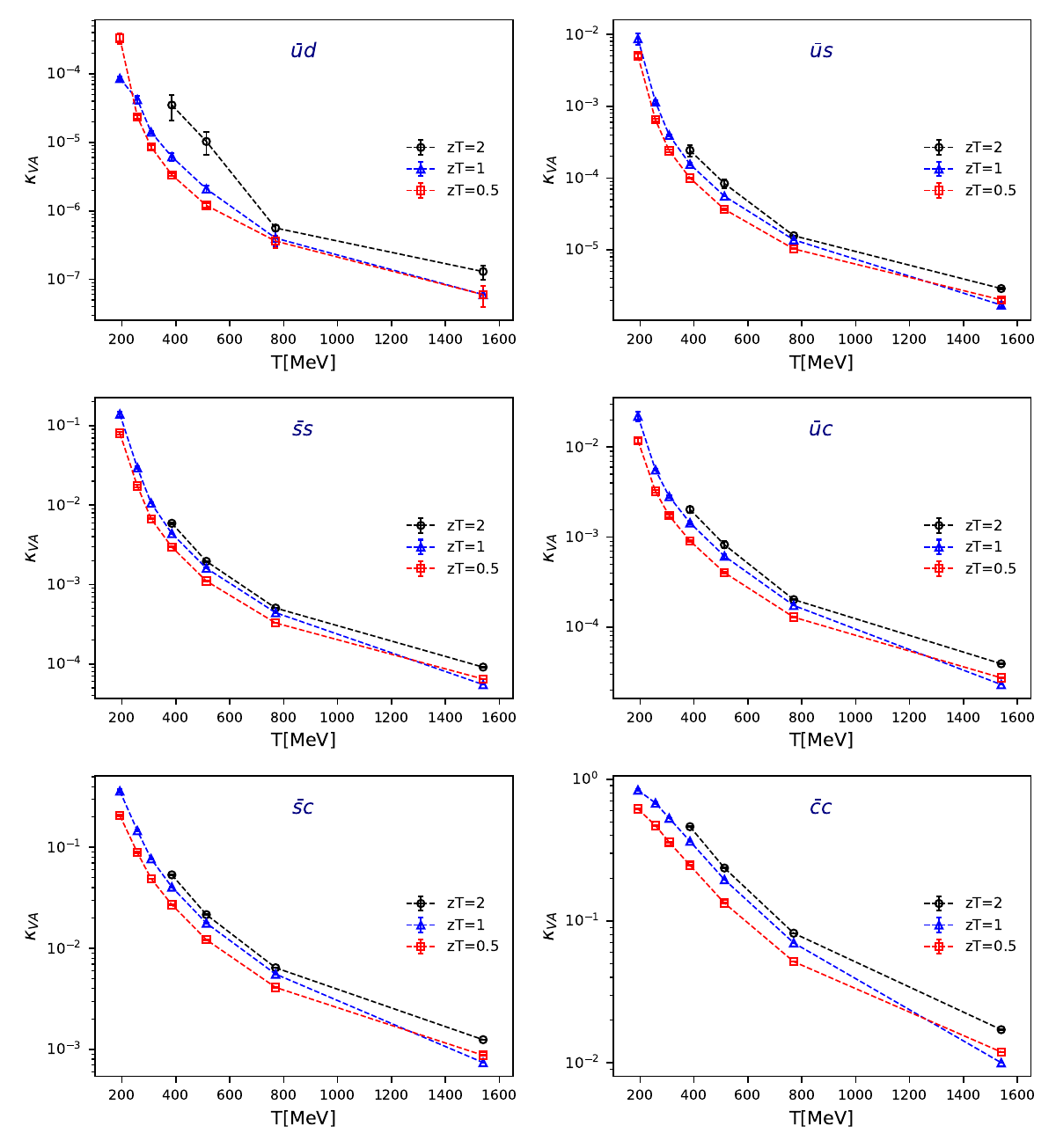}
    \caption{
    The $SU(2)_L \times SU(2)_R $ chiral symmetry breaking parameter $\kappa_{VA}$ at $zT = (0.5, 1, 2)$, 
    for seven temperatures in the range $T \sim 190$--$1540$~MeV and six flavor combinations 
    ($\bar u d$, $\bar u s $, $\bar s s $, $\bar u c$, $\bar s c $, $\bar c c $).
  }
\label{fig:kVA_z}
\end{figure*}

\begin{figure*}[tbp]
  \centering
  \includegraphics[width=0.9\linewidth,clip=true]{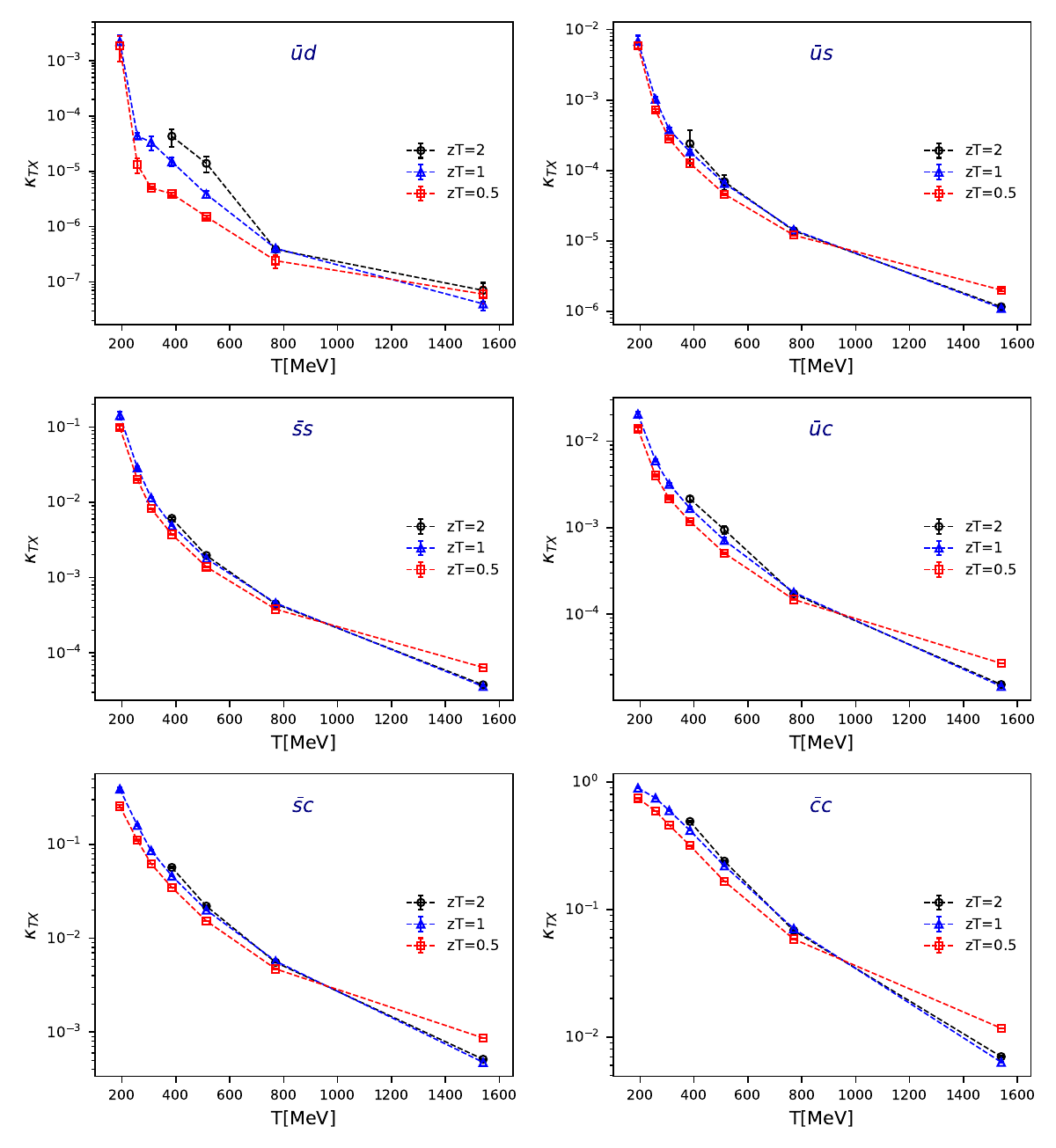}
    \caption{
    The $U(1)_{A}$ symmetry breaking parameter $\kappa_{TX}$ at $zT = (0.5, 1, 2)$,  
    for seven temperatures in the range $T \sim 190$--$1540$~MeV and six flavor combinations 
    ($\bar u d$, $\bar u s $, $\bar s s $, $\bar u c$, $\bar s c $, $\bar c c $).
  }
\label{fig:kTX_z}
\end{figure*}

\subsection{Hierarchical restoration of chiral symmetries}
\label{HRCS}

Now we proceed to investigate the restoration of chiral symmetries     
in $N_f=2+1+1$ lattice QCD at the physical point. 
We use the criteria (\ref{eq:SU2_crit_z}) and (\ref{eq:U1_TX_crit_z}) to 
obtain $T_c$ and $T_1$ for each flavor combination.
To this end, we collect the data of $\kappa_{VA}(zT)$ and $\kappa_{TX}(zT)$ 
at the same $zT=(0.5, 1, 2)$, and plot them as a function of $T$, as shown 
in Figs. \ref{fig:kVA_z} and \ref{fig:kTX_z}. 
According to (\ref{eq:k_VA_PS_TX}), it follows that 
for any $\epsilon_{VA}$ in (\ref{eq:SU2_crit_z}) 
and any $\epsilon_{TX}$ in (\ref{eq:U1_TX_crit_z}),    
the flavor dependence of $T_c$ and $T_1$ is in the order of  
\bea
\label{eq:Tc_order}
&& T_c^{\bar u d} < T_c^{\bar u s} < T_c^{\bar u c} < T_c^{\bar s s} < T_c^{\bar s c} < T_c^{\bar c c}, \\ 
&& T_1^{\bar u d} < T_1^{\bar u s} < T_1^{\bar u c} < T_1^{\bar s s} < T_1^{\bar s c} < T_1^{\bar c c}, 
\label{eq:T1_order}
\eea 
which immediately gives 
\bea
\label{eq:Tc1_order}
T_{c1}^{\bar u d} < T_{c1}^{\bar u s} < T_{c1}^{\bar u c} < T_{c1}^{\bar s s} < T_{c1}^{\bar s c} 
< T_{c1}^{\bar c c}.  
\eea 
Equations (\ref{eq:Tc_order})--(\ref{eq:Tc1_order}) are the first results of lattice QCD. 
They immediately give the hierarchical restoration of chiral symmetries in $N_f=2+1+1$ QCD, 
i.e., from the restoration of $SU(2)_L \times SU(2)_R \times U(1)_A$ chiral symmetry 
of $(u, d)$ quarks at $ T_{c1}^{\bar u d}$ to the restoration of $SU(3)_L \times SU(3)_R \times U(1)_A$ chiral symmetry 
of $(u, d, s)$ quarks at $ T_{c1}^{\bar s s} > T_{c1}^{\bar u d}$, then to  
the restoration of $SU(4)_L \times SU(4)_R \times U(1)_A$ chiral symmetry 
of $(u, d, s, c)$ quarks at $ T_{c1}^{\bar c c} > T_{c1}^{\bar s s}$.

In the following, we demonstrate the hierarchical restoration of chiral symmetries explicitly, 
for $\epsilon_{VA}=(0.05, 0.01)$ and $\epsilon_{TX}=(0.05, 0.01)$, respectively.

\begin{table}[tbp]
\begin{center}
\caption{The temperature $T_c^{\bar q_1 q_2}$[MeV] satisfying 
         the criterion (\ref{eq:SU2_crit_z}) at $zT=(0.5, 1, 2)$,  
         for $\epsilon_{VA} = (0.05, 0.01)$, respectively. 
}
\setlength{\tabcolsep}{2.5pt}
\vspace{2mm}
\begin{tabular}{|c|cc|cc|cc|}
\hline
\multirow{2}{*}{  } & \multicolumn{2}{c|}{$zT=0.5$} 
                    & \multicolumn{2}{c|}{$zT=1$} 
                    & \multicolumn{2}{c|}{$zT=2$}   \\
\hline
    $\epsilon_{VA}$ & 0.05 & 0.01 & 0.05 & 0.01 & 0.05 & 0.01 \\
\hline
\hline
 $T_c^{\bar u d}$ & $< 190$ & $< 190$  & $< 190$ & $< 190$  & $< 190$ & $< 190$   \\
 $T_c^{\bar u s}$ & $< 190$ & $< 190$  & $< 190$ & $< 190$  & $< 190$ & $< 190$   \\
 $T_c^{\bar s s}$ & 210(5)  & 285(5)   &  235(5) & 315(5)   & 260(10) & 345(10)   \\ 
 $T_c^{\bar u c}$ & $< 190$ & 210(5)   & $< 190$ & 230(5)   & $< 190$ & 260(10)   \\
 $T_c^{\bar s c}$ & 305(5)  & 560(5)   &  360(5) & 640(5)   &  395(5) & 675(5)    \\
 $T_c^{\bar c c}$ & 785(5)  & 1640(10) &  900(5) & 1540(10) & 990(5) & 1785(10)  \\
\hline
\end{tabular}
\label{tab:Tc_all}
\end{center}
\end{table}

\begin{table}[tbp]
\begin{center}
\caption{The temperature $T_1^{\bar q_1 q_2}$[MeV] satisfying 
         the criterion (\ref{eq:U1_TX_crit_z}) at $zT=(0.5, 1, 2)$,  
         for $\epsilon_{TX} = (0.05, 0.01)$, respectively. 
}
\setlength{\tabcolsep}{2.5pt}
\vspace{2mm}
\begin{tabular}{|c|cc|cc|cc|}
\hline
\multirow{2}{*}{  } & \multicolumn{2}{c|}{$zT=0.5$} 
                    & \multicolumn{2}{c|}{$zT=1$} 
                    & \multicolumn{2}{c|}{$zT=2$}   \\
\hline
    $\epsilon_{TX}$ & 0.05 & 0.01 & 0.05 & 0.01 & 0.05 & 0.01 \\
\hline
\hline
 $T_1^{\bar u d}$ & $< 190$ & $< 190$  & $< 190$ & $< 190$  & $< 190$ & $< 190$   \\
 $T_1^{\bar u s}$ & $< 190$ & $< 190$  & $< 190$ & $< 190$  & $< 190$ & $< 190$   \\
 $T_1^{\bar s s}$ & 220(5)  &  295(5)  &  235(5) & 320(5)   & 255(10) &  350(10)  \\
 $T_1^{\bar u c}$ & $< 190$ &  200(5)  & $< 190$ & 230(5)   & $< 190$ &  250(10)  \\
 $T_1^{\bar s c}$ & 335(5)  &  605(5)  &  375(5) & 655(5)   &  400(5) &  660(5)   \\
 $T_1^{\bar c c}$ & 835(5)  &  1610(10) &  875(5) & 1395(5) &  865(5) &  1420(5) \\
\hline
\end{tabular}
\label{tab:T1_all}
\end{center}
\end{table}

Using linear interpolation or extrapolation of the data points in 
Figs. \ref{fig:kVA_z} and \ref{fig:kTX_z},  
we obtain the results of $T_c$ and $T_1$ for six flavor combinations, as listed in 
Tables \ref{tab:Tc_all} and \ref{tab:T1_all}, for $\epsilon_{VA}=(0.05, 0.01)$ and 
$\epsilon_{TX}=(0.05, 0.01)$, respectively. 

In Tables \ref{tab:Tc_all} and \ref{tab:T1_all}, for both $\bar u d$ and $\bar u s$ sectors,    
both $T_c$ and $ T_1$ are less than 190 MeV, 
for any combinations of $\epsilon_{VA}=(0.05, 0.01)$, $\epsilon_{TX}=(0.05, 0.01)$, and $zT=(0.5, 1, 2)$.  
For these cases, $SU(2)_L \times SU(2)_R \times U(1)_A $ 
is restored at a temperature lower than 190 MeV, for both $\bar u d$ and $\bar u s$ sectors,    
However, for the $\bar u c$ sector, only for $\epsilon_{VA}=\epsilon_{TX} = 0.05$,   
$SU(2)_L \times SU(2)_R \times U(1)_A $ is restored at a temperature lower than 190 MeV. 

Now we investigate the hierarchical restoration of chiral symmetries 
with $\epsilon_{VA}=\epsilon_{TX}=0.05$ and $zT=1$. 
From Tables \ref{tab:Tc_all} and \ref{tab:T1_all}, one immediately sees that 
the $SU(3)_L \times SU(3)_R \times U(1)_A $ chiral symmetry of $(u,d,s)$ quarks 
is restored at $T_{c}^{\bar s s} \simeq T_1^{\bar s s} \simeq 235(5)$~MeV, since 
$T_{c}$ and $T_1$ of both $\bar u d$ and $\bar u s$ sectors are lower than 190 MeV.
Moreover, for the $\bar c c$ sector, $T_{c}^{\bar c c} \simeq 900(5)$~MeV, 
$T_{1}^{\bar c c} \simeq 875(5)$~MeV, 
and $T_c$ and $T_1$ of other flavor sectors are at lower temperatures, 
in the order of (\ref{eq:Tc_order}) and (\ref{eq:T1_order}). 
Thus, the $SU(4)_L \times SU(4)_R \times U(1)_A $ chiral symmetry of $(u,d,s,c)$ quarks 
is restored at $T_{c1}^{\bar c c} \sim 900(5)$~MeV. 

Next, we study how $T_c$ ($T_1$) depends on $\epsilon_{VA}$ ($\epsilon_{TX}$). 
Since $\kappa_{VA}^{\bar q_1 q_2} (\kappa_{TX}^{\bar q_1 q_2})$ at fixed $zT$ 
is a monotonic decreasing function of $T$, it follows  
that $T_c$ ($T_1$) is monotonically increased as $\epsilon_{VA}$ ($\epsilon_{TX}$) 
is decreased (i.e., the precision of the chiral symmetry becomes higher).  
For example, if we set $\epsilon_{VA}=\epsilon_{TX}=0.01$, then at $zT=1$, 
the $SU(3)_L \times SU(3)_R \times U(1)_A $ chiral symmetry of $(u,d,s)$ quarks 
is restored at $T_{c1}^{\bar s s} \simeq 320(5)$~MeV 
[since $T_c^{\bar s s} \lesssim T_1^{\bar s s} \simeq 320(5)$~MeV],   
and the $SU(4)_L \times SU(4)_R \times U(1)_A $ chiral symmetry of $(u,d,s,c)$ quarks 
is restored at $T_{c1}^{\bar c c} \simeq 1540(10)$~MeV 
[since $T_c^{\bar c c} \simeq 1540(10)$~MeV and $T_1^{\bar c c} \simeq 1395(5)$~MeV]. 
Obviously, no matter how small the values of $\epsilon_{VA}$ and $\epsilon_{TX}$ become,
the hierarchical restoration of chiral symmetries in QCD with 
physical $(u,d,s,c)$ quarks will occur at higher temperatures. 
%
%
%

\subsection{$SU(2)_{\text CS}$ chiral-spin symmetry}
\label{SU2_CS}

%

\begin{figure*}[tbp]
  \centering
\begin{tabular}{@{}c@{}c@{}}
  \includegraphics[width=8.0cm,clip=true]{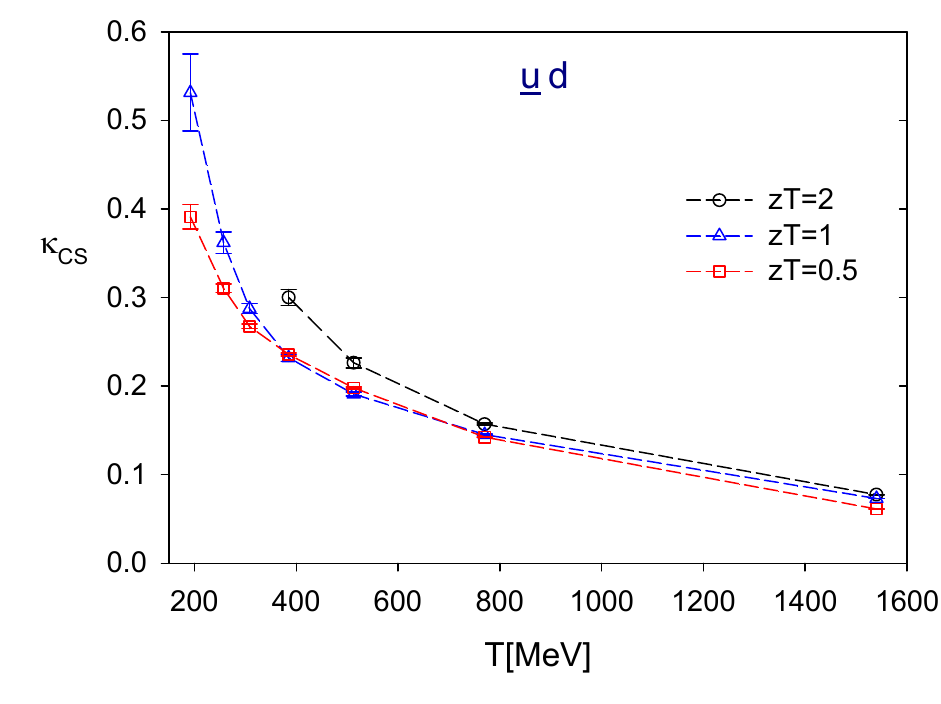}
&
  \includegraphics[width=8.0cm,clip=true]{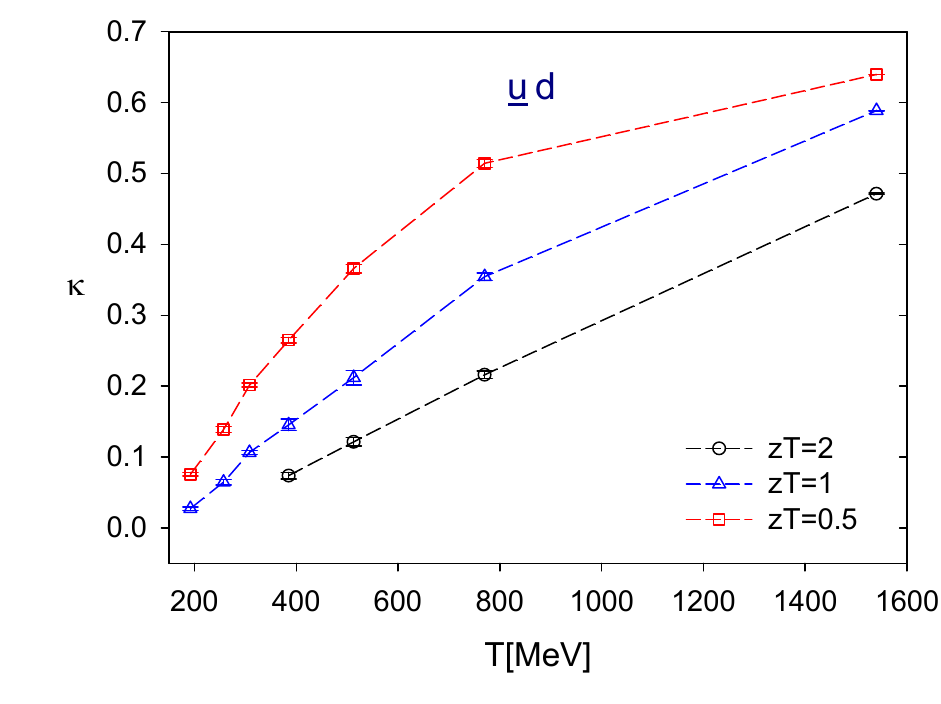}  
\\
  \includegraphics[width=8.0cm,clip=true]{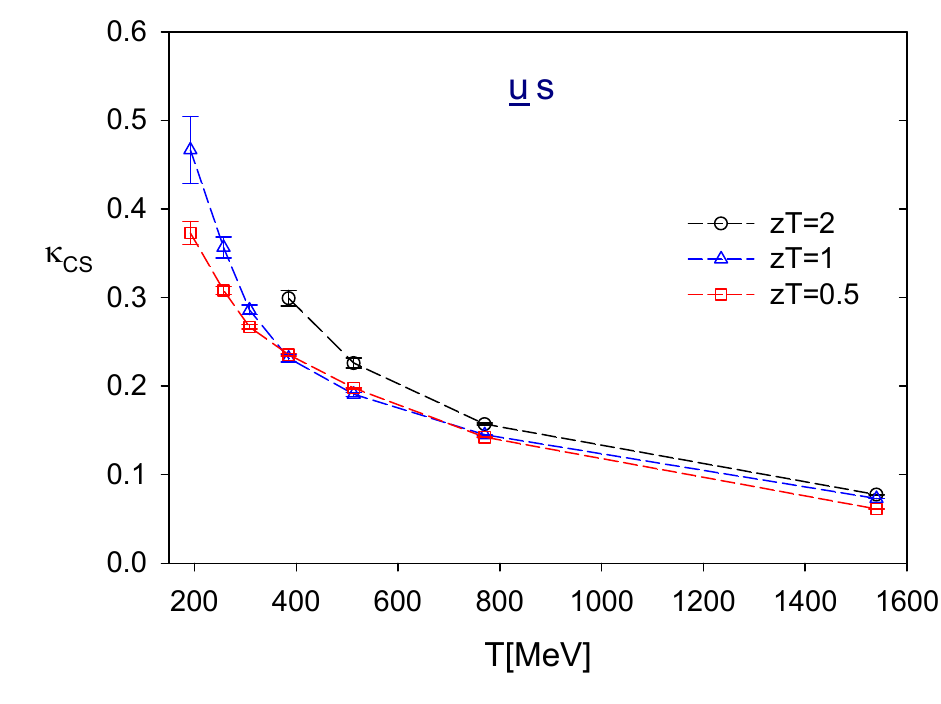}
&
  \includegraphics[width=8.0cm,clip=true]{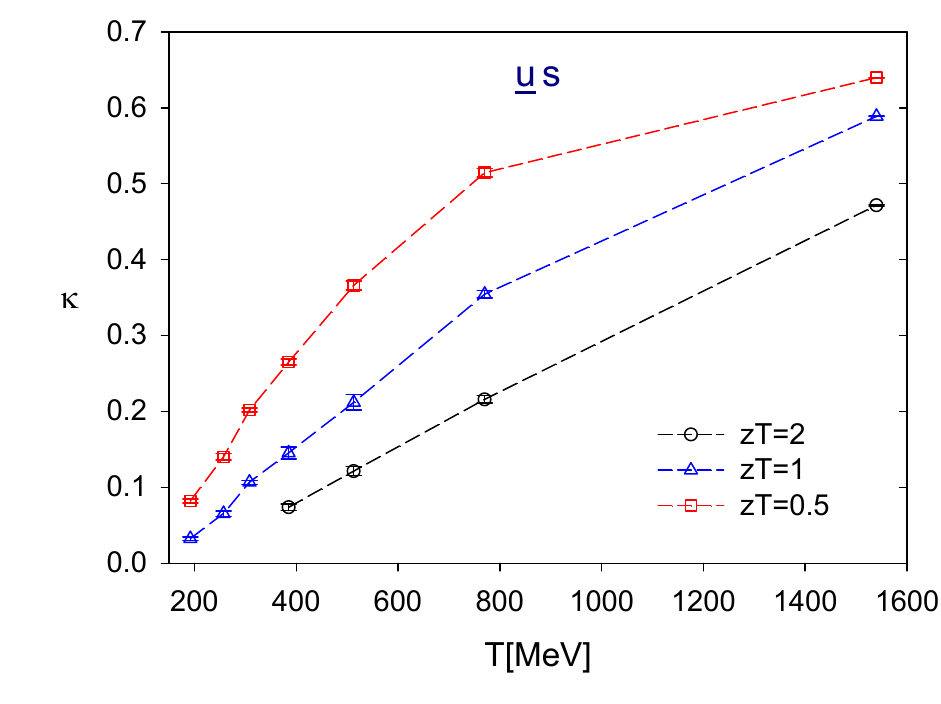}  
\\
  \includegraphics[width=8.0cm,clip=true]{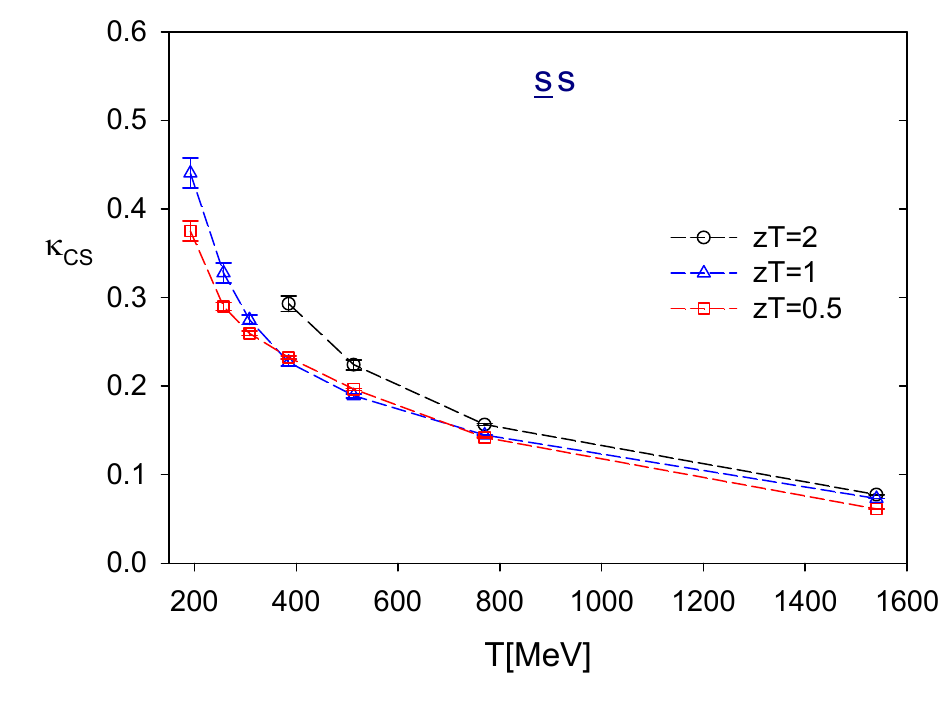}
&
  \includegraphics[width=8.0cm,clip=true]{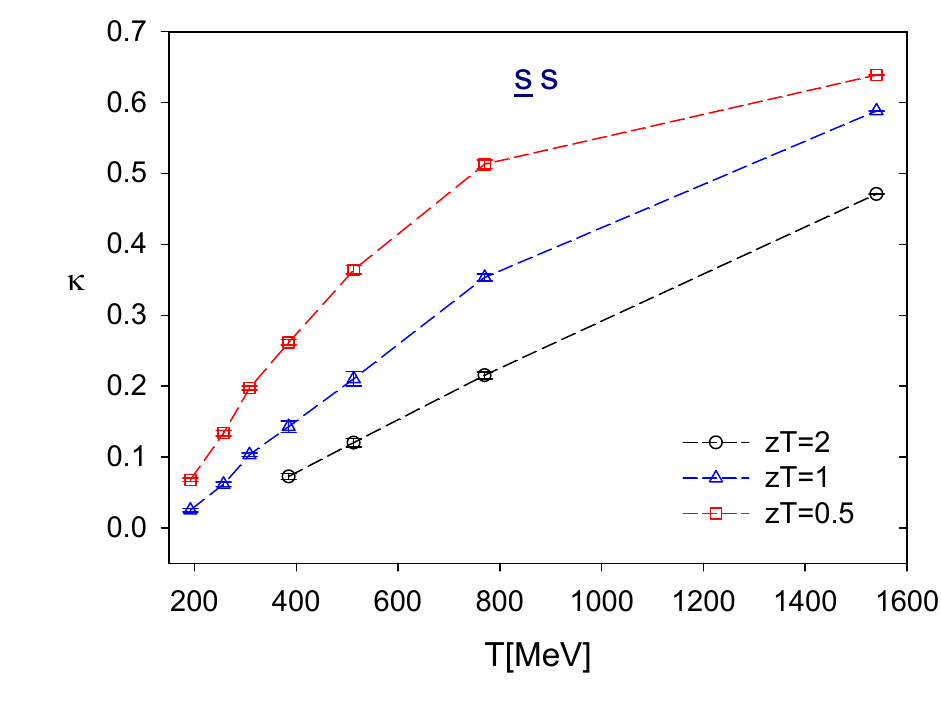}  
\end{tabular}
  \caption{
    The $SU(2)_{CS}$ symmetry breaking and fading parameters ($\kappa_{AT}$, $\kappa$) 
    at $zT = (0.5, 1, 2)$, for flavor combinations ($\bar u d$, $\bar u s$, $\bar s s $), 
    and $T \sim 193$--$1540$~MeV.   
  }
\label{fig:kCS_k_light}
\end{figure*}

\begin{figure*}[tbp]
  \centering
\begin{tabular}{@{}c@{}c@{}}
  \includegraphics[width=8.0cm,clip=true]{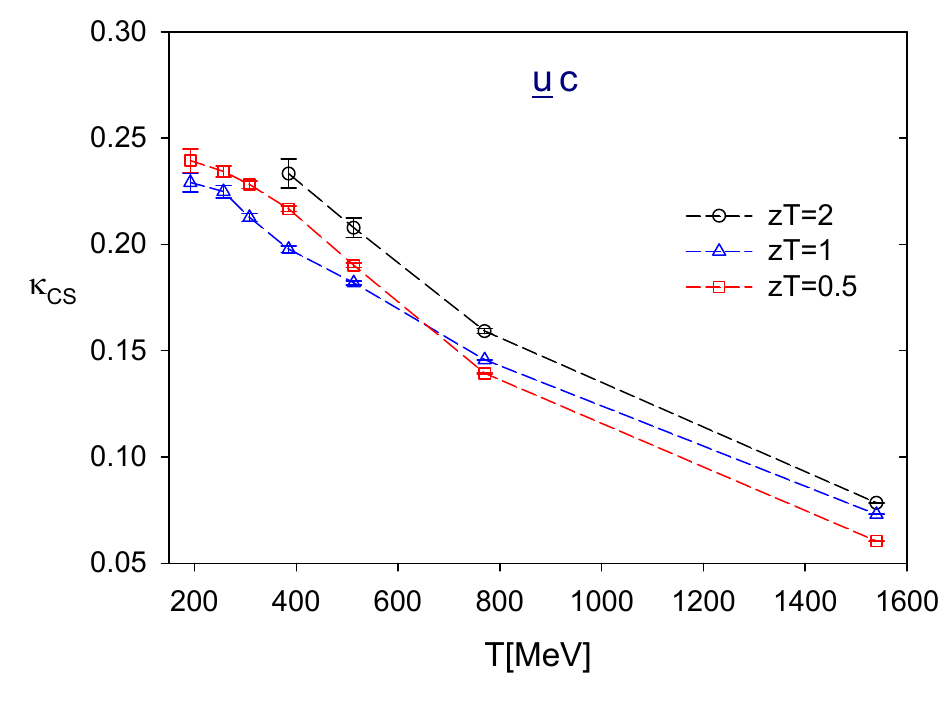}
&
  \includegraphics[width=8.0cm,clip=true]{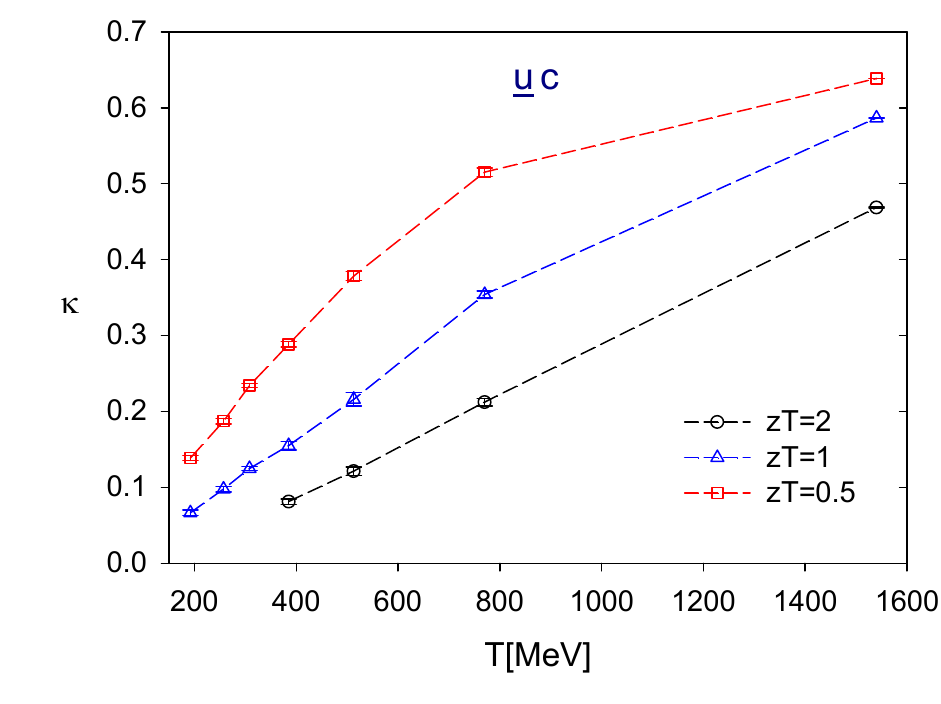}  
\\
  \includegraphics[width=8.0cm,clip=true]{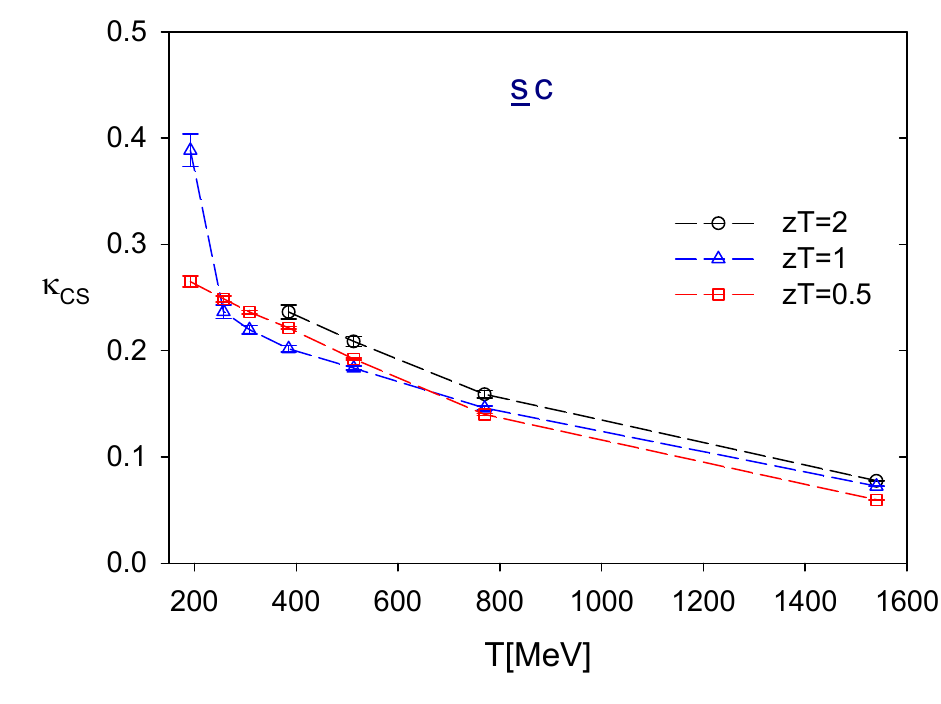}
&
  \includegraphics[width=8.0cm,clip=true]{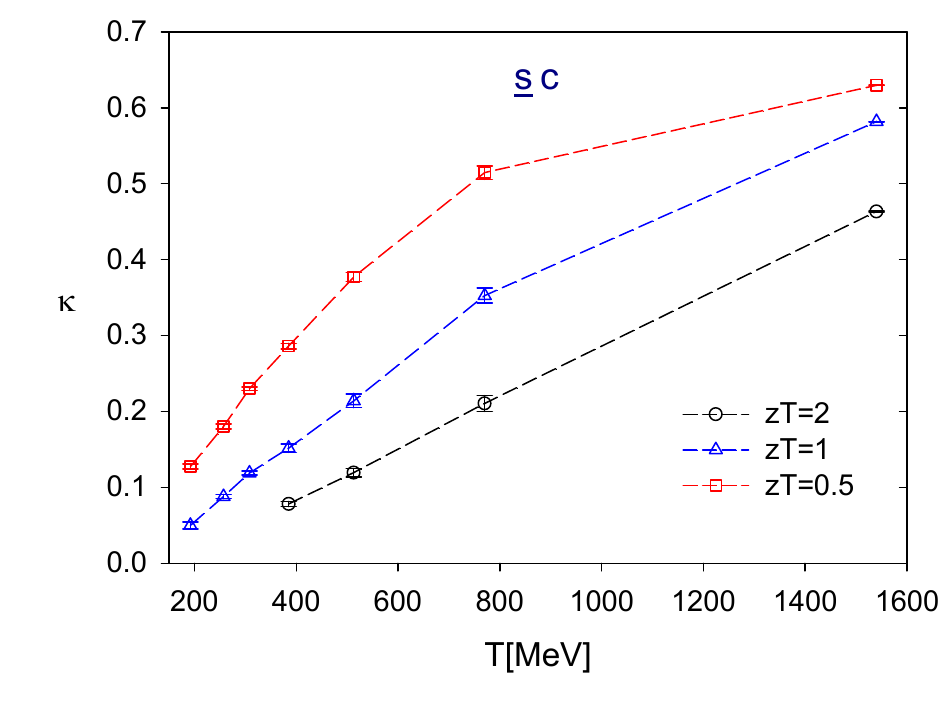}  
\\
  \includegraphics[width=8.0cm,clip=true]{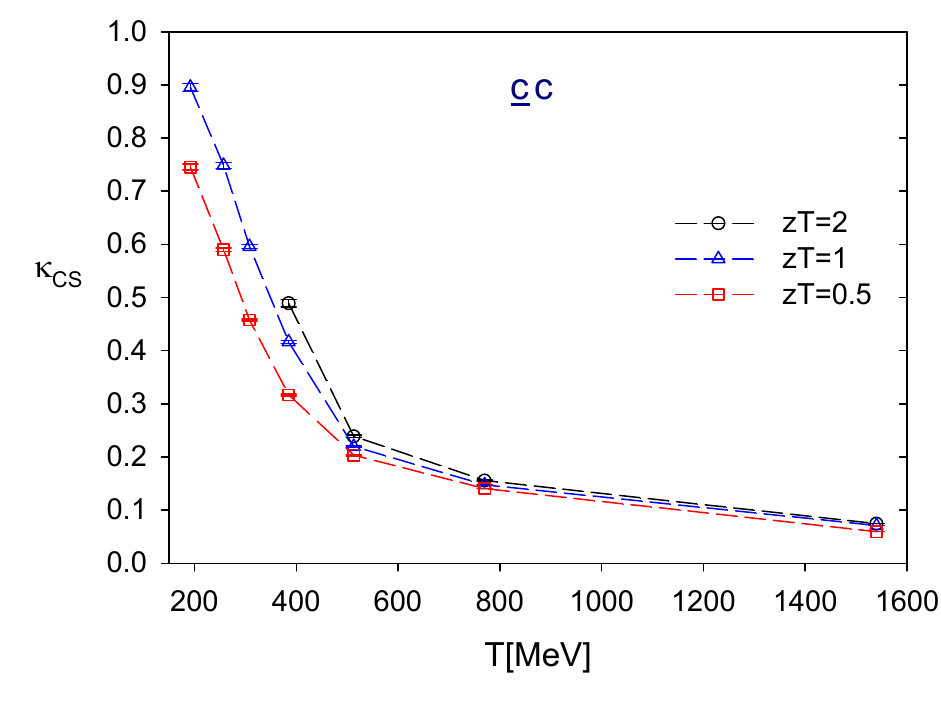}
&
  \includegraphics[width=8.0cm,clip=true]{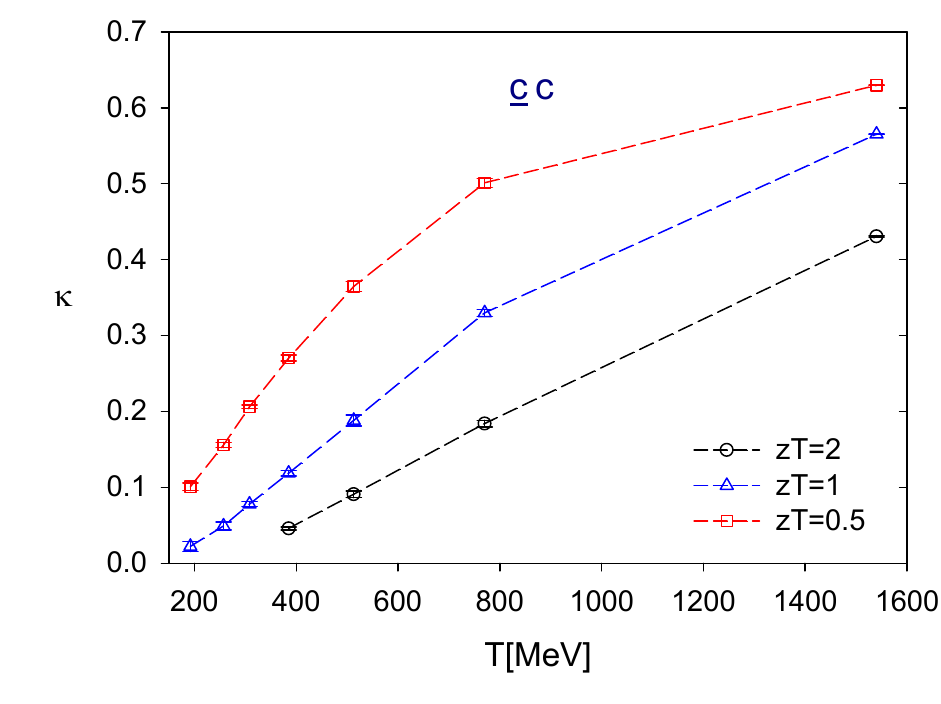}  
\end{tabular}
    \caption{
    The $SU(2)_{CS}$ symmetry breaking and fading parameters ($\kappa_{AT}$, $\kappa$) 
    at $zT = (0.5, 1, 2)$, for flavor combinations ($\bar u c$, $\bar s c$, $\bar c c $), 
    and $T \sim 193$--$1540$~MeV.  
  }
\label{fig:kCS_k_heavy}
\end{figure*}

\begin{figure*}[tbp]
  \centering
\begin{tabular}{@{}c@{}c@{}}
  \includegraphics[width=8.0cm,clip=true]{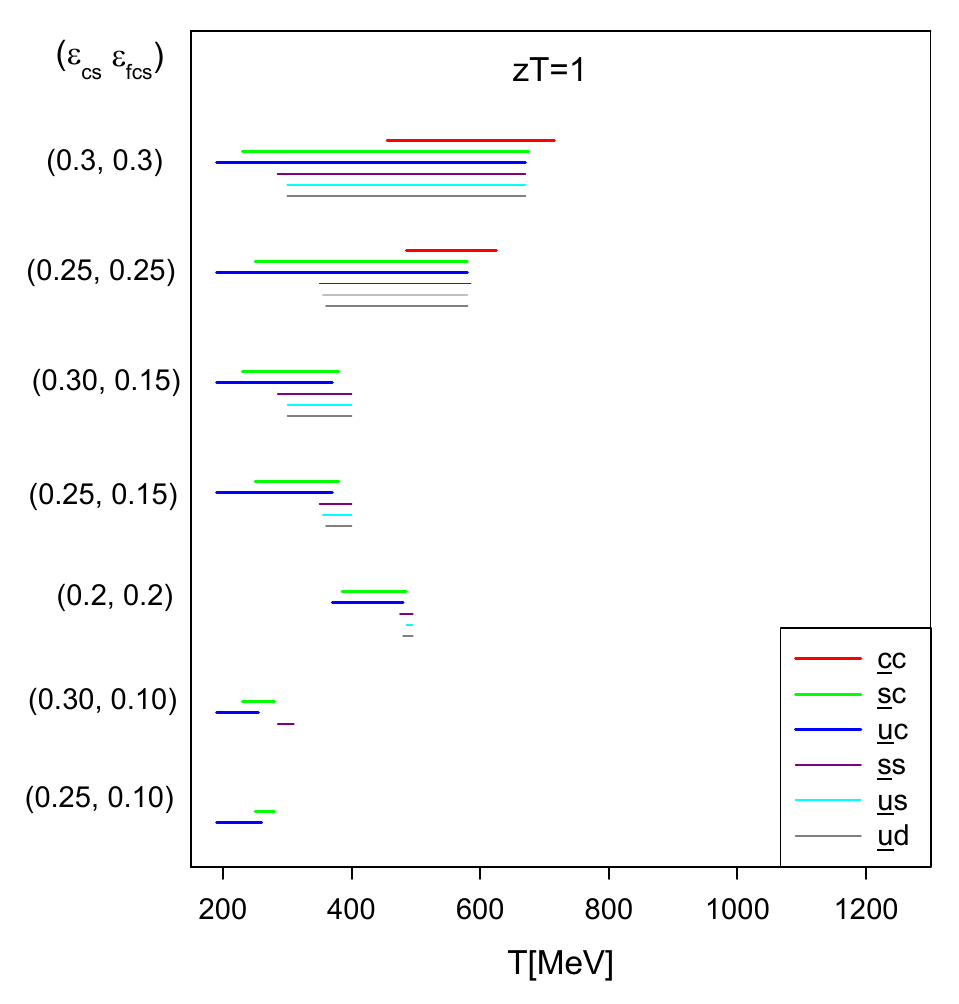}
&
  \includegraphics[width=8.0cm,clip=true]{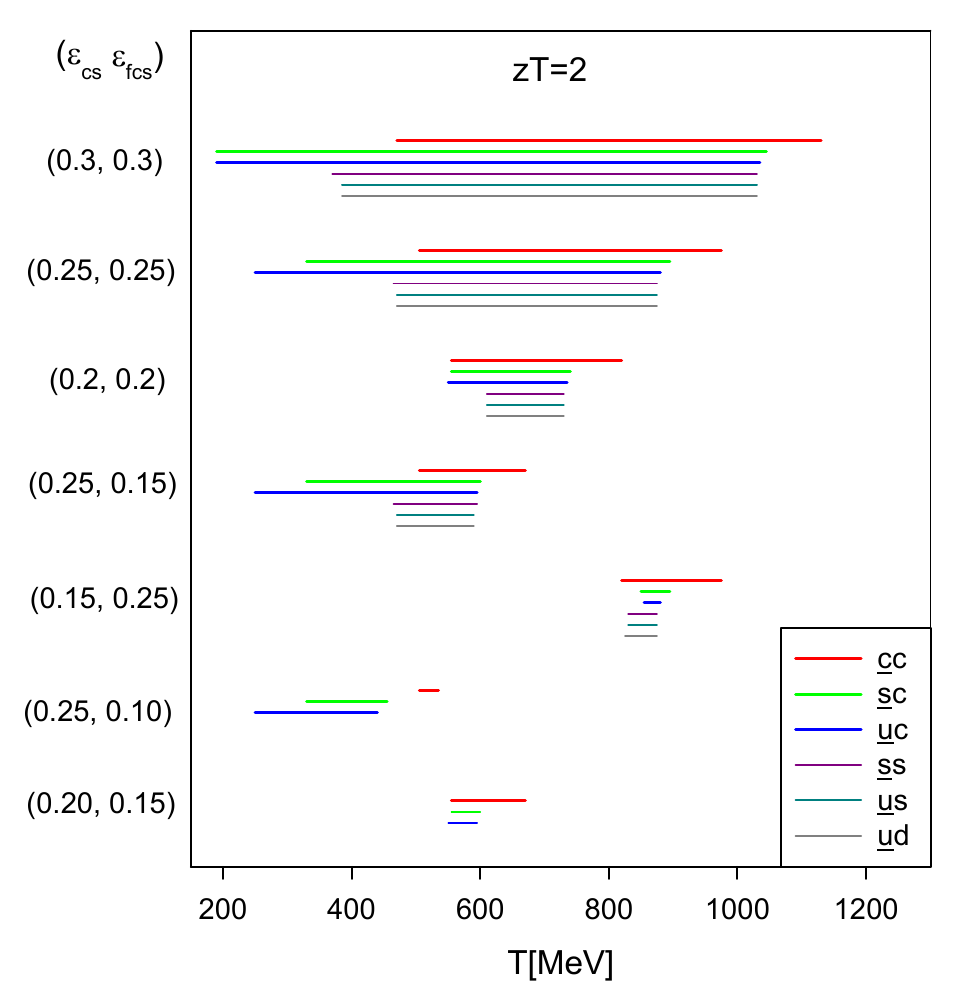}  
\end{tabular}
  \caption{
    The windows of $T$  
    satisfying the criterion (\ref{eq:SU2_CS_crit_z}) 
    for the $SU(2)_{CS}$ symmetry are plotted for six flavor contents and  
    a range of $(\epsilon_{cs}, \epsilon_{fcs})$, 
    according to the data in Tables \ref{tab:TCS_all_zT10000} and \ref{tab:TCS_all_zT20000}.    
  }
\label{fig:TCS_zT12}
\end{figure*}

Next we study the approximate $SU(2)_{CS}$ symmetry 
of $N_f=2+1+1$ lattice QCD at the physical point.
We use the criterion (\ref{eq:SU2_CS_crit_z}) to determine 
the window of $T$ of the approximate $SU(2)_{CS}$ symmetry,  
for six flavor combinations.  
To this end, we collect the data of $SU(2)_{CS}$ symmetry-breaking parameter 
$\kappa_{CS} = \max(\kappa_{AT}, \kappa_{TX})$ and symmetry-fading parameter $\kappa$ 
at $zT=(0.5, 1, 2)$, and plot them as a function of $T$,  
in Figs. \ref{fig:kCS_k_light} and \ref{fig:kCS_k_heavy},  
for light mesons $(\bar u d, \bar u s, \bar s s)$ and heavy mesons $(\bar u c, \bar s c, \bar c c)$ 
respectively. In general, for any flavor content, at fixed $zT$, 
$\kappa_{CS}$ is a monotonic decreasing function of $ T $, 
while $\kappa$ is a monotonic increasing function of $T$. 
Thus, for any $\epsilon_{cs}$ and $\epsilon_{fcs}$, 
the window of $T$ satisfying the criterion (\ref{eq:SU2_CS_crit_z}) can be determined. 
Note that, if $\epsilon_{cs}$ or $\epsilon_{fcs}$ becomes too small, 
the window of $T$ would shrink to zero (null). 
Using linear interpolation and extrapolation of the data points in  
Figs. \ref{fig:kCS_k_light} and \ref{fig:kCS_k_heavy},  
we obtain the results of $T$ window in Tables \ref{tab:TCS_all_zT10000} and \ref{tab:TCS_all_zT20000} 
at $zT=(1, 2)$, respectively, each for six flavor combinations, and for all combinations 
of $\epsilon_{cs} $ and $\epsilon_{fcs}$ sampling from $(0.1, 0.15, 0.20, 0.25, 0.30)$.
For visual comparison, we plot the windows of $T$ in Fig. \ref{fig:TCS_zT12}, 
for a range of values of $(\epsilon_{cs}, \epsilon_{fcs})$ from large to small ones. 
Tables \ref{tab:TCS_all_zT10000} and \ref{tab:TCS_all_zT20000} and Fig. \ref{fig:TCS_zT12} are
the first results of lattice QCD.  

It is interesting to see that the $T$ windows of the approximate $SU(2)_{CS}$ symmetry 
are dominated by the channels of heavy vector mesons of ($\bar u c$, $\bar s c$, $\bar c c$). 
As the precision of $SU(2)_{CS}$ symmetry gets higher with smaller $\epsilon_{cs}$ or $\epsilon_{fcs}$, 
the $T$ windows of the light vector mesons ($\bar u d$, $\bar u s$, $\bar s s$) 
shrink to zero, only those of heavy vector mesons survive. 
This suggests that the most attractive vector meson channels to detect
the emergence of approximate $SU(2)_{CS}$ symmetry are in the  
($\bar u c$, $\bar s c$, $\bar c c$) sectors, which may have phenomenological implications 
to the observation of the approximate $SU(2)_{CS}$ symmetry in 
relativistic heavy ion collision experiments such as those at LHC and RHIC. 
Moreover, the results of Tables \ref{tab:TCS_all_zT10000} and \ref{tab:TCS_all_zT20000} 
and Fig. \ref{fig:TCS_zT12} also suggest that the hadronlike objects, in particular, 
in the channels of vector mesons with $c$ quark,  
are likely to be predominantly bound by the chromoelectric interactions into color singlets  
at the temperatures inside their $T$ windows of the approximate $SU(2)_{CS}$ symmetry,   
since the noninteracting theory with free quarks does not possess the $SU(2)_{CS}$ symmetry at all.


%
%
\begin{table*}[tbp]
\begin{center}
\caption{The approximate ranges of $T$ satisfying the criterion (\ref{eq:SU2_CS_crit_z}) at $zT=1$ 
         for six flavor contents. 
         The table lists all nonzero windows of $T$ for all possible combinations  
         of $\epsilon_{cs}$ and $\epsilon_{fcs}$ sampling from $(0.1, 0.15, 0.20, 0.25, 0.30)$.
         Each $T$ window is in units of MeV, with uncertainties $\pm 5$~MeV on both ends of the window.
}
\setlength{\tabcolsep}{4pt}
\vspace{2mm}
\begin{tabular}{|cc|cccccc|}
\hline
    $\epsilon_{cs}$
  & $\epsilon_{fcs}$
  & $\bar u d$
  & $\bar u s$ 
  & $\bar s s$ 
  & $\bar u c$ 
  & $\bar s c$ 
  & $\bar c c$ \\
\hline
\hline
0.30 & 0.30 &   300--670   &  300--670  &  285--675   & 190--670   &  230--675  &  455--715  \\
0.30 & 0.25 &   300--580   &  300--580  &  285--585   & 190--580   &  230--580  &  455--625  \\
0.25 & 0.30 &   360--670   &  355--670  &  350--675   & 190--670   &  250--675  &  485--715  \\
0.25 & 0.25 &   360--580   &  355--580  &  350--585   & 190--580   &  250--580  &  485--625  \\
0.30 & 0.20 &   300--495   &  300--495  &  285--495   & 190--480   &  230--485  &  455--530  \\
0.20 & 0.30 &   480--670   &  485--670  &  475--675   & 370--670   &  385--675  &  535--715  \\
0.25 & 0.20 &   360--495   &  355--495  &  350--495   & 190--480   &  250--485  &  485--530  \\
0.20 & 0.25 &   480--580   &  485--580  &  475--585   & 370--580   &  385--580  &  535--625  \\
0.30 & 0.15 &   300--400   &  300--400  &  285--400   & 190--370   &  230--380  &  NULL     \\
0.25 & 0.15 &   360--400   &  355--400  &  350--400   & 190--370   &  250--380  &  NULL     \\
0.20 & 0.20 &   480--495   &  485--495  &  475--495   & 370--480   &  385--485  &  NULL     \\
0.30 & 0.10 &   NULL      &  NULL     &  285--310   & 190--260   &  230--280  &  NULL     \\
0.25 & 0.10 &   NULL      &  NULL     &  NULL      & 190--260   &  250--280  &  NULL     \\ 
\hline
\end{tabular}
\label{tab:TCS_all_zT10000}
\end{center}
\end{table*}

%
%
\begin{table*}[tbp]
\begin{center}
\caption{The approximate ranges of $T$ satisfying the criterion (\ref{eq:SU2_CS_crit_z}) at $zT=2$ 
         for six flavor contents. 
         The table lists all nonzero windows of $T$ for all possible combinations  
         of $\epsilon_{cs}$ and $\epsilon_{fcs}$ sampling from $(0.1, 0.15, 0.20, 0.25, 0.30)$.
         Each $T$ window is in units of MeV, with uncertainties $\pm 5$~MeV on both ends of the window.
}
\setlength{\tabcolsep}{4pt}
\vspace{2mm}
\begin{tabular}{|cc|cccccc|}
\hline
    $\epsilon_{cs}$
  & $\epsilon_{fcs}$
  & $\bar u d$
  & $\bar u s$ 
  & $\bar s s$ 
  & $\bar u c$ 
  & $\bar s c$ 
  & $\bar c c$ \\
\hline
\hline
0.30 & 0.30 &   385--1030  &  385--1030 &  370--1030  & 190--1035  &  190--1045 &  470--1130  \\
0.30 & 0.25 &   385--875   &  385--875  &  370--875   & 190--880   &  190--895  &  470--975   \\
0.30 & 0.20 &   385--730   &  385--730  &  370--730   & 190--735   &  190--740  &  470--820   \\
0.30 & 0.15 &   385--590   &  385--590  &  370--595   & 190--595   &  190--600  &  470--670   \\
0.30 & 0.10 &   385--455   &  385--455  &  370--460   & 190--440   &  190--455  &  470--535   \\
0.25 & 0.30 &   470--1030  &  470--1030 &  465--1030  & 250--1035  &  330--1045 &  505--1130  \\
0.25 & 0.25 &   470--875   &  470--875  &  465--875   & 250--880   &  330--895  &  505--975   \\
0.25 & 0.20 &   470--730   &  470--730  &  465--730   & 250--735   &  330--740  &  505--820   \\
0.25 & 0.15 &   470--590   &  470--590  &  465--595   & 250--595   &  330--600  &  505--670   \\
0.20 & 0.30 &   610--1030  &  610--1030 &  610--1030  & 550--1035  &  555--1045 &  555--1130  \\
0.20 & 0.25 &   610--875   &  610--875  &  610--875   & 550--880   &  555--895  &  555--975   \\
0.20 & 0.20 &   610--730   &  610--730  &  610--730   & 550--735   &  555--740  &  555--820   \\
0.15 & 0.30 &   825--1030  &  830--1030 &  830--1030  & 855--1035  &  850--1045 &  820--1130  \\
0.15 & 0.25 &   825--875   &  830--875  &  830--875   & 855--880   &  850--895  &  820--975   \\
0.25 & 0.10 &   NULL      &  NULL     &  NULL      & 250--440   &  330--455  &  505--535   \\
0.20 & 0.15 &   NULL      &  NULL     &  NULL      & 550--595   &  555--600  &  555--670   \\
0.15 & 0.20 &   NULL      &  NULL     &  NULL      & NULL      &  NULL     &  NULL      \\
\hline
\end{tabular}
\label{tab:TCS_all_zT20000}
\end{center}
\end{table*}

\section{Concluding remarks}
\label{conclusions}

In this study, we have generated seven gauge ensembles of $N_f=2+1+1$ lattice QCD with $(u/d,s,c)$ optimal
domain-wall quarks at the physical point, on the $32^3 \times (16, 12, 10, 8, 6, 4, 2)$ lattices
with lattice spacing $a \sim 0.064$~fm,
for seven temperatures in the range of 190-1540 MeV, as summarized in Table \ref{tab:7_ensembles}.
Our plan is to complete 21 gauge ensembles with
three lattice spacings $a \sim (0.064, 0.069, 0.075)$~fm, which can be used to extract
the continuum limit of the observables, for temperatures in the range of 160--1540 MeV.

Using seven gauge ensembles with $a \sim 0.064$~fm, we computed the meson $z$ correlators for 
the complete set of Dirac bilinears (scalar, pseudoscalar, vector, axial vector, tensor vector,
and axial-tensor vector), and each for six combinations of quark flavors
($\bar u d$, $\bar u s$, $\bar s s$, $\bar u c$, $\bar s c$, $\bar c c$).
Then we use the criteria (\ref{eq:SU2_crit_z}) and (\ref{eq:U1_TX_crit_z})  
to determine $T_c$ and $T_1$ for each flavor combination, and obtain the 
hierarchy of restoration of chiral symmetries, in the order of
\BAN
&& T_c^{\bar u d} < T_c^{\bar u s} < T_c^{\bar u c} < T_c^{\bar s s} < T_c^{\bar s c} < T_c^{\bar c c}, \\ 
&& T_1^{\bar u d} < T_1^{\bar u s} < T_1^{\bar u c} < T_1^{\bar s s} < T_1^{\bar s c} < T_1^{\bar c c}. 
\EAN 
These are the first results in lattice QCD. They immediately give 
the hierarchical restoration of chiral symmetries in $N_f=2+1+1$ QCD, i.e.,  
from the restoration of $SU(2)_L \times SU(2)_R \times U(1)_A$ chiral symmetry 
of $(u, d)$ quarks at $ T_{c1}^{\bar u d}$ 
to the restoration of $SU(3)_L \times SU(3)_R \times U(1)_A$ chiral symmetry 
of $(u, d, s)$ quarks at $ T_{c1}^{\bar s s} > T_{c1}^{\bar u d}$, then to  
the restoration of $SU(4)_L \times SU(4)_R \times U(1)_A$ chiral symmetry 
of $(u, d, s, c)$ quarks at $ T_{c1}^{\bar c c} > T_{c1}^{\bar s s}$. 

Obviously, the hierarchical restoration of chiral symmetries is expected to be realized in QCD with 
physical $(u, d, s, c, b)$ quarks, adding the restoration of 
$SU(5)_L \times SU(5)_R \times U(1)_A$ chiral symmetry 
of $(u, d, s, c, b)$ quarks at $ T_{c1}^{\bar b b} > T_{c1}^{\bar c c}$. 
We will study the restoration of chiral symmetries in $N_f = 2+1+1+1$ lattice QCD with 
physical $s$, $c$, and $b$ quarks, but unphysical $u/d$ quarks 
with $M_{\pi^{\pm}} \sim 700$~MeV \cite{Chiu:2020tml}, for eight ensembles with temperatures 
in the range of 300--3250 MeV, on $40^3 \times (20, 16, 12, 10, 8, 6, 4, 2)$ lattices
with lattice spacing $a \sim 0.03$~fm. 

In this work, we observe that for mesons with quark contents ($\bar q q$, $\bar q Q$, $\bar Q Q$) 
and $m_q < m_Q $ in $N_f=2+1+1$ QCD, the temperatures of the restoration 
of $SU(2)_L \times SU(2)_R \times U(1)_A $ chiral symmetry in these three sectors 
satisfy the hierarchy:  
\BAN
T_{c1}^{\bar q q} \le T_{c1}^{\bar q Q} \le T_{c1}^{\bar Q Q}, \hspace{4mm} m_q < m_Q.
\EAN
However, we do not know the hierarchy of $T_{c1}$ for mesons 
involving more than two quarks. For example, for three quarks $(q_1, q_2, q_3)$ with $m_1 < m_2 < m_3$,   
we do not know whether $T_{c1}^{\bar q_1 q_3} < T_{c1}^{\bar q_2 q_2}$ or vice versa. 
In reality, for physical $(u, s, c)$ quarks, we observe that 
\bea
\label{eq:Tc1_usc}
T_{c1}^{\bar q_1 q_3} < T_{c1}^{\bar q_2 q_2}.   
\eea 
Yet, in general, it is unclear to what extent (\ref{eq:Tc1_usc}) depends on the ratios of quark masses.  

One of the phenomenological implications of the hierarchical restoration of chiral symmetries is 
the pattern of hadron dissolution at high temperatures, which leads to the hierarchical dissolution 
of hadrons, and the hierarchical suppression of hadrons in the quark-gluon plasma. 
Theoretically, the meson with quark content $\bar q Q $ dissolves 
completely as $\bar q$ and $Q$ become deconfined, i.e., when the screening mass 
of $\bar q Q$ is larger than its counterpart in the noninteracting theory with free quarks of the 
same masses. Presumably, $ m_{\text{scr}}^{\bar q Q} \ge m_{\text{scr}}^{\bar q Q(\text{free})} $   
happens at the temperature $T_d^{\bar q Q} \gtrsim T_{c1}^{\bar q Q}$, 
after the $SU(2)_L \times SU(2)_R \times U(1)_A $ chiral symmetry of $\bar q \Gamma Q$ 
has been effectively restored. Thus, for $N_f=2+1+1$ lattice QCD at the physical point,   
one expects that the hierarchy of dissolution of mesons is exactly the same as that of 
the restoration of chiral symmetries (\ref{eq:Tc1_order}), i.e.,    
\bea
\label{eq:Td_order}
T_{d}^{\bar u d} < T_{d}^{\bar u s} < T_{d}^{\bar u c} < T_{d}^{\bar s s} < T_{d}^{\bar s c} 
< T_{d}^{\bar c c}.   
\eea 
This leads to the hierarchical suppression of mesons in quark-gluon plasma, which could be 
observed in the relativistic heavy ion collision experiments such as those at LHC and RHIC. 
Here we recall the seminal paper by Matusi and Satz \cite{Matsui:1986dk}, 
in which it was proposed that the dissolution of J/$\psi$ in the quark-gluon plasma would 
result in the suppression of their production in heavy ion collision experiments. 
To investigate whether (\ref{eq:Td_order}) holds in $N_f=2+1+1$ lattice QCD at the physical point 
is beyond the scope of this paper. 

Besides the meson $z$ correlators, the restoration of chiral symmetry in high temperature QCD
can also be observed in the baryon $z$ correlators \cite{Detar:1987kae}.
For QCD with $N_f=2(3)$ massless quarks, the chiral multiplets of baryon operators have been obtained 
by the group theoretical methods, see e.g., Ref. \cite{Cohen:1996sb} and the references therein.
Now, for QCD with physical $(u, d, s, c, b)$ quarks, with quark masses ranging from a few MeV
to a few GeV, we expect that the hierarchical restoration of chiral symmetries 
can be observed from the degeneracies of $z$ correlators of baryon chiral multiplets. 
It would be interesting to see whether the hierarchy of chiral symmetry restoration 
from the baryon $z$ correlators is compatible with that from the meson $z$ correlators.

About the approximate $SU(2)_{CS}$ chiral spin symmetry, it is interesting to see that 
the $T$ windows satisfying the criterion (\ref{eq:SU2_CS_crit_z}) are dominated by the 
the channels of heavy vector mesons with flavor contents $(\bar u c, \bar s c, \bar c c)$,  
as shown in Tables \ref{tab:TCS_all_zT10000} and \ref{tab:TCS_all_zT20000} and Fig. \ref{fig:TCS_zT12}. 
These are the first results of lattice QCD.  
They suggest that the hadronlike objects, in particular, 
in the channels of vector mesons with $c$ quark, at the temperatures inside their $T$ windows,  
are likely to be predominantly bound by the chromoelectric interactions into color singlets, 
since the noninteracting theory with free quarks does not possess the $SU(2)_{CS}$ symmetry at all.
Moreover, they provide hints to look for the approximate emergent $SU(2)_{CS}$ symmetry 
in the relativistic heavy ion collision experiments such as those at LHC and RHIC, e.g.,  
to focus on the channels of vector mesons with $c$ quark. 

Obviously, it is interesting to find out the $T$ windows of the approximate 
$SU(2)_{CS}$ symmetry for heavy vector mesons involving the $b$ quark, 
in lattice QCD with $(u, d, s, c, b)$ quarks.
To this end, we will investigate the approximate $SU(2)_{CS}$ symmetry in $N_f = 2+1+1+1$ lattice QCD 
with physical $s$, $c$, and $b$ quarks, but unphysical $u/d$ quarks 
with $M_{\pi^{\pm}} \sim 700$~MeV \cite{Chiu:2020tml}, for eight ensembles with temperatures 
in the range of 300--3250 MeV, on $40^3 \times (20, 16, 12, 10, 8, 6, 4, 2)$ lattices  
with lattice spacing $a \sim 0.03$~fm. 

Finally, it is necessary to clarify the nature of these mesonlike objects in 
the $J=1$ channels (i.e., $V_k$, $A_k$, $T_k$, and $X_k$) which are relevant to the 
approximate $SU(2)_{CS}$ symmetry, for all six flavor contents 
$(\bar u d, \bar u s, \bar s s, \bar u c, \bar s c, \bar c c)$,  
by examining how their spectral functions evolve as $T$ is increased.
If bound-state peaks exist in the $T$ windows of the approximate $SU(2)_{CS}$ symmetry,  
and also the widths of these peaks gradually broaden as $T$ is increased, 
and the peaks eventually disappear as $T$ is increased above the windows, 
then the degrees of freedom in these mesonlike objects can be asserted to be color-singlet (melting)
mesons rather than deconfined quarks and gluons.  
To this end, one may consider the approach of 
Refs. \cite{Bros:1992ey,Bros:2001zs,Lowdon:2022xcl} for $J=0$ mesons, and generalize it to $J=1$ mesons. 
Also, the spatial $z$ correlators of vector mesons have to be evaluated to high precision 
even at large distances such that the damping factor $D_{m,\beta}(\vec{u})$ \cite{Bros:2001zs}
of each $J=1$ meson channel can be extracted reliably.
The proposed prescription in Ref. \cite{Chiu:2023hnm} provides a viable way to attain this goal.

\textit{Note added.}---This version corresponds to the published version
[Phys. Rev. D \textbf{110}, 014502 (2024)], with the corrections of the Erratum incorporated:
the six data points of the $\bar s c$ sector at $T \simeq 770$~MeV in Figs.~\ref{fig:kVA_z} and \ref{fig:kTX_z},
and the entries of $T_c^{\bar s c}$ and $T_1^{\bar s c}$ for $\epsilon_{VA}=\epsilon_{TX}=0.01$
in Tables~\ref{tab:Tc_all} and \ref{tab:T1_all}. All other results and conclusions are unchanged.

\section*{Acknowledgments}

The author is grateful to Academia Sinica Grid Computing Center
and National Center for High Performance Computing for the computer time and facilities.
This work is supported by the National Science and Technology Council
(Grants No.~108-2112-M-003-005, No.~109-2112-M-003-006, and No.~110-2112-M-003-009),
and Academia Sinica Grid Computing Centre (Grant No.~AS-CFII-112-103). 
This paper was completed while visiting Nuclear Theory Program of Nuclear Science Division 
in Lawrence Berkeley National Laboratory. 
The author thanks the members of Nuclear Theory Program for kind hospitality 
and interesting discussions.


\end{document}